\documentclass[]{elsarticle}
\usepackage{fullpage}

\usepackage{hyperref}
\usepackage{lineno}
\modulolinenumbers[5]

\usepackage{natbib}
\usepackage{graphicx}

\begin{document}
\newcommand{\secp}{\mbox{\rlap{.}$''$}}
\newcommand{\juc}{\mbox{$J$=1$-$0}}
\newcommand{\jdu}{\mbox{$J$=2$-$1}}
\newcommand{\jtd}{\mbox{$J$=3$-$2}}
\newcommand{\jcc}{\mbox{$J$=5$-$4}}
\newcommand{\jsc}{\mbox{$J$=6$-$5}}
\newcommand{\jdn}{\mbox{$J$=10$-$9}}
\newcommand{\jdsq}{\mbox{$J$=16$-$15}}
\newcommand{\doceCO}{\rm $^{12}$CO}
\newcommand{\treceCO}{\rm $^{13}$CO}
\newcommand{\mloss}{\mbox{$\dot{M}$}}
\newcommand{\Msun}{\mbox{$M_{\odot}$}}
\newcommand{\Lsun}{\mbox{$L_{\odot}$}}
\newcommand{\ms}{\mbox{$M_{\odot}$}}
\newcommand{\ls}{\mbox{$L_{\odot}$}}
\newcommand{\my}{\mbox{$M_{\odot}$~yr$^{-1}$}}
\newcommand{\lsim}{\raisebox{-.4ex}{$\stackrel{\sf <}{\scriptstyle\sf \sim}$}}
\newcommand{\gsim}{\raisebox{-.4ex}{$\stackrel{\sf >}{\scriptstyle\sf \sim}$}}
\newcommand{\aprop}{\raisebox{-.3ex}{$\stackrel{\propto}{\scriptstyle\sf \sim}$}}
\newcommand{\s}{\mbox{$''$}}
\newcommand{\h}{$^{\rm h}$}
\newcommand{\m}{$^{\rm m}$}
\sf 
\newcommand{\ai}{\mbox{\'{\i}}}
\newcommand{\n}{\~n}
\newcommand{\kms}{\mbox{km~s$^{\sf -1}$}}

\def\mnras{MNRAS}
\def\apj{ApJ}
\def\apss{ApSS}
\def\apjl{ApJL}
\def\apjs{ApJS}
\def\aj{AJ}
\def\aap{A\&A}
\def\pasj{PASJ}
\def\nat{Nature}

\def\wat{H$_{2}$O}
\def\meth{CH$_{3}$OH}
\def\nh3{NH$_{3}$}
\def\kms{km~s$^{-1}$}
\def\kmsy{km~s$^{-1}$~yr$^{-1}$}
\def\kmo{km~s$^{-1}$~mas$^{-1}$}
\def\kmt{km~s$^{-1}$~mas$^{-2}$}
\def\Vlsr{$V_{\rm LSR}$}
\def\Jyb{Jy~beam$^{-1}$}
\def\Vsys{$V_{\rm sys}$}
\def\mum{$\mu$m}
\def\Jcal{J1834$-$0301}
\def\HII{H{\sc ii}}
\newcommand{\msyr}{$M_{\odot}$~yr$^{-1}$}
\newcommand{\pas}{$\rlap{.}^{\prime\prime}$}
\newcommand{\degree}{$^{\circ}$}
\def\farcs{\hbox{$.\!\!^{\prime\prime}$}}
\def\arcsec{\hbox{$^{\prime\prime}$}}
\def\fs{\hbox{$.\!\!^{s}$}}
\def\la{\mathrel{\mathchoice {\vcenter{\offinterlineskip\halign{\hfil $\displaystyle##$\hfil\cr<\cr\sim\cr}}}
{\vcenter{\offinterlineskip\halign{\hfil$\textstyle##$\hfil\cr <\cr\sim\cr}}}
{\vcenter{\offinterlineskip\halign{\hfil$\scriptstyle##$\hfil\cr <\cr\sim\cr}}}
{\vcenter{\offinterlineskip\halign{\hfil$\scriptscriptstyle##$\hfil\cr <\cr\sim\cr}}}}}
\def\ga{\mathrel{\mathchoice {\vcenter{\offinterlineskip\halign{\hfil $\displaystyle##$\hfil\cr>\cr\sim\cr}}}
{\vcenter{\offinterlineskip\halign{\hfil$\textstyle##$\hfil\cr >\cr\sim\cr}}}
{\vcenter{\offinterlineskip\halign{\hfil$\scriptstyle##$\hfil\cr >\cr\sim\cr}}}
{\vcenter{\offinterlineskip\halign{\hfil$\scriptscriptstyle##$\hfil\cr >\cr\sim\cr}}}}}

\newcommand{\et}    {et al.}
\newcommand{\eg}    {e.\,g.,}
\newcommand{\ie}    {i.\,e.,}
\newcommand{\supa}  {$^\mathrm{a}$}
\newcommand{\supb}  {$^\mathrm{b}$}
\newcommand{\supc}  {$^\mathrm{c}$}
\newcommand{\supd}  {$^\mathrm{d}$}
\newcommand{\supe}  {$^\mathrm{e}$}
\newcommand{\supf}  {$^\mathrm{f}$}
\newcommand{\supg}  {$^\mathrm{g}$}
\newcommand{\phnn}  {\phantom{00}}
\newcommand{\phnnn}  {\phantom{000}}
\newcommand{\phpnn} {\phantom{.00}}

\begin{frontmatter}

\title{The Science Case for Simultaneous mm-Wavelength Receivers in Radio Astronomy}

   \author[iuwa]{Richard {Dodson}}
   \author[iuwa,cass,oan]{Mar\'{\i}a J. {Rioja}}
   \author[kasi,ust]{Taehyun {Jung}}
   \author[iaas]{Jos\'e Luis {Gom\'ez}}
   \author[oan]{Valentin {Bujarrabal}}
   \author[iaai]{Luca {Moscadilli}}
   \author[icur]{James C. A. {Miller-Jones}}
   \author[alb]{Alexandra J. Tetarenko}
   \author[alb]{Gregory R. Sivakoff}

 \address[iuwa]{
 International Centre for Radio Astronomy Research, The University of Western Australia, 35 Stirling Hwy, Western Australia}
 \address[cass]{CSIRO Astronomy and Space Science, 26 Dick Perry Avenue, Kensington WA 6151, Australia }
 \address[oan]{Observatorio Astron\'omico Nacional (IGN), Alfonso XII, 3 y 5, 28014 Madrid, Spain}
 \address[kasi]{Korea Astronomy and Space Science Institute 776, Daedeokdae-ro, Yuseong-gu, Daejeon, 34055, Republic of Korea}
 \address[ust]{University of Science and Technology, 217, Gajeong-ro, Yuseong-gu, Daejeon, 34113, Korea}
 \address[iaas]{Instituto de Astrof\'{\i}sica de Andaluc\'{\i}a-CSIC, Glorieta de la Astronom\'{\i}a s/n, E-18008 Granada, Spain}
 \address[iaai]{INAF-Osservatorio Astrofisico di Arcetri, Largo E. Fermi 5, 50125, Firenze, Italy}
 \address[icur]{International Centre for Radio Astronomy Research, Curtin University, GPO Box U1987, Perth, WA 6845, Australia}
 \address[alb]{Department of Physics, CCIS 4-183, University of Alberta, Edmonton, AB T6G 2E1, Canada}


 
  \begin{abstract}
   This review arose from the European Radio Astronomy Technical Forum (ERATec) meeting held in Firenze, October 2015, and aims to highlight the breadth and depth of the high-impact science that will be aided and assisted by the use of simultaneous mm-wavelength receivers. 
   
   \noindent Recent results and opportunities are presented and discussed from the fields of: continuum VLBI (observations of weak sources, astrometry, observations of AGN cores in spectral index and Faraday rotation), spectral line VLBI (observations of evolved stars and massive star-forming regions) and time domain observations of the flux variations arising in the compact jets of X-ray binaries. 
  
   \noindent Our survey brings together a large range of important science applications, which will greatly benefit from simultaneous observing at mm-wavelengths. Such facilities are essential to allow these  applications to become more efficient, more sensitive and more scientifically robust. In some cases without simultaneous receivers the science goals are simply unachievable. 
   Similar benefits would exist in many other high frequency astronomical fields of research.
    \end{abstract}

   \begin{keyword}
   Astronomical instrumentation, methods and techniques -- Instrumentation: interferometers -- Methods: observational -- Telescopes --
   quasars: general -- Astrometry -- quasars: jets -- 
   Stars: AGB and post-AGB -- Stars: formation -- X-rays: binaries  
   \end{keyword}

\end{frontmatter}






\section{Introduction}






Very Long Baseline Interferometry (VLBI) studies at cm wavelengths is a well-established field, with advanced technological developments and analysis techniques 
that result in superb quality images, including those of very weak  $\mu$Jy sources \citep[e.g.,][]{garret_micro} and with micro-arcsecond ($\mu$as) astrometry \citep[e.g.,][]{reid_micro}, using phase referencing ({\it hereafter} PR) techniques. 

VLBI at mm and sub-mm wavelengths ({\it hereafter} mm-VLBI) can result in the highest angular resolutions achieved in astronomy and has a unique access to emission regions that are inaccessible with any other approach or at longer wavelengths, because the compact areas of interest are often self-absorbed or scatter-broadened.
Therefore it holds the potential to increase our understanding of the physical processes in, for example, Active Galactic Nuclei (AGN) and in the vicinity of super-massive black holes, and for studies of molecular transitions at high frequencies. 

Nevertheless the scientific applications of mm-VLBI have to date been much less widespread,
the reason being that the observations become progressively more challenging as the wavelengths gets shorter because of the: limited telescope surface accuracy and efficiency, higher receiver system temperatures and lower sensitivity and shorter atmospheric coherence times. Also most compact extra-galactic sources used as cm-wave phase references are intrinsically weaker, if not resolved-out, at shorter wavelengths.
These in turn prevent the use of phase referencing calibration techniques, which are routinely used in cm-VLBI, and all benefits resulting from them, beyond $\sim$43\,GHz (with a single exception (to date) of PR at 86\,GHz by \citeauthor{porcas_02} \citeyear{porcas_02}).

Continuous development and technical improvements have led to a sustained 
increase of the high frequency threshold for VLBI observations in the last two decades \citep[e.g.,][]{krichbaum_mmvlbi}. Regular observations up to 86\,GHz are being carried out 
with well established networks such as the Very Long Baseline Array (VLBA) and Global mm-VLBI Array (GMVA), most recently with the Korean VLBI Network (KVN) up to 130\,GHz, and ad-hoc observations at the highest frequencies up to 240\,GHz with the Event Horizon Telescope (EHT) \citep{Doeleman:2008ek}.  The field of mm-VLBI will benefit from the arrival of phased-up Atacama Large Millimeter and submillimeter Array (ALMA) \citep{alma_pp,tilanus_14} for joint VLBI observations.



The benefit of multi-frequency based calibration techniques in observations at high frequencies, which are dominated by non-dispersive fast tropospheric fluctuation errors,   
has long been known. It relies on using the correctly scaled calibration derived at lower frequencies to correct the higher frequencies; 
see, for example, \cite{carilli_99,asaki_fpt_96} for connected interferometers. This has been extended to VLBI observations using the frequency agility capability (i.e. fast frequency switching) of the VLBA, initially offering an increased coherence \citep{middelberg_05} and, after further development, bona-fide astrometric measurements \citep{vlba_31,rioja_11a} at the highest frequency of 86-GHz. 
The first step is known as Frequency Phase Transfer (FPT), where the solutions for the low frequency $\nu_{\rm low}$ are applied to the high frequency $\nu_{\rm high}$, scaled by the ratio of the frequencies. 
This does not provide astrometry because of dispersive terms, such as the ionosphere, and also the inherent phase ambiguities in the phase solutions at $\nu_{\rm low}$. 
The latter is not a problem if the ratio is an integer, as then the ambiguities continue to cancel, or if the number of ambiguities are zero, which requires the  positions to be well known. 
The FPT calibration step eliminates all the common non-dispersive residual errors on the target source.
The second step provides Source Frequency Phase Referencing (SFPR) by including an additional `conventional' phase referencing step on a second source, which eliminates all other common residual errors, which are mainly dispersive and/or instrumental. This switching can be over long cycle times and large angular separations, as these terms are slow changing and have weak angular dependence.

%
The use of this calibration technique on the VLBA has not been widespread, the reason being that the data reduction was challenging because of the fast temporal variations arising from the troposphere. Nevertheless there are a few examples of such analysis \citep{rioja_11a,rioja_14,marti-vidal_16}.
The KVN has introduced a new engineering development in receiver technology \citep{kvn_optics} to address this issue, with the capability for simultaneous multi-frequency observations at four bands. Using KVN observations and multi-frequency based calibration techniques it is possible to phase reference the observations at frequencies up to 130-GHz. 
In this document we will explore some of the scientific topics and current issues that will benefit from this technical capability.

Simultaneous multi-frequency observing enhances mm-VLBI in two main areas: massively increasing the coherence time, for observing weak sources at high frequencies, and enabling accurate astrometric registration between frequency bands.
The raw increase in coherence time, even at 130-GHz, has been shown to be easily enlarged from tens of seconds to a few tens of minutes, using observations of the target source only \citep{rioja_15}. 
This increase in coherence time would be the equivalent to a three-fold increase in the dish diameter, or an increase of two orders of magnitude in the recorded bandwidth. Furthermore the latter would only apply for continuum science, as this does not assist the important spectral line science cases. The coherence time can be further increased by including the observations of a calibrator source.

Accurate astrometry is required to allow bona-fide spectral index or Faraday rotation measure investigations in continuum sources and the measurements of the frequency-dependent position of the continuum cores in extragalactic radio sources \citep{1993AJ....106..444W,Hovatta:2012jv}.
Spectral line models enable high-precision measurements of the structure and kinetics of the various components of the studied sources, using different tracers \citep[e.g.][]{dodson_17,soria-ruiz_04}, or by using absorption of continuum flux by intervening gas \citep{2005Ap&SS.295..249V,momjian2003sensitive}.  
%
VLBI is the most developed area driving the demand for simultaneous mm-wavelength receivers, but it is not the only field. Time domain radio astronomy is a growth area, and for rapidly changing sources with strong mm-wavelength emission simultaneous mm-wavelength receivers add an important new capability to aid the interpretation of observations. The study of pulsars is a well-known example and is a target of the BlackHole Cam ERA project, but the examples we review in this paper are results from the spectral evolution of the strong and rapidly changing mm emission from the jets in compact X-ray binaries. 

Therefore there is a wealth of possibilities opened up by the application of simultaneous multi-frequency mm-wavelength observing, using demonstrated methods and technology. In this paper we present some of the headline cases for continuum and spectral line VLBI and time-domain observations. 


\subsection{Current achievable sensitivities for mm-wave observatories}

We review the current status and sensitivities of facilities capable of supporting simultaneous mm-wave observing. 
VLBA is capable of switching between receivers in 10--20 seconds, depending on the pairing.
KVN was built for simultaneous observing, and this capability is now been extended to all of KaVA (the KVN and VERA Array).
Table \ref{tab:mm-fac} lists the expected baseline sensitivities\footnote{from the EVNCalc tool http://www.evlbi.org/cgi-bin/EVNcalc.pl} for these arrays with current standard recording bandwidths and observing modes (i.e. fast frequency switching or simultaneous), using the lowest frequency in the range (with typical coherence times) and the highest frequency in the range (with a typical post-FPT coherence time of 30min). Overheads for FFS are included. 
Post-SFPR coherence time is essentially infinite, theoretically providing infinite sensitivity. A typical SFPR image for a long observation (5 hours of on-source time) would be about three times more sensitive than the FPT image (and astrometrically registered).

We note that these results are for current (2017) facilities and bandwidths, and facilities are being upgraded and/or built around the world: ALMA offers a generational leap in sensitivity and maximum frequency for interferometric arrays, additionally a number of large mm-wave single dishes have been built (Yebes 40m, LMT 50m, the ALMA 12m prototypes now located in Arizona and Greenland, KVN, Metsahovi 14m, and several others proposed). Single dish antennas can be easily upgraded to host simultaneous multi-frequency receivers, as has been done at VERA and Yebes. Discussions are on-going as to whether this should be done at ALMA, largely to aid in the long baseline observing at their highest frequencies.
In addition Australia Telescope Compact Array (ATCA), ALMA and the Jansky Very Large Array (JVLA) are all capable of sub-array mode observing, which does allow for - at the very least - Frequency Phase Transfer and should allow SFPR in compact configurations. 

\begin{table}[ht]
    \centering
    \begin{tabular}{c|c||c|c}
    Facility & Freq.  & Baseline Sensitivity & FPT Baseline Sensitivity\\
     &  Range &  mJy ($\nu_{\rm low})$ & mJy ($\nu_{\rm high}$)\\
     \hline
    VLBA     &  22-86 (switched) & 4$^a$
                                 & 8$^b$\\
    KaVA     &  22-43 & 10$^c$
                      & 4$^d$\\
    KVN      &  43-130& 22$^e$
                      & 12$^f$\\
    \end{tabular}
    \caption{Baseline sensitivities for the lower observing frequency in the range and for the higher observing frequency, after Frequency Phase Transfer to increase the coherence time, for current mm-wave facilities capable of this style of observing. $^a${1/3 cycle time of 2min at 2Gbps} $^b${1/3 cycle time of 30min at 2Gbps}, $^c${2min at 0.5Gbps}, $^d$ {30min at 0.5Gbps}, $^e$ {1min at 0.5Gbps}, $f$ {30min at 0.5Gbps}.}
    \label{tab:mm-fac}
\end{table}

\section{Continuum Studies}\label{sec:cont}
   \subsection{Weak Sources}\label{sec:weak}

The high angular resolution provided, combined with the opacity effect in high electron density regions that smother the radiation at cm-frequencies, require us to use mm-VLBI to explore the details of the innermost regions of galaxies and AGNs, by reaching a spatial resolution of a few tens of Schwarzschild radii (R$_{\rm s}$). The Galactic Centre Sgr A*, for example, has a R$_{\rm s}$ of 10 $\mu$-as \citep{sgra_size}, which can only be seen by overcoming the synchrotron self-absorption barriers. Although the physical mechanisms of energetic processes, such as jet formation and acceleration in AGNs have been widely studied at cm-wavelengths, our understanding at mm-wavelengths is still limited due to several considerations, such as the number of available sources, sensitivity, suitable telescopes etc.. 
Since radio sources become generally weaker as the observing frequency increases, most radio sources are difficult to detect and image using mm-VLBI, as compared to cm-wavelengths. Most VLBI surveys of continuum sources have therefore been conducted at lower frequencies. In particular, the advent of the VLBA greatly facilitated large VLBI surveys addressing various statistical properties of AGNs; for example, their structure, compactness, brightness temperatures and spectral indices, gamma-ray connection, polarimetry and cosmological evolution (see Table \ref{tab:weak}).
However, weak sources can always be imaged as long as a calibration scheme to compensate for the atmospheric disturbances is included in the experiment. This compensation does not need to be as perfect as for phase referencing, where all errors are corrected, for the simple detection of weak sources.  For this, we only require an increase in the coherence time, which, combined with the large bandwidths of modern recording systems, is sufficient for this science case. For this we use FPT to remove all fast changing non-dispersive contributions \citep{jung_11,rioja_11a}.

As shown in Table \ref{tab:weak}, there is also a remarkable difference in the number of calibrator sources between cm- and mm-wavelengths. 
In the most recent update on the radio fundamental catalog (RFC 2016c) by \citet{petrov_16}, a total of 11,426 compact radio sources has been compiled from S/C/X/K-band observations, and of those more than 3,400 sources are listed in the VLBA calibrator surveys (e.g. VCS1 to VCS6, \citep{petrov_08} and references therein). In comparison only 121 sources are catalogued at 3 mm \citep{sslee_86}, and these are limited to the very brightest radio sources. We note, however, that ALMA is undertaking a number of surveys \citep{Fomalont_14,2016SPIE.9906E..5UA} at 3mm and above, which will significantly improve the situation particularly in the Southern Hemisphere, albeit at shorter baselines.

Together with atmospheric fluctuations, the small number of suitable telescopes and their sensitivity at mm-wavelengths make it very challenging to produce high-dynamic range and high-fidelity images of radio sources. If we assume a typical Allan standard deviation of the atmospheric delay in the troposphere as $\sim10^{-13}$ \citep[Chpt. 9]{TMS}, the expected coherence time for 3 mm VLBI is only about 20 seconds, and the sensitivity of radio telescopes in GMVA stations generally ranges from 1000 to 5000 Jy in system equivalent flux density (SEFD)\footnote{http://www3.mpifr-bonn.mpg.de/div/vlbi/globalmm/}, which leads to a $\sim$0.1\,Jy 5$\sigma$ sensitivity in that time.  In such situations, VLBI observations using the simultaneous mm-wavelength receivers have demonstrated that one can compensate for phase fluctuations effectively, resulting in an order of magnitude sensitivity enhancement \citep{jung_11,rioja_15}. As a result, a large number of radio sources that have never previously been detected or imaged at mm-wavelengths can now be targeted. This is crucial because our current understanding of physical processes at the centers of galaxies and AGNs has only been studied in a small number of examples, mostly bright radio sources, which may introduce statistical biases.

\begin{figure}[htbp]
\begin{center}
\includegraphics[width=0.8\textwidth]{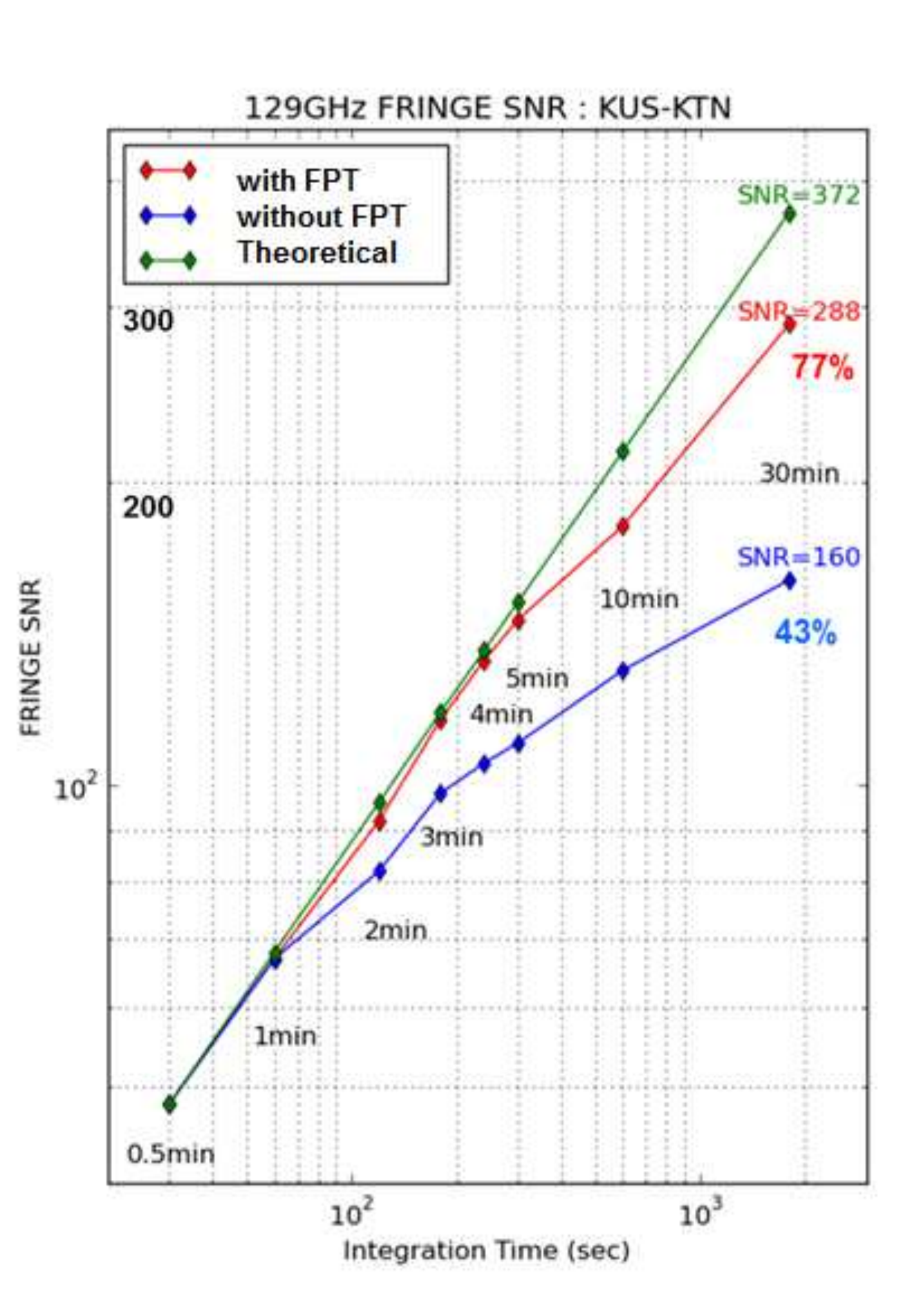}
\end{center}
\caption{The {\sc FRING} SNR for 3C279 at 129GHz, over a range of integration times, from the single KVN baseline between Ulsan and Tamna, observed on 09 Apr. 2012. 
Plotted in blue are the direct detections, in red are the detections once calibrated with the scaled phase solutions from 86GHz and in green are the theoretical sensitivities, extrapolated from the SNR achieved in 0.5 min. The FPT calibrated data achieves 77\% of the theoretical sensitivity even with an integration time of 30 minutes \citep{jung_12}.}
\label{fig:weak}
\end{figure}

According to the results achieved by the KVN (where the baselines are short and the atmosphere seems particularly stable), 30 minutes integration of visibilities is possible even at 129-GHz, giving a 5 sigma detection sensitivity as low as $\sim$25 mJy. This is shown in Fig. \ref{fig:weak} where the detected source SNR is plotted as a function of integration time for no calibration, with FPT calibration, and compared to the theoretical SNR assuming no coherence losses. 

\begin{table*}[t]
    \centering
    \begin{tabular}{lr|rr}
\hline
\hline
Survey ID & Wavelength & No. Sources & Reference \\
\hline
CJF survey & 6 cm & 293 & \citet{taylor_96} \\
VSOP VLBApls & 6 cm & 374 & \citet{fomalont_00} \\
CJF Polarimetry survey & 6 cm & 177 & \citet{pollack_03} \\
ICRF & 3.6 cm & $\sim$500 & \citet{ojha_04,ojha_05}\\
& & & and references therein \\
MOJAVE & 2 cm & $>$133 & \citet{lister_05} \\
2cm Survey & 2 cm & 250 & \citet{kovalev_05} \\
VLBA Calibrator Survey & 13 \& 3.6 cm & $>$3400 & \citet{kovalev_07} \\
VIPS & 6 cm & 1127 & \citet{helmboldt_07} \\
VERA FSS / GaPS & 1.35 cm & 500 & \citet{petrov_07} \\
VSOP Survey & 6 cm & $\sim$300 & \citet{dodson_08} \\
TANAMI & 3.5 \& 1.3 cm & 80 & \citet{ojha_10} \\
mJIVE-20 & 20 cm & $>$4300 & \citet{deller_14} \\
\hline
GMVA 3mm & 3 mm & 123 & \citet{sslee_86}\\
ICRF 22 \& 43-GHz & 13.7 \& 7 mm  & $\sim$100 & \citet{lanyi_10} \\
KVN Q-CAL survey & 7 mm & 638 & \citet{petrov_12}  \\
\hline
    \end{tabular}
    \caption{Summary of VLBI Surveys, their wavelengths and the number of sources catalogued. The difference in scale of the number of sources in the cm and mm surveys are clear.
    }
    \label{tab:weak}
\end{table*}

Recently, KVN has commenced a Multi-frequency AGN Survey with the KVN (MASK), which is under way in order to densify the grid of available radio sources at 22, 43, 86, and 129\,GHz. The early detection statistics for the first 10\% of the total sample ($\sim$1500) with a 1 Gbps recording rate (64 MHz bandwidth per band) and using FPT from 22\,GHz shows around a 97, 97, 90, and 60\% detection rate at 22, 43, 86, and 129\,GHz, respectively. After transferring calibration solutions from K to Q/W/D with a total 30 min integration at 1-Gbps, we find a 5-sigma sensitivity of 10/20/30 mJy, respectively. This result is very promising for finding many more weak sources by utilising simultaneous mm-wavelength receivers. In addition, it should be noted that simultaneous flux measurements at different frequencies give better constraints on spectral analysis, especially for AGNs that are highly variable on time scales similar to the observation length, and whose flares are associated with structural changes in the early stages of their evolution, such as Sgr A* \citep[e.g.][]{fish_11}.  

Nevertheless, it is very difficult to obtain high-fidelity images of compact radio sources, such as the immediate vicinity of supermassive black holes, with the few KVN baselines (max $\sim$500 km), and thus, studying the structure and evolution of (sub-)pc scale emissions is demanding. In addition, the statistical study of relativistic jets using the observed distribution of brightness temperatures ($T_{\rm b}$) and/or the intrinsic $T_{\rm b}$ of mm-VLBI cores will be challenging. Accordingly, increasing the number of mm-VLBI telescopes with simultaneous mm-wavelength receivers is a crucial requirement to deepen our understanding of galaxies and AGNs. With such a capability, mm-VLBI could observe a similar number of sources and obtain comparable sensitivity to that achieved currently with cm-VLBI.

   \subsection{Astrometry}\label{sec:astrometry} 

%
Astrometry  provides fundamental information for a large number of fields in astronomy. 
An accurate (i.e. astrometric) registration of images (be they total intensity, polarised emission, spectral line or others) obtained at different frequencies is crucial to form a meaningful interpretation
based on multi-frequency comparisons, in a similar manner as is required between epochs for multi-epoch temporal studies.
Bona fide astrometry can be achieved with a suitable calibration strategy that removes the dominant contribution of the medium through which the signal propagates, while retaining the intrinsic astrometric signature of the source in the phase observable.
If so, the astrometric accuracy is ultimately limited by the uncertainty in the precise phase observable and reaches the thermal limit of the instrument ($\sigma_{\rm pos} \sim \theta_{\rm B}$ /SNR, where $\sigma_{\rm pos}$ is the positional accuracy, $\theta_{\rm B}$ is the synthesised beam and SNR is the Signal to Noise Ratio), which leads to $\mu$-as astrometry. Phase referencing does this, by interleaving observations of a second source, but this fails for frequencies beyond $\sim$43-GHz, defeated by the fast tropospheric fluctuations.

There are a number of image-based methods in AGN studies which have been used to align images between frequencies, but these are very dependent on a-priori assumptions and results will always be open to question. 
The core of the jet, defined as the upstream end of the jet, would seem a priori to be the best choice, since it is usually the strongest jet feature. However, opacity effects at centimetre wavelengths can move its position with observing frequency \citep[e.g.,][]{Blandford:1979eg}; an effect known as opacity core-shifts. 
Furthermore, the blending of multiple components close to the VLBI core at longer wavelengths further complicates the alignment of the images. It is therefore more convenient to choose jet components that are known to be optically thin, strong, and compact, for a more accurate determination of their position. Finding such components is not always easy, specially when comparing observations at relatively distant observing wavelengths, due to the vastly different convolving beams. An alternative method for the alignment of the images can be obtained by performing a cross-correlation of the optically thin jet emission \citep{Walker:2000gb}, which has provided a very simple approach to derive results \citep[e.g.,][]{Croke:2008bn,Fromm:2013en,2014A&A...566A..26M}. 
However these image-based non-bona fide methods are prone to producing results with reduced accuracy and precision. \citet{hovatta_14} argued that any results derived without accurate phase referencing are questionable close to the core, which are the regions of most interest in mm-VLBI.

The KVN \citep{early_kvn,sslee_14} is the first dedicated mm-VLBI array and addresses one of the fundamental limitations of the field, the atmospheric stability, with an innovative multi-channel receiver design.
KVN currently consists of three antennas spread across South Korea, located in the campus of the Universities of Yonsei and Ulsan on the mainland, and in Tamna, on Jeju island. 
The observing frequencies are  centred at 22, 43, 87 and 130\,GHz. 
The multi-channel receiver \citep{kvn_optics} of KVN is designed for atmospheric compensation using simultaneous observations at multiple bands, which results in an effective increase of the coherence time, well beyond that imposed by atmospheric propagation.
The KVN backend \citep{kvn_backend}, combined with the SFPR calibration strategy \citep{vlba_31,rioja_11a,rioja_11b, rioja_14}, allows high precision astrometric measurements even at the highest frequencies of 130-GHz \citep{rioja_15}.
%
The SFPR step of the calibration removes the remaining dispersive residual errors (i.e. instrumental and ionospheric propagation effects) using the interleaving observations of another source. This two-step calibration  retains the astrometric signature of any source position shifts between the two frequencies in the interferometric phase. This is done for all frequency pairs which have an integer frequency ratio R (with R = $\nu_{high} / \nu_{low}$).
We know of no demonstrated upper frequency limit and the method would be expected to work as long as the tropospheric propagation effects were non-dispersive. The baseline lengths between the KVN antennas range between 300 and 500\,km, providing a spatial resolution $\sim$1\,mas at the highest frequency band. Work is on-going to develop a global network of KVN-compatible antennas \citep{jung_15}, which will provide tens of $\mu$-as resolution.

\begin{figure}[htbp]
\begin{center}
\includegraphics[width=1.0\textwidth]{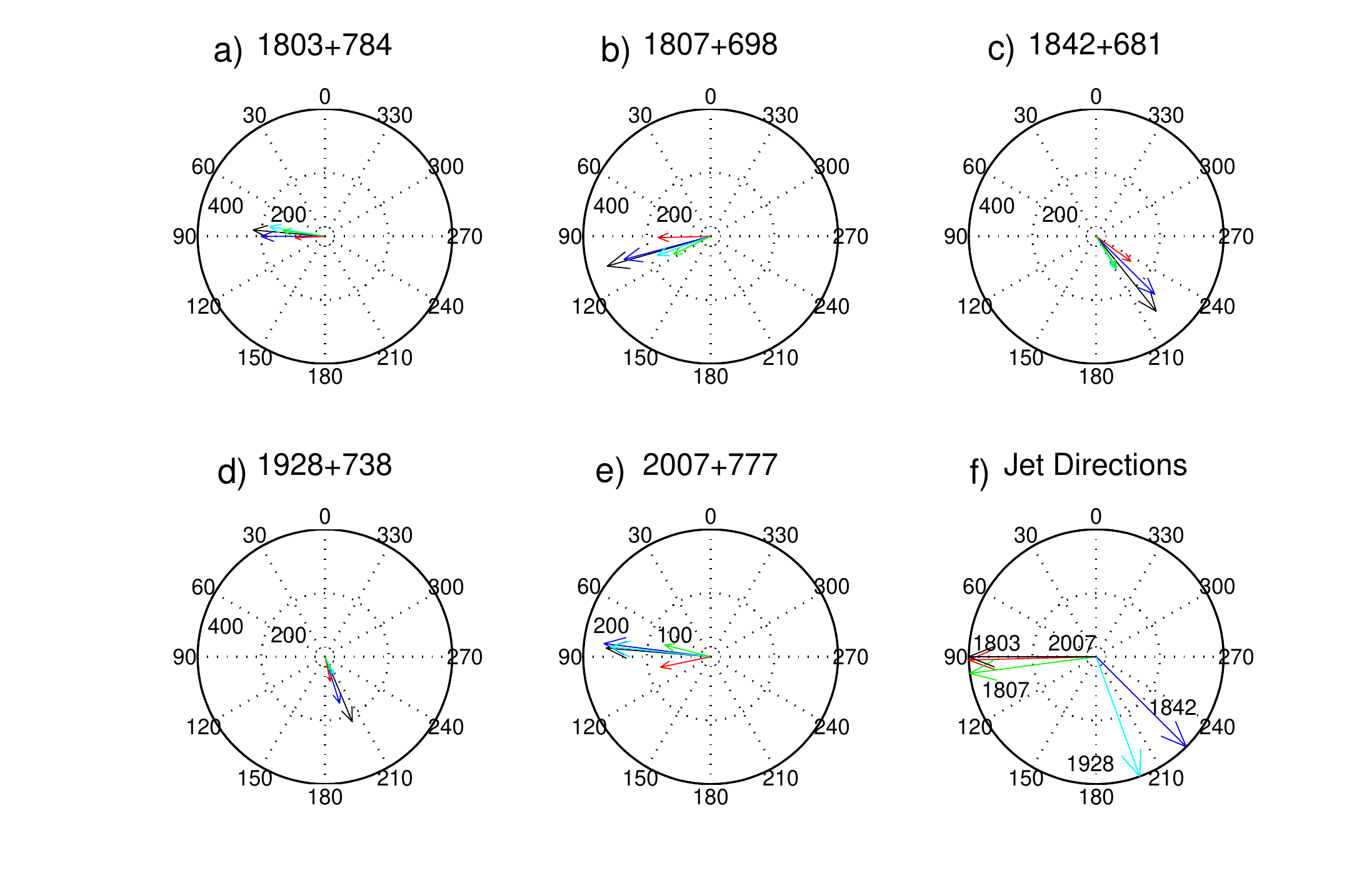}
\end{center}
\caption{The derived individual core-shifts for a subset of five sources from the S5 polar cap sample, between 22, 43, 86 and 130-GHz. (a)–(e) Polar plots of the decomposed absolute single-source position shifts between two frequencies, for five frequency pairs for the AGNs. These are shown in different colors (${\rm{K}}\to {\rm{Q}}$ (red), ${\rm{K}}\to {\rm{W}}$ (blue), ${\rm{K}}\to {\rm{D}}$ (black), ${\rm{Q}}\to {\rm{W}}$ (green), and ${\rm{Q}}\to {\rm{D}}$ (cyan)). In the polar plots the position angles are shown outside the largest circle and are 0$^o$ and 90$^o$ toward north and east, respectively, and the magnitude units, as specified in the concentric circles, are in $\mu$as. f) shows the expected jet directions on an arbitrary scale. These results have been derived from the SFPR pairwise measurements, using SVD plus the alignment constraint between the jet and the position shift directions, to break the degeneracy. See Fig. 10 of \citet{rioja_15} for details.}
\label{fig:dband}
\end{figure}

%
In \citet{rioja_14} we presented results of simultaneous SFPR astrometric measurements with KVN at 22 and 43\,GHz for continuum sources, along with a detailed comparative study with contemporaneous observations using fast frequency switched SFPR observations, made with the VLBA. \citet{dodson_14} presented the demonstration for non-integer spectral line observations, with the registration of maser emission of \wat\ and SiO masers of the AGB star R-LMi at 22 and 43\,GHz. 
In \citet{rioja_15} astrometric VLBI measurements were successfully extended to all the four frequency bands supported by the KVN, that is up to 130\,GHz. 
Fig. \ref{fig:dband} shows the results from this study. 
More recently \citet{cho_17} have successfully applied SFPR for registration of the four bands in spectral line observations, comprising the water masers at 22-GHz, and 4 SiO maser transitions at 42.8, 43.1, 86.2 and 129.3-GHz, in VX Sgr. 

\begin{figure}[htbp]
\begin{center}
\includegraphics[width=0.7\textwidth,angle=-90]{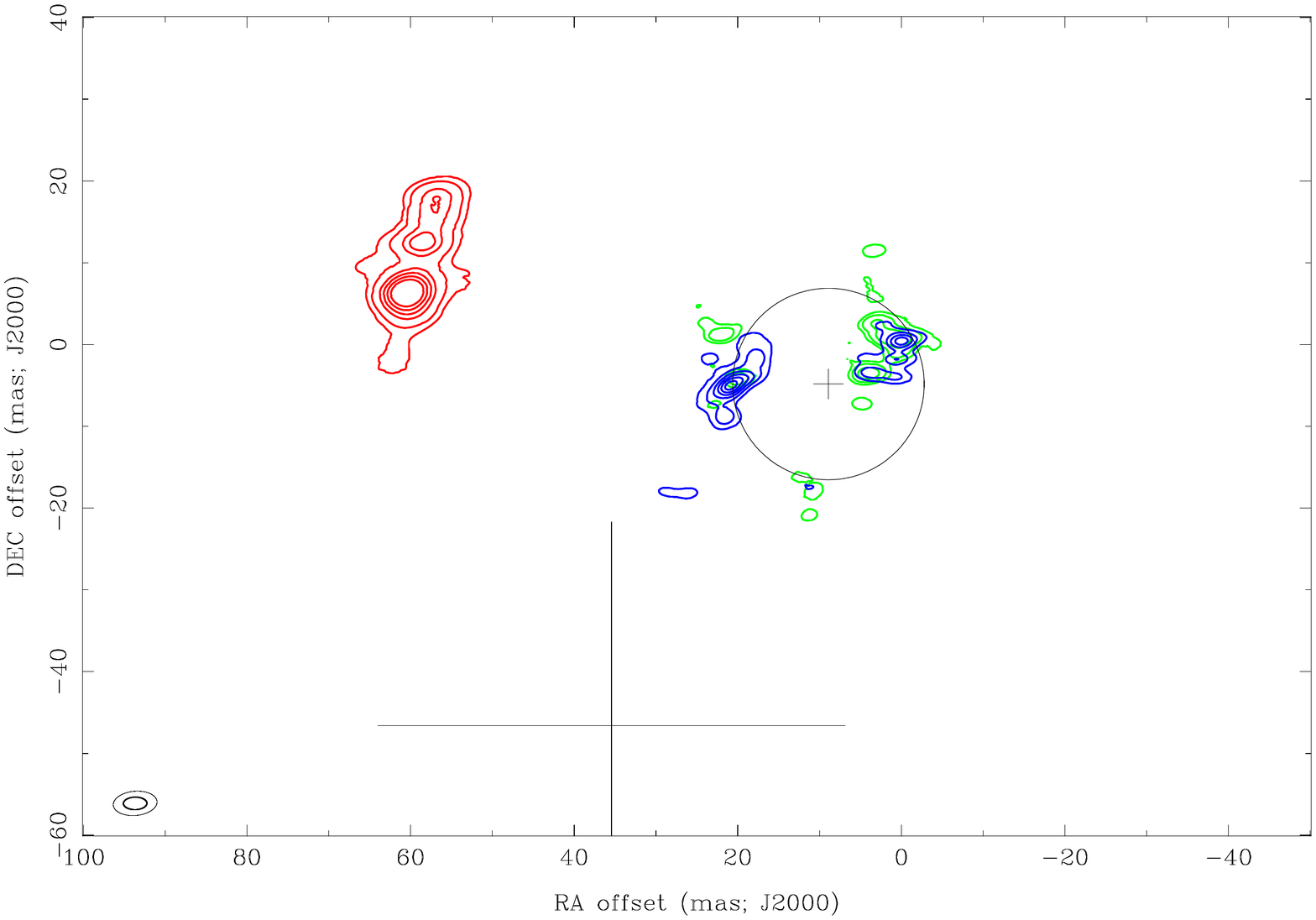}
\end{center}
\caption{The astrometric position of the AGB star 
R-LMi (small cross), as derived from the centroid of the SiO J=0$\rightarrow$1 maser emission, with blue ($v$=1) and green ($v$=2) contours. The SiO masers are astrometrically registered to the \wat\ maser emission (red contours), which is conventionally phase referenced to a known extra-galactic calibrator. The large cross indicates the Hipparcos position for this source, with the bar length being the proper  motion error over the 21 years since the optical positional epoch.  For details see \citet[Fig. 5][]{dodson_14}.}
\label{fig:rlmi}
\end{figure}

SFPR is primarily designed for astrometric registration between frequencies, but can produce positional astrometry with respect to an external reference point at high frequencies when combined with conventional phase referencing observations at the lower frequency. An example of such an analysis was the astrometric measurement of R-LMi \citep{dodson_14}, see Fig. \ref{fig:rlmi}. 
In this case the two decades time span between our observations and the Hipparcos measurements \citep{hip07} lead to large positional uncertainties. The conventional phase referencing astrometry of the \wat\ masers with respect to a quasar, then the SFPR referencing of the emission from SiO masers with respect to that from \wat\ masers, allowed the positional astrometry of the star (at 43-GHz) to be determined with significantly better accuracy, reducing the uncertainty in the proper motion for R-LMi by an order of magnitude.

\subsection{Faraday Rotation studies}\label{sec:faraday}

VLBI polarimetric observations provide a unique tool to obtain information about the strength, degree of ordering, and orientation of the magnetic field on the plane of the sky in AGN jets. Simultaneous polarimetric VLBI observations at two or more frequencies allow, also, for the determination of the Faraday rotation of the plane of polarisation, since it is proportional to the square of the observing wavelength. (In practise more than two are required, to resolve ambiguities.)  This can be used to determine the direction of the line-of-sight magnetic field -- and therefore the three-dimensional structure of the field, given that the amount of Faraday rotation depends on the line-of-sight magnetic field strength. It provides also a way to probe the thermal plasma around the radio-emitting jet regions, since the amount of rotation depends also on the thermal electron density. This has been used to study, for example, the interaction of the jet with the external medium \citep{Gabuzda:2001cu,2000Sci...289.2317G,2008ApJ...681L..69G}.

These Faraday rotation studies are predicated on a proper alignment of the VLBI images across different observing wavelengths, but self-calibration of the visibility data during the imaging process results in the loss of absolute positioning of the source (unless a starting model is used with an accurate position, itself derived from phase referencing). 
As discussed previously, registration of the VLBI images is commonly performed by looking for optically thin jet features (components) that can be unambiguously identified in all of the images, or be obtained by performing a cross-correlation of the optically thin jet emission.
The concerns raised in \citet{hovatta_14} about the accuracy of these methods, particularly close to the core region, are discussed in Section \ref{sec:astrometry}. A {\em bona-fide} astrometric registration of the images can only be obtained through phase-referenced observations, and for mm-VLBI the best method available is SFPR. Nevertheless a number of the results discussed here are based on these non bona-fide astrometry techniques, demonstrating the science that can be extracted. 
  
 
\begin{figure}[htbp]
\begin{center}
\includegraphics[width=0.9\textwidth]{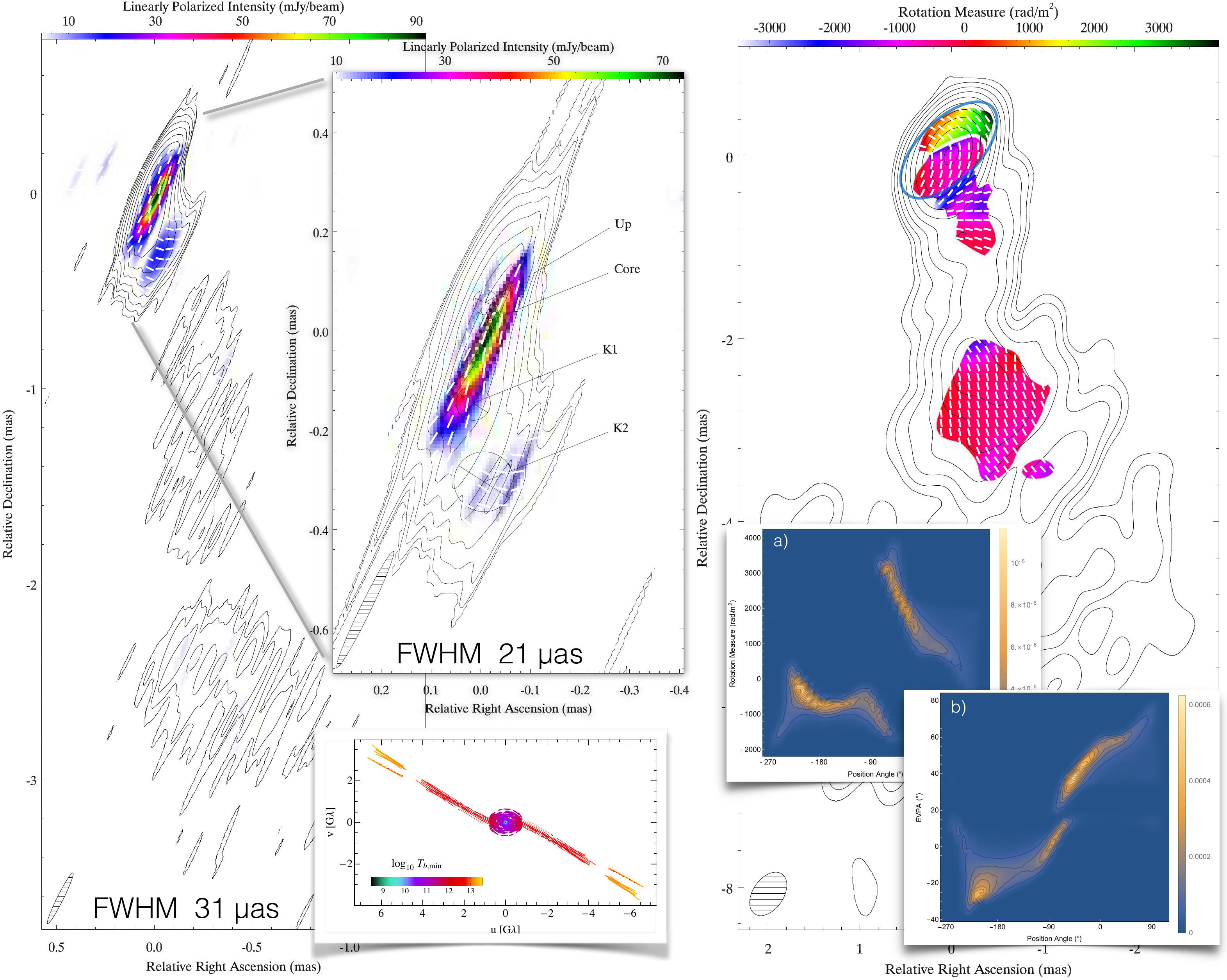}
\end{center}
\vspace{-0.2cm}
\caption{{\em Left}: RadioAstron images of BL~Lac at 1.3~cm from November 11, 2013. Total intensity is shown in contours, linearly polarised intensity in color scale, and white bars indicate the EVPAs. The inset panel shows the uv-coverage. {\em Right} shows the rotation measure map and a) two-dimensional histograms of the rotation measure and b) Faraday-corrected EVPA in the core area. For details see figures in \citet{2016ApJ...817...96G}.}
\label{BLLac_RA}
\end{figure}

Faraday rotation analysis can be used to look for helical magnetic fields in AGN jets, which are expected to arise from the differential rotation of the accretion disk, and are thought to have an important role in the actual jet formation and collimation processes \citep[e.g.,][]{McKinney:2013fd,Zamaninasab:2014hs}. Faraday rotation measure (RM) gradients across the jet width should appear naturally if the jets are threaded by a helical magnetic field due to the systematic change in the net line-of-sight component of the magnetic field across the jet, with increasing values toward the jet boundaries \citep{Laing:1981bx}. The first observational evidence for the existence of such gradients in RM across the jet was given by \cite{Asada:2002to} based on near-simultaneous multi-wavelength VLBA observations of the quasar 3C~273. Further observations have confirmed the transverse RM gradient in 3C~273 \citep{Zavala:2005im,Asada:2008dq,Hovatta:2012jv}, as well as in other sources \citep{Gabuzda:2004dt,Asada:2008gb,Asada:2010dn,OSullivan:2009bx}.
  
One of the most comprehensives studies of Faraday rotation in AGN jets was performed by \cite{Hovatta:2012jv}, in which 149 sources from the MOJAVE sample were observed with the VLBA at four frequencies between 8 and 15~GHz. Four sources, CTA~102, 4C~39.25, 3C~454.3, and 3C~273 were found to contain clear RM gradients across the jet width. The jet in 3C~273 also displayed variations in the RM screen on a time scale of months, which may require some internal Faraday rotation, produced in the jet emitting region.
However, it is not clear whether this can be a universal jet launching mechanism as in some cases the required rotation measure gradients are not detected, or imply too weak a magnetic field, and an alternative cause in the shape of a Faraday screen external to the jet cannot be ruled out \citep{Taylor:2010ih, Zavala:2004ga}.

Studying the magnetic field configuration in the vicinity of the central black holes in AGN jets requires the highest angular resolution possible, which involves VLBI observations at either progressively shorter wavelengths or larger baseline distances, such as those provided by the RadioAstron space VLBI mission. The first polarimetric RadioAstron 1.3~cm observations were performed in November 11, 2013, in which BL~Lac was observed in combination with a ground array of 15 antennas (see Fig.~\ref{BLLac_RA}). Correlated visibilities between the ground antennas and the space radio telescope have been found extending up to a projected baseline distance of 7.9 Earth diameters, yielding a maximum angular resolution of 21 $\mu$as, the highest achieved to date in astronomy \citep{2016ApJ...817...96G}. Fig. \ref{BLLac_RA} shows the RM image obtained by combining the RadioAstron 1.3~cm observations of BL~Lac with simultaneous ground array 7~mm and 2~cm observations. The RM and Faraday-corrected EVPAs in the core area display a clear point symmetry around its centroid. 
This is better observed in the probability distribution function of the two-dimensional histogram for the RM and Faraday-corrected EVPAs as a function of position angle with respect to the centroid of the core, as shown in Fig.~\ref{BLLac_RA}. This suggests that the core region in BL~Lac is threaded by a large-scale helical magnetic field, as shown in relativistic magnetohydrodynamic simulations \citep{Porth:2011ev}.
Whilst we are confident in the solidity of these results, the lack of bona-fide astrometry prevents a robust characterization of the errors associated with the alignment of the images. Furthermore, for other weaker sources a proper astrometric analysis may be required to prevent unambiguous interpretation of the results

\subsection{Opacity core-shifts, 
\texorpdfstring{$\gamma$}{gamma}-ray flares, and the nature of the VLBI core}\label{sec:core}

The standard Blandford \& K\"onigl conical jet model hypothesises that the core is not a physical feature in the jet, but corresponds to the location at which the jet becomes optically thin, and therefore its position shifts with observing frequency \citep{Blandford:1979eg,Konigl:1981kx,1993A&A...274...55G,Lobanov:1998vr}.
Multi-frequency VLBI observations at centimeter wavelengths have measured this core frequency shift in multiple sources, albeit without phase-referencing \citep[e.g.,][]{kovalev_08,osullivan_09,sokolovsky_11,fromm_15}. Nevertheless phase-referenced VLBI observations have confirmed that the cm-wavelength radio core indeed is consistent with the optically thick-thin transition, in a smaller number of targets, such as 3C\,395, 4C\,39.25, 1038+528, 3C\,390.1, M\,81, M\,87 and 3C454.3 \citep[respectively]{lara_94,guirado_95,rioja_98,ros_01,martividal_11,hada_11,kutkin_14}.

On the other hand results of over seven years of monthly monitoring of 36 blazars (the most luminous and variable BL Lac objects and flat-spectrum radio quasars), with the VLBA at 7mm by the VLBA-BU-BLAZAR program, show that most $\gamma$-ray flares are simultaneous within errors with the appearance of a new superluminal component or a major outburst in the core of the jet, which is defined as the bright, compact feature in the upstream end of the jet \citep{Marscher:2008ii,2013ApJ...773..147J,2015ApJ...813...51C}. A burst in particle and magnetic energy density is therefore required when jet disturbances cross the radio core in order to produce $\gamma$-ray flares, which can naturally be explained by identifying the radio core with a recollimation shock \citep{gomez_95,1997ApJ...482L..33G,daly_88}.
Multi-wavelength observations of blazars therefore suggest that the radio core is a physical feature (recollimation shock) in the jet at a fixed location. 

\begin{figure}[htbp]
\includegraphics[width=1.0\textwidth]{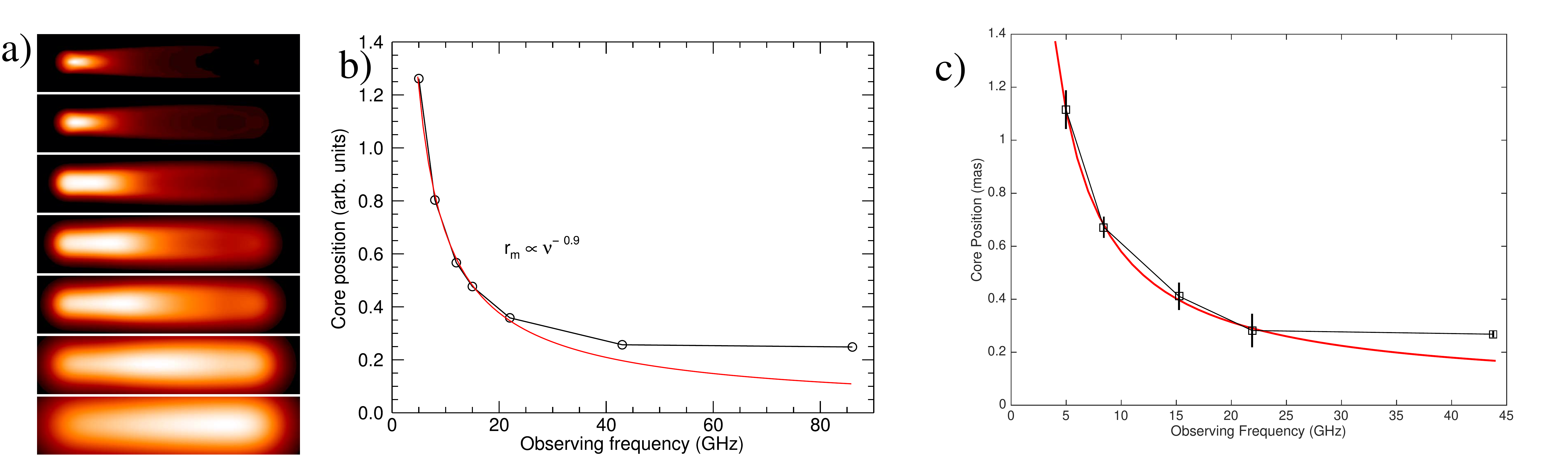}
\caption{{\it a)} A sequence of simulated synchrotron total intensity images computed at, from the top, 86, 43, 22, 15, 12, 8, and 5-GHz, using a relativistic hydrodynamical model of a jet with a recollimation shock. {\it b)} Position of the simulated core as a function of frequency (black circles and line). The red curve indicates the best fit to the core positions between 5 and 22-GHz, which follows the conical Blandford \& K\"onigl jet model. 
However the 43 and 86-GHz data clearly deviate from the opacity core-shift curve, revealing the  recollimation shock. {\it c)} Astrometric core-shifts of BL-Lac between 4.8 and 43-GHz, plotted as a function of frequency in black with 1-$\sigma$ errors, adapted from \citet{dodson_16}. The Blandford \& Koenigl model fitted to the cm-wavelength data (5 to 22-GHz, with $\kappa$ equal to -0.99 and $r_0$ equal to 5.3 mas\,GHz$^{\kappa}$) is overlaid in red. 
\label{Core-shift}}
\end{figure}

We have therefore two sets of results, one suggesting that the radio core marks the transition between the optically thick-thin jet regimes while the other implies that it corresponds to a recollimation shock.
A possible solution to reconcile these apparently contradicting observational results is to consider that the Blandford \& K\"onigl core-shift at centimeter wavelengths is produced because the foot of the jet becomes optically thick at these low frequencies. On the other hand, at millimeter wavelengths one can see closer to the Black Hole, revealing a recollimation shock that was previously hidden.
While the Blandford \& K\"onigl model holds for our best studied case, M87 \citep{Hada:2011im}, simultaneous observations at $\gamma$/X-rays, optical, infrared, and radio, together with mm-VLBI imaging, have shown that in several other radio galaxies and blazars the core indeed is inferred to lie parsecs away from the central black hole \citep{2002Natur.417..625M,2010ApJ...710L.126M,2011ApJ...734...43C,2015A&A...576A..43F}.
  
  
To test this scenario we have performed numerical simulations using the finite-volume code {\sc Ratpenat}, which solves the equations of relativistic hydrodynamics \citep[and references therein]{perucho_10}. The jet is launched with a Lorentz factor of 7 and an initial over-pressure of 1.5 times that of the external medium, in order to obtain a recollimation shock. Using the hydrodynamical results as input, we have then computed the synchrotron emission at different observing frequencies \citep[for details of the numerical model used see][]{gomez_95,Gomez:1997gq,mimica_09}, adjusting the model parameters so that the jet is optically thin at 86 and 43-GHz, and becomes optically thick at lower frequencies. A magnetic field in equipartition with the particle energy density is assumed. The results are shown in Fig. \ref{Core-shift}a and b.
At cm-wavelengths (5 to 22~GHz) numerical simulations reproduce the opacity core-shift of a Blandford \& K\"onigl conical jet model, while at millimeter wavebands (43 and 86GHz) the core position clearly departs from this behaviour, revealing the core as an optically thin recollimation shock at a fixed jet location.
Testing this scenario requires astrometric observations at millimeter wavelengths, for which the SFPR technique can provide the necessary high precision. Recent observations, see Fig. \ref{Core-shift}c, suggest that this effect has been detected for the first time, in BL-Lac, reproducing the simulations extremely closely \citep{dodson_16}.

\section{Spectral Studies} \label{sec:line}
\subsection{Multifrequency VLBI observations of maser emission in evolved stars} \label{sec:evolved}


\subsubsection{AGB and post-AGB evolutionary phases}

Most stars in the sky, with initial masses between about 0.5 and 8 \ms, reach the AGB phase (or a similar one) at the end of their lives. After this phase, most of the initial mass has been ejected to form planetary nebulae (PNe), after the short phase of protoPN. Finally, this ejected material will return to the interstellar medium, enriching it (and the new generations of stars to be formed from it) with heavy elements.

The copious mass-loss of AGB stars forms thick circumstellar envelopes (CSEs) around them. Mass-loss is a basic phenomenon in their evolution, the ejection rate tends to increase with time and, by the end of the AGB phase, it is so strong (reaching up to 10$^{-3}$ \my) that most of the initial stellar mass is ejected in a relatively short time. Then, the AGB phase ends, the stellar core becomes exposed, becoming the new central star. This new star is very compact and shows an increasingly high temperature, rapidly evolving to the blue and white dwarf phase.

The circumstellar nebula is also evolving very fast: from the nearly spherical CSE in the AGB phase, which is in relatively slow expansion (at, say, 10-15 \kms), to the strongly axisymmetric PNe around the dwarfs, which show very fast bipolar outflows, with axial velocities as high as several hundred \kms, \citep[etc]{bujarrabal_01, balick_02}. This metamorphosis is
a very fast and spectacular phenomenon. Within about 1000 yr, the star becomes a blue dwarf able to significantly ionize the nebula, which has already developed a wide bipolar shape. 

The mass-loss process in AGB stars behaves, basically, in two phases, \citep[e.g.][]{hofner_03}. In inner circumstellar regions, the stellar pulsation (all these old stars are strongly pulsating, in more or less regular modes) propagates in the outer atmospheric layers, leading to 
quite high densities during some pulsational phases, out to a few times the optical/NIR photospheric radius from the photosphere. 
When the gas temperature drops to about 1000 K, at a few photospheric radii, the formation of dust grains becomes very efficient and refractory material condenses. Radiation pressure acts very efficiently on dust grains, which are dynamically coupled with gas. The result is that the whole circumstellar layers start expanding, quickly reaching (at 10-20 stellar radii) their final expansion velocity. The circumstellar dynamics, which is in fact driving the evolution of the AGB stars and the ejection of PNe, is really active only in those inner regions. The study of the inner layers of CSEs is therefore crucial to understand these phases of the stellar evolution.

\subsubsection{SiO and \texorpdfstring{\wat\ }{Water}  masers from AGB stars}

SiO (at 43, 86, 130, 215, ... GHz) and \wat\ (22-GHz) masers are known to be very useful tools for studying the detailed structure and dynamics of the AGB CSEs. SiO masers come from regions at about 2-4 stellar radii, where pulsation is propagating and dust is not yet completely formed. 
The SiO maser lines in general come from a ring of spots, centered on the star, very probably due to the dominant tangential amplification \citep[etc]{diamond_94, desmurs_00}.  On the other hand, \wat\ masers form shortly after grain formation, when the final expansion velocity is being reached, e.g.\ \citet{richards_12}. 
The very high angular resolution of VLBI imaging at these frequencies allows very detailed mapping of such crucial shells, providing also valuable information on their dynamics.  
However, the complex pumping of the masers, which is not well understood (particularly for SiO), and the often poor information about the relative positions of the spots detected at the different frequencies strongly limit our studies of circumstellar masers from VLBI data.

\begin{figure}
\centering
\includegraphics[width=1.0\textwidth]{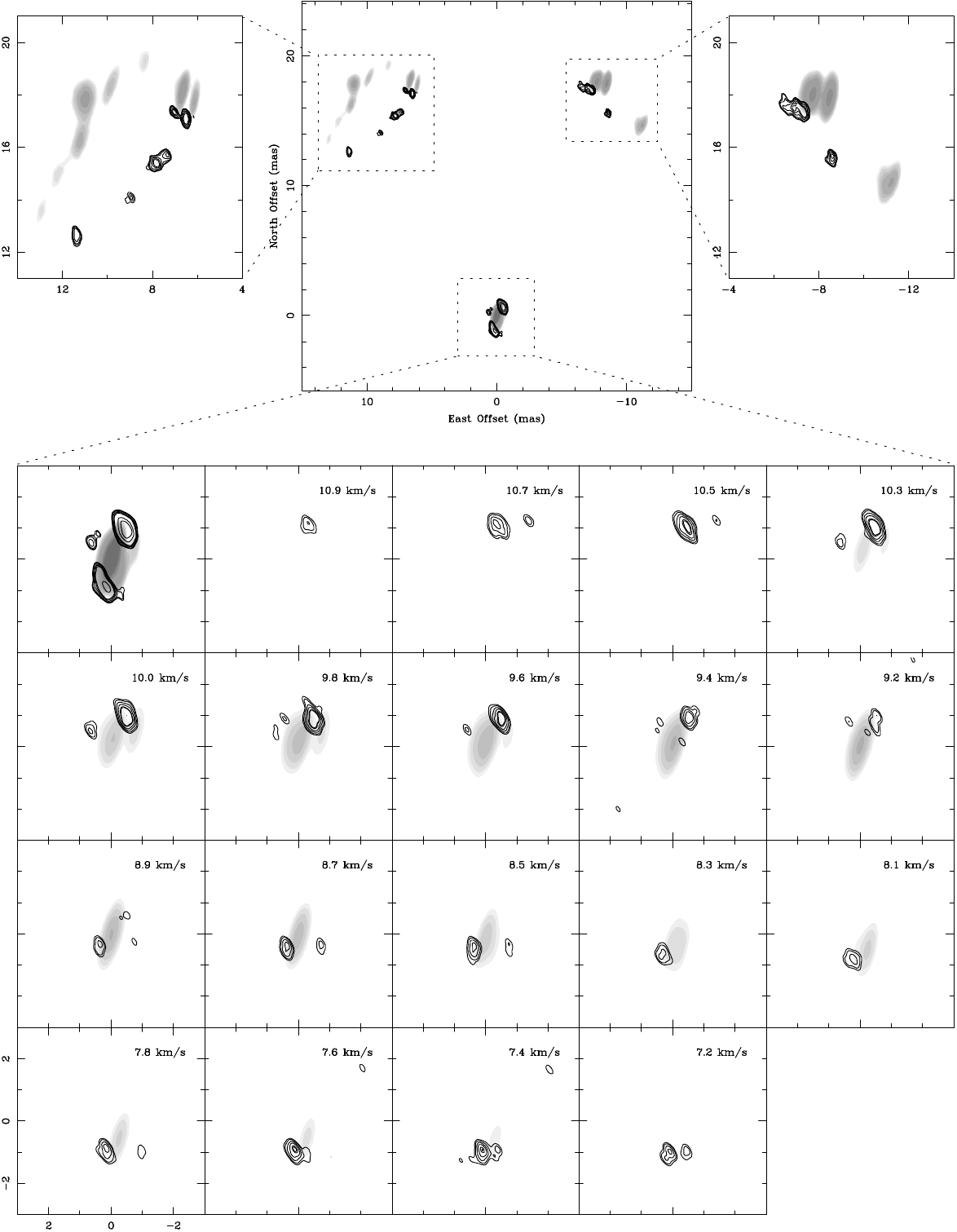}
{\caption
{Observations of SiO \juc\ $v$=1 (greys) and $v$=2 (contours) in
  IRC\,+10011, \cite[Fig. 9][]{soria-ruiz_04}. Note that both lines occupy
  similar regions but that, in fact, the spots are never coincident, in
spite of the uncertain relative astrometry. \label{fig:irc10011}}}
\end{figure}

SiO masers appear in rotational transitions (\juc, \jdu, \jtd, ...)
within vibrationally excited states ($v$ = 1, 2, 3, ...). The various proposed
pumping models agree in identifying the inversion mechanism: line trapping
in the rovibrational transitions ($v \rightarrow v-1$), which decreases the
original radiative probabilities to efficiency probabilities that are
relatively lower for higher-$J$ levels; see e.g.\ \citet{bujarrabal_81, lockett_92, gray_09}. This leads to a systematic overpopulation of levels with relatively high $J$-values and to the inversion of the $J$
$\rightarrow$ $J-1$ transitions in the $v>$0 states, which is particularly
efficient for lower frequency lines. What is not yet well known is the
source of the energy, i.e. the way in which we excite molecules from
the ground $v$=0 state, which can be collisional or radiative \citep{humphreys_02, bujarrabal_94}. The
issue is important, because the relationship between the physical conditions for the maser emission must
depend on the pumping mechanism. If we wish to probe the physical conditions based on the observed radiation we must understand how the former generates the latter. 
Radiative pumping tends to discriminate
more clearly the conditions required for the pumping of the different
$v$-states. The fact that the maser spots of the $v$=1 and $v$=2
\juc\ lines tend to form clusters that are similar for both lines (the
ring radius being slightly smaller for the $v$=2 line) has been argued
to favour collisional pumping \citep[etc]{miyoshi_94, desmurs_00, rioja_08, kamohara_10}. 
The \juc\ masers are intense and their frequencies, around 43-GHz ($\lambda$
=7\,mm), are easily reachable at present in VLBI, therefore, they are
the best observed SiO masers.  However, high-resolution ($<$mas) observations
have shown that, curiously, the spots of the two lines are practically
never coincident, which is interpreted as supporting radiative pumping
(Fig. \ref{fig:irc10011}). Of course, such a comparison is hampered by the often
poor bona-fide astrometry in VLBI experiments at 7mm wavelength. Of those listed above only \citet{rioja_08} and \citet{kamohara_10} were phase referenced, and both observed with VERA. At
higher frequencies (higher $J$-values), conventional VLBI phase referencing experiments are more difficult and astrometry is still not yet possible (although we note that a significant number of four-band KVN observations, albeit with greater than mas resolutions, are in preparation, and that the first KaVA results
have been recently published, \citet{yun_16}).

The problem also appears when comparing 22-GHz \wat\ with SiO masers. SiO
masers are clearly placed in a more central region, but we are not sure exactly how they are placed in relation to the \wat\ spots, as these in general show a complex distribution with a less obvious center. 
When the transitions are astrometrically registered, however, this issue is addressed, as in \citet{dodson_14}.

The publication of maps of other SiO lines has raised new problems. The
$v$=3 \juc\ line is placed at more than 5000 K above the ground state,
requiring a high excitation, and one expects it to be placed in a
different region, probably closer to the star. But observations place
it at about the same distance, in very similar
distributions to that for $v$ = 1 and 2 and always with very rare
coincidences at the smallest scale. On the other hand, the $v$=1
\jdu\ maser, which theoretically shares the same pumping mechanism and
excitation requirements as $v$=1 \juc, appears in clearly different
regions and at further distances than the 7mm lines. These new observations
seem in clear contradiction with all the simple pumping
models. Needless to say, the $v$=3 \juc\ maser is weaker than
the $v$=1,2 \juc\ ones and the $v$=1 \jdu\ line frequency, 86-GHz, is
hard to observe in VLBI, therefore, astrometry in such
observations has hitherto been impossible. 

A possible solution to the puzzle is the presence of line
overlap. Photons emitted by a rovibrational transition of \wat,
$v_2$=1 11$_{6,6}$ -- $v$=0 12$_{7,5}$, can be absorbed by the $v$=1
$J$=0 -- $v$=2 J=1 component, because both frequencies are almost
exactly the same \citep{olofsson_81, soria-ruiz_04}. It can
be shown that \wat\ often emits strongly at that frequency in our
sources. Therefore, the absorption of this radiation by SiO leads to an
underpopulation of $v$=1 $J$=0 and an overpopulation of $v$=2 $J$=1 line,
quenching the (otherwise expected to be intense) $v$=2 \jdu, which is
known to be anomalously weak; in fact the effects of overlap were first
invoked to explain this observational fact. At the same time, this
phenomenon introduces a strong over-excitation of the two 7mm masers,
$v$=1,2 \juc, whose excitations are now strongly coupled. Calculations
show that the excitation conditions of both lines are now very similar
and also similar to those of the $v$=3 \juc\ line \citep{v3_reference, desmurs_14}. Meanwhile, $v$=2
\jdu\ now appears to require different conditions, in particular lower
densities that should appear at larger distances from the
star. Radiative models including line overlap can explain, at least
qualitatively, all the observational results. 
This is not the only possible explanation; an alternative could be non-monotonic temperature gradients, e.g. due to shocks or dust formation \citep{richards_14}.
However, we stress the scarcity of available data with high sensitivity and with any meaningful
astrometric registration, particularly at high frequencies.  
We do not yet have accurate and systematic information on the relative distributions of
these lines, for which not only do we need good absolute astrometry but we
would also need to recover (almost) all the maser flux, to better
understand the complete brightness distribution.  In particular, we lack
VLBI data of lines at higher frequencies, necessary to be sure that line
overlap is the main (or unique) major modification of the basic pumping
mechanisms.

\subsubsection{Maser lines in proto-PNe}

As soon as the star+nebula system leaves the AGB phase, the observation
of the masers associated with them becomes more and more difficult. The reason is that the
inner circumstellar masers are becoming more and more diffuse at the
same time as the mass-loss rate decreases. SiO masers are very rarely
observed. There is however a paradigmatic example showing both SiO and
\wat\ masers: the strongly bipolar nebula OH\,231.8+4.2,
the Calabash Nebula. 

\begin{figure}[htbp]
\includegraphics[width=1.0\textwidth]{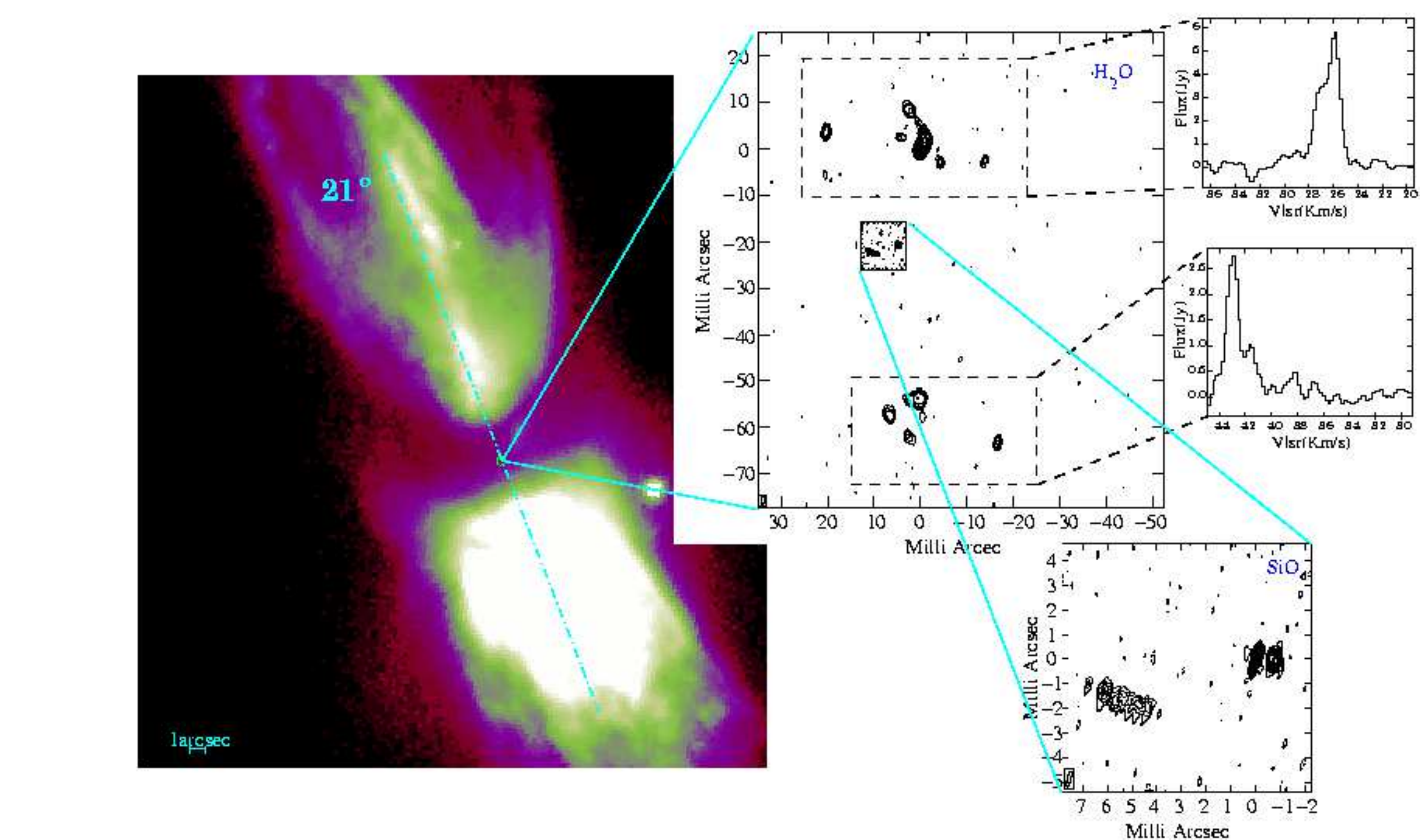}
{\caption
{Observations of SiO and \wat\ masers in OH\,231.8+4.2, compared with
  the optical image. The uncertainty in the SiO maser position is
  about 200 mas, too large to definitively determine its relationship with the
  \wat\ clumps. From \citet[][, Fig. 1]{desmurs_07}}
  \label{fig:oh231}}
\end{figure}

The evolution from the AGB to the PN phase is as spectacular as it is rapid. It
is thought that the strong axial symmetry typical of post-AGB nebulae is
due to the ejection, during the first protoPN phases, of very fast and
collimated stellar jets. These jets shock the fossil slow-moving AGB envelope, generating a series of axial shocks that cross the massive CSE,
inducing high axial velocities in it \citep[see e.g.,][]{balick_02,bujarrabal_01}. The very large amounts of energy
and momentum observed in young PNe impose severe limitations on the jet
launching mechanism. At present, the only way to explain the origin of
such energetic flows is to assume that a fraction of the ejected CSE is
reaccreted by the central star or a companion through a rapidly
rotating disk \citep{soker_01,frank_04}. The jet would then be powered by a magnetocentrifugal
launching process, similar to that at work in forming stars. For that
to be efficient, the presence of a stellar companion
(or at least a massive planet) is necessary, since otherwise the circumstellar
material lacks the angular momentum to form an accretion
disk. OH\,231.8+4.2 is strongly bipolar, shows a very fast and massive
bipolar outflow, and is known to harbor a binary star in its
center. It is therefore a textbook candidate to study these crucial
phenomena.

VLBI observations of SiO and \wat\ masers have yielded very promising
results (see \citet{sanchez_02}, \citet{desmurs_07},
and Fig. \ref{fig:oh231}). Water vapor emission comes from two regions in opposite
directions along the nebula axis and their velocity field is fully
compatible with the general velocity field in that direction, suggesting that
the \wat\ clumps represent the inner nebula, at the base of the
bipolar flow. SiO masers occupy smaller regions and are placed almost
exactly perpendicular to the axis. The movements depicted by the
observations of SiO are compatible with a disk orbiting the central
star(s). In principle, we are seeing in this object the whole central
structure of disk plus outflow that we would like to see to confirm our
ideas on the post-AGB nebular dynamics, which has not been
observed to date in other sources. But the relative astrometry is poor
and we are not sure that really the SiO structure is placed at the
center of the nebula, between the two \wat\ clusters. The problem is then
serious, as we lack this crucial datum to draw definitive conclusions on the disk/outflow
association. For the last 10 years we have been trying to improve
the astrometry, 
but, up to now, the best SiO maser position obtained is uncertain by about 200 mas,
completely insufficient for our purposes. Furthermore
the SiO masers in this source have been fading over the past several years, so new VLBI
measurements are more and more difficult. The VLBI data on
OH\,231.8+4.2 constitute a self-explanatory example of the effects of
the lack of accurate and sensitive VLBI data in our studies of the
inner post-AGB nebulae. We note that SFPR observations of this source with the KVN have recently been performed, and appear promising \citep{dodson_17}.

\subsubsection{Summary: Evolved Stars}

We have seen how important the VLBI observations of various maser
lines are in the study of the most interesting layers of nebulae around AGB
stars: the inner regions in which the dynamics are still active and the
relevant phenomena to understand these evolutionary phases are actually
taking place. In AGB stars, they harbor the grain formation and gas+dust
acceleration layers; in post-AGB objects, this inner nebula includes
the outflow acceleration regions and may hold
the key to understanding the formation and shaping of PNe. In both
cases, however, the existing VLBI data are still insufficient to
address these problems, because of the lack of systematic
observations with accurate astrometry and high sensitivity. Good VLBI
data are also necessary to understand the SiO maser mechanism, which is still  under
debate.

Accurate astrometry is still difficult at 7mm. At higher
frequencies, VLBI experiments are very difficult and conventional astrometry is almost impossible. In all cases, sensitivity limitations are
significant and an important fraction of the total flux is lost in the
interferometric process. Thanks to phase transfer between the SiO lines
and from 22-GHz (where the frequency ratios are close to integer), sensitive multifrequency observations will strongly
help to map several transitions, out to the sub-mm regime. They will
also help to systematically obtain good relative astrometry for the different
transitions.

\subsection{Massive Star Formation}\label{sec:massive}





\subsubsection{Introduction}

The formation of isolated low-mass (M $\sim$ 1~\ms) stars is now understood
quite well. It proceeds through
\ (i)~mass accretion onto the protostar through a Keplerian disk, an expected
result of angular momentum conservation, and 
\ (ii)~ejection of material via a jet collimated along the disk axis, which
removes angular momentum from the disk allowing matter to accrete onto the star.
Jets are generally modeled as magneto-centrifugally driven winds, 
powered by the rotation and gravitational energy and channeled along the magnetic field lines
either from the disk inner edge \citep[``X-wind'',][]{hirota_17}, or across a much larger 
(up to 100~AU) portion of the disk \citep[``Disk-Wind'',][] {Pud05}.

The formation of more massive stars might require substantially different accretion/ejection processes.
For the mass accretion rates predicted by theory \citep[e.g.][]{Shu85}, 
the accretion time becomes longer than the protostar contraction time for masses $\ge$8~\ms. At this point,
the star reaches the Zero Age Main Sequence (ZAMS) and, as a consequence, its radiation pressure would be sufficient to halt the inflow
\citep{Wol87}, thus preventing further increase of the stellar mass. 
A number of scenarios have been put forward
to solve this ``radiation pressure'' problem for high-mass star formation:
much higher mass accretion rates, competitive accretion \citep{Bon03} 
and/or stellar mergers \citep{Bon98} in clusters. 

However, recent 3-D, radiative hydrodynamic simulations of high-mass star formation (SF) seem to indicate that accretion disks 
and collimated outflows could also be an effective mechanism for assembling very massive stars, up to
$\approx$140~\ms\  \citep[e.g.][]{Kui10}.
The disk geometry focuses the accretion flow onto the equatorial plane, 
allowing the accreting material to overcome the strong radiation pressure and reach the stellar surface. 
These models also predict beaming of the stellar photons into the lower-density outflow cavity, 
which helps to alleviate the radiation pressure in the equatorial plane.
The size of accretion disks around high-mass YSOs is predicted to be on the order of only a few 100~AU, whereas clustering of multiple forming stars should occur on scales of \ 10$^3$~AU \citep[e.g.][]{Kru09}.

From the above it is clear that to improve our understanding of the accretion and ejection processes in high-mass YSOs, 
it is essential to image linear scales as small as
10--100~AU (requiring angular resolutions $<$0\farcs1 
at the typical distances $>$ 1~kpc of massive star forming regions) to 
resolve the gas kinematics around single objects.
To this purpose very useful diagnostic tools are the intense molecular 
(in particular SiO, \wat, and \meth) masers often observed in proximity to high-mass YSOs.
Thanks to their high brightness temperature ($\ge$10$^9$~K), molecular masers can be targets of
VLBI observations, which, achieving sub-milliarcsec angular resolutions, 
permit us to derive the proper motions of the maser ``spots'' 
(i.e., the single maser emission centers) and access the full 3-D gas kinematics. 
However for exact reconstruction of the kinematics the different observations have to be astrometrically registered. The only way to register mm-VLBI reliably is with the SFPR method, and the best observing efficiency for SFPR is provided with simultaneous mm-VLBI receiver systems \citep{rioja_14}.
Combining maser VLBI data with (sub-arcsecond) interferometric observations of thermal (continuum and line) 
emissions, one can get a very detailed view of the gas kinematics and physical conditions near the forming star.
Such a study has been performed so far only towards a small number of objects \citep[see discussion in][]{moscadelli_11}. In the following we illustrate
one of the most remarkable results of maser VLBI applied to high-mass SF.  

\begin{figure}[htbp]

\includegraphics[width=1.0\textwidth]{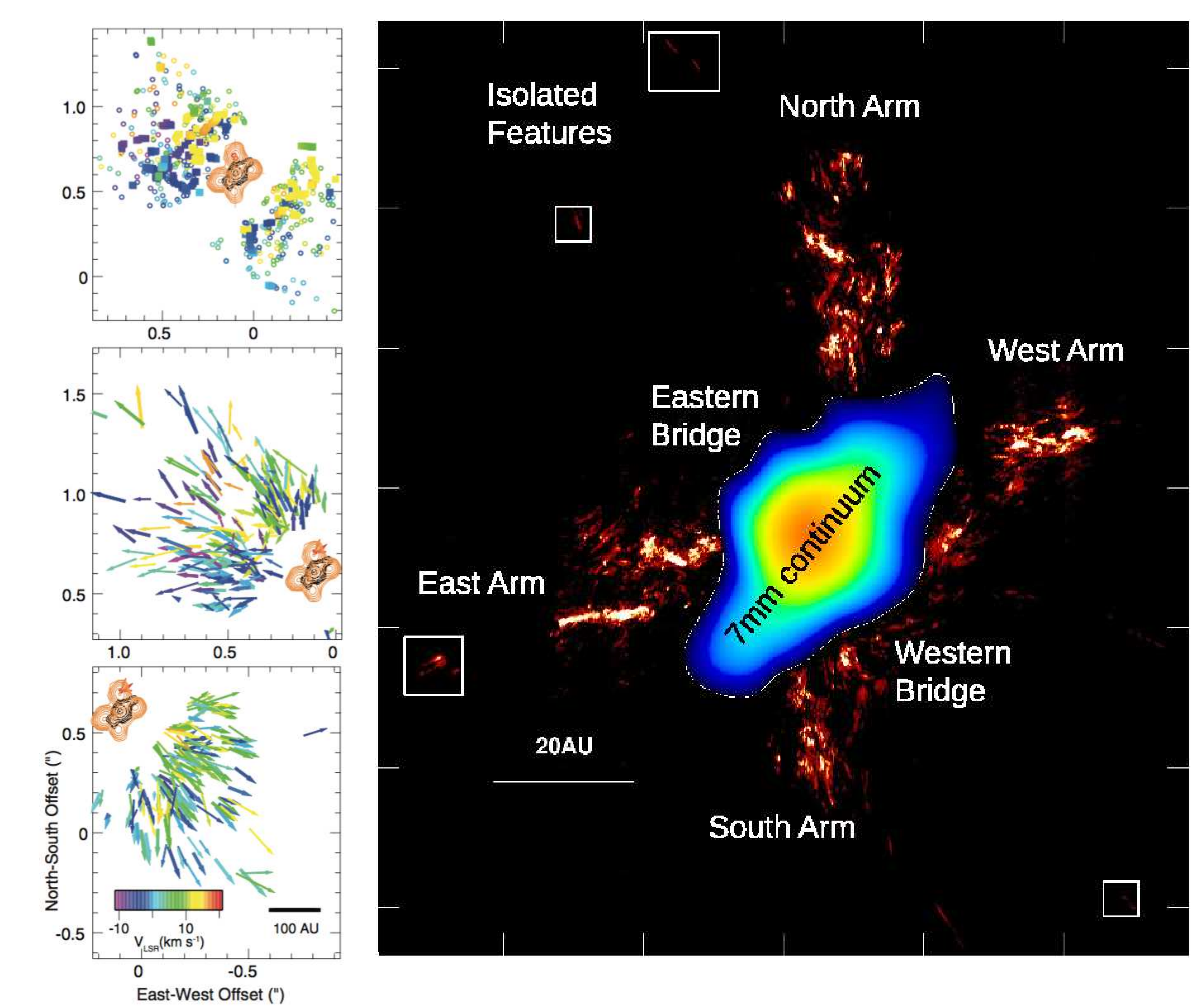} 
\caption{Maser kinematics in Source~I. {\bf Left Panel)} Top:~SiO v=0 and \wat\ masers ({\em open
circles} and {\em squares}, respectively -- \cite[Fig. 3][]{Gre13}), velocity-integrated SiO v=1 masers ({\em red
contours} -- \cite{God09}), and 7~mm continuum ({\em black contours} -- \cite{Rei07}), as mapped with
the JVLA. Middle:~expanded view of SiO v=0 maser proper motions in the
northeastern lobe of Source I. Bottom:~expanded view of the southwest lobe. 
Colors denote maser \Vlsr\ as indicated in the wedge on the bottom left corner of the bottom panel.
\ {\bf Right Panel)} Combined (velocity-integrated) SiO v=1 and v=2 maser emission 
distribution ({\em red-tone image}) observed with the VLBA  \citep[Fig. 5][]{Mat10}, and 7~mm continuum ({\em color map}) 
imaged with the JVLA \citep{Rei07}. \label{fig:SourceI}}
\label{SourceI}
\end{figure}

\subsubsection{Source I in Orion~BN/KL}
The closest (415 pc) and best studied high-mass YSO is Source~I in the Orion~BN/KL region. 
The continuum emission of Source~I, imaged at 7~mm with the JVLA by \cite{Rei07},
is highly elongated and consistent with an accretion disk ionised by a YSO of $\approx$10~\ms.
This source has been monitored by \cite{Mat10} with the Very Long Baseline Array (VLBA)
 in vibrationally-excited SiO maser
transitions every month for over three years, and a movie of the 3-D molecular gas
flow was created. The SiO masers are distributed symmetrically around the ionised disk,
outlining four radial arms connected by two tangential bridges. The pattern of SiO maser proper motions
shows that material in the bridges rotates about the disk axis, whereas gas in the arms streams radially 
away from the star. Thus, the vibrationally-excited SiO masers are tracing both a compact rotating disk 
and a wide-angle wind emanating from the disk at radii $<$ 100~AU (Fig.~\ref{fig:SourceI}, right panel), 
which can be modeled in terms of a magnetocentrifugally driven wind \citep{Vai13}.
JVLA imaging of ground-state SiO maser transitions,
probing larger scales, shows that the wide-angle wind
collimates into a bipolar outflow at radii of 100--1000~AU (Fig.~\ref{fig:SourceI}, left panel)
\citep{Gre13}. This study has provided direct evidence for the formation of a massive star via disk-mediated
accretion and revealed for the first time the launch and
collimation region of an outflow from a rotating compact
disk on scales comparable with the Solar system.
SiO masers are ideal candidates for SFPR astrometric registration, as the frequency ratio between the transitions is extremely close to integer. 

\subsubsection{Advantages of multi-frequency maser observations}

\begin{figure}
\centering
\includegraphics[width=1.0\textwidth]{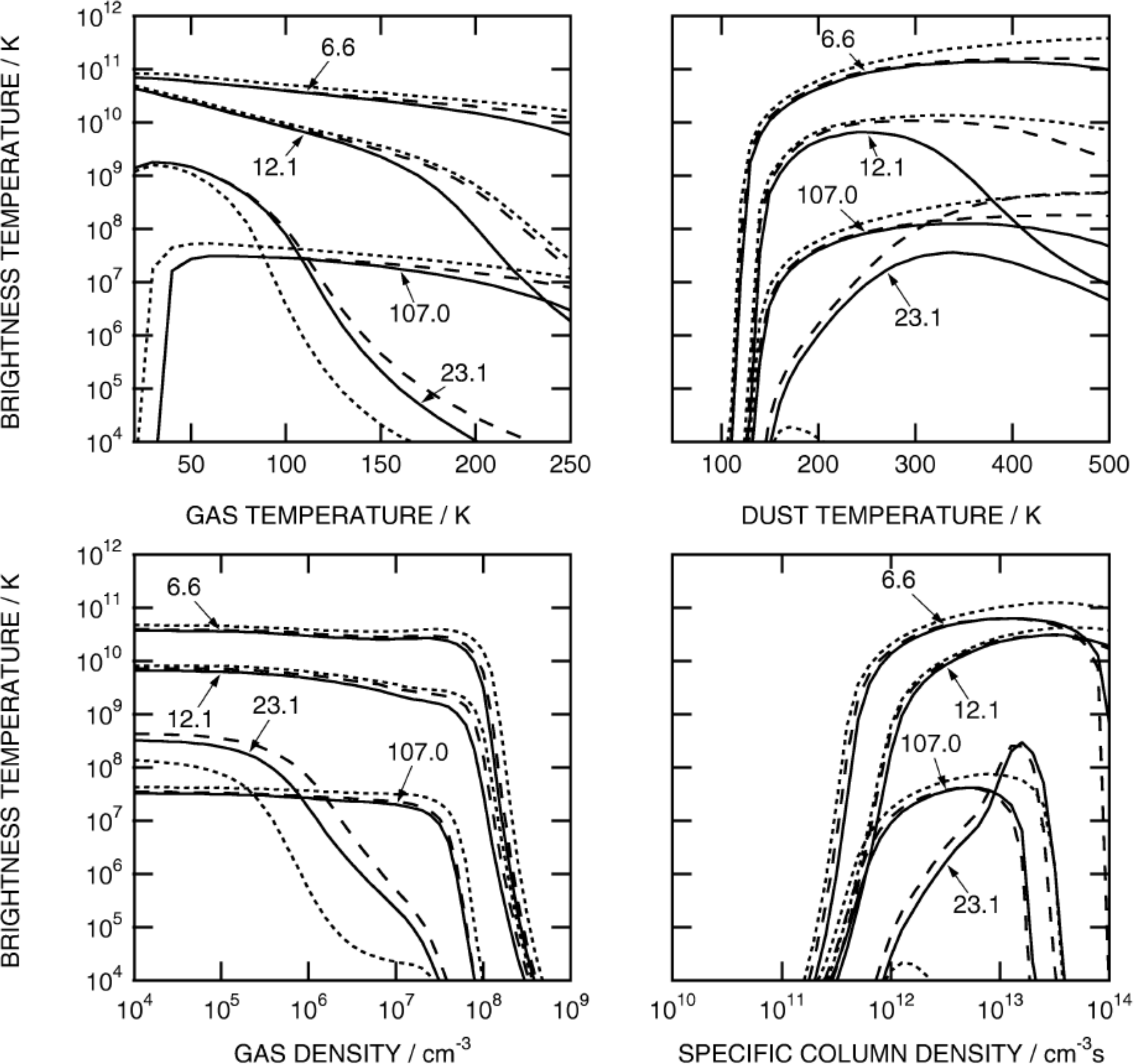}  
\caption{Effect of parameter (gas and dust temperature, gas and specific column density) variations 
on the brightness temperature of four different methanol masers.
The four curves are labelled with the frequency of the maser transition to which they refer: 6.6, 12.1,  23.1 and 107~GHz (reproduced from Fig. 2 in \citet{cragg_05}).
\label{fig:cragg}}
\end{figure}

High-frequency, single-dish studies have shown that the commonly observed, centimeter wavelength, masing molecules emit many masing transitions at millimeter wavelengths, too. For instance, for methanol, 16 distinct maser transitions have been detected over the frequency range \ 6--240~GHz \citep[see, for instance,][]{ellingsen_12}, 
and for water, the 16 (as of 2016, \citet[Tab. 6]{2016MNRAS.456..374G})  maser lines cover the frequency range \ 22--660~GHz \citep[see, for instance,][]{neufeld_13}. 
These ``millimeter wavelength" masers open up new perspectives for star-formation studies as, in general, for a given molecule, models of maser pumping predict that transitions at different frequencies are strongly inverted and become intense masers under varying physical and chemical conditions (described, in particular, in terms of (n$_{H_2}$)~density, gas and dust temperature, molecular abundance; see, for instance, \citet{neufeld_91}, for water, and \citet{cragg_05}, for methanol masers).
As an example, Fig. \ref{fig:cragg} shows the dependence of the brightness temperature of four different methanol masers on various physical parameters, as predicted by the models of \citet{cragg_05}.
The intensity ratios of different maser transitions  from the same molecule can thus be used to probe the physical and chemical conditions of the gas, as long as the observations are sufficiently well registered to ensure that the measurements arise from the same emission regions. The precision with which the physical parameters of the gas can be determined with this method clearly increases with the number of observed maser transitions. To this purpose the role of  the newly-detected millimeter wavelength masers will be fundamental, as will the precision alignment of the observations in different wave bands. VLBI observations of maser emissions over an extended frequency range could in principle allow us to map the density and temperature of the gas surrounding the high-mass YSO with milliarcsecond accuracy. 
Whilst these observations hitherto have not been observed simultaneously, given that we can now register different transitions together using the SFPR method, such capabilities would greatly enhance the scientific return from the observations. 
The basic assumption of the maser models, which can be observationally verified, is that the various maser lines are effectively emerging from the same volume of gas. If applied to a large enough sample of sources, multi-frequency maser VLBI, by constraining in a consistent way the physical conditions and kinematics of the gas (see below)  in the proximity of (within 10--100~AU from) the high-mass YSOs, will be very useful to track the evolutionary phase of the high-mass YSOs, which is presently poorly determined owing to insufficient angular resolution and/or ambiguous interpretation of the observables. 

\subsubsection{Prospects for the Future}
 
Maser VLBI offers a unique tool to investigate high-mass SF at the smallest ($\sim$10~AU) scales, unaccessible even to the new generation of millimeter interferometers. In the few objects studied so far using both maser VLBI and thermal interferometric observations, the maser 3-D kinematics and the \Vlsr\ distribution of the (thermal) tracers of high-density gas complement each other, providing a picture of the gas motion around the forming star at radii from 10 to 10$^4$~AU. The results obtained seem to indicate that disks and jets could play an important role in the formation of more massive stars, as well. However, maser VLBI reveals also some new interesting kinematic features, whose significance for the stellar formation process has still to be deciphered.

In the case of maser emission, owing to the constraints imposed by the maser pumping, one can expect that a particular region of gas around the YSO offers the most suitable conditions for producing maser action in different lines (from the same molecule). 
In general, this expectation has been mostly confirmed by previous multi-line VLBI observations, but recent interferometric studies have also demonstrated that maser lines  of very different excitation  can effectively be used to trace different physical conditions, \citet[e.g. discussions in][]{2016MNRAS.456..374G}, which predicts 50 further \wat\ maser transitions which will fall in the ALMA bands.
As an example, \cite{hirota_14b} have recently performed ALMA observations of two millimeter, water lines, the 321~GHz (1862~K above the ground state) and the 336~GHz (at 2956~K) lines, towards Source~I in Orion~KL. While the spatial and velocity distribution of the 321~GHz emission is elongated and traces the root of the outflow close to the surface of the accretion disk, the 336~GHz line has a compact structure and is most likely emerging from hot ($\approx$3000~K) neutral material orbiting the YSO in an edge-on small (radius of 50~AU) ring. 
Therefore, VLBI observations of different maser lines from the same YSO could turn to be even more strategic for kinematic studies, if the spatial associations are registered, to determine the physical parameters of the gas.

\section{Time domain observations of X-ray binary jets}\label{sec:xrayb}

Black hole X-ray binaries (BH XRBs) show much of the same physics as observed in AGN, exhibiting relativistic jets powered by an accretion flow.  However, while they are typically factors of $10^2$--$10^5$ times closer than nearby AGN, their black hole masses are $10^5$--$10^8$ times smaller.  This implies that the physical size scales that we can probe with high-frequency VLBI correspond to a larger number of gravitational radii in BH XRBs than in AGN.  Nonetheless, since timescales close to a black hole scale with mass, BH XRBs afford us the unique advantage of studying the physical processes of jet launching and evolution in real time.


In the current phenomenological picture describing XRB outbursts \citep{Fender:2004dt}, compact, steady jets are seen in the hard X-ray spectral state at the beginning and end of an outburst. Such compact jets have size scales of a few AU, and have been directly resolved along the jet axis in two systems, using VLBI \citep{dhawan_00,stirling_01}.  While these compact jets are believed to have a relatively low Lorentz factor ($\Gamma <$ 2), {\em this has never been directly measured on VLBI scales via the relative proper motions of approaching and receding jet components}. The detection of a counterjet in GRS\,1915+105 was used to suggest a compact jet speed to 0.3--0.5$c$ via the jet/counterjet brightness ratio \citep{ribo_04}, but the variations between different frequencies and epochs, coupled with the uncertainty in both the jet structure (discrete or continuous flow) and the distance and inclination angle of the source makes it hard to draw definitive conclusions on the true Lorentz factor.

At the peak of the outburst, there is a transition to a softer X-ray spectral state, upon which discrete, transient ejecta are launched.  These are much brighter than the compact jets, and move away from the central source with proper motions of up to tens of milliarcseconds per day, in some cases showing apparent superluminal motion \citep[e.g.][]{Mirabel:1994gl,tingay_95}.  While the proper motions of the ejecta can be used to place constraints on fundamental jet parameters such as speed, inclination angle, and the associated Lorentz and Doppler factors \citep[e.g.][]{1999ARA&A..37..409M}, the typical large uncertainties in source distance imply that we can only place lower limits on the true Lorentz factors \citep{fender_03}. VLBI studies of these transient ejecta are complicated by the strong changes in jet flux density and morphology over the course of an observing run, which violate the fundamental assumptions of aperture synthesis.  Such issues become a particular problem at mm-wavelengths, where the amplitude variability is stronger and the angular resolution is higher.  However, for typical BH XRBs the jet structure is relatively simple, being made up of a few unresolved components.  Thus, an observation can be broken up into multiple smaller chunks, which do not individually violate the key tenets of interferometry, and can be imaged and analysed separately \citep[e.g.][]{dhawan_00}.  Alternatively, high time resolution photometry at multiple radio frequencies can be used to glean some information on the jet structure.  The characteristic variability timescale at any given frequency provides a measure of the jet size scale at the surface where the optical depth is unity at that frequency (i.e.\ the radius of the jets).  As an example, \citet{miller-jones_09} used the minimum variability timescale seen in simultaneous 15 and 43-GHz JVLA observations of Cygnus X-3 to determine the characteristic size of the jets during a period of minor flaring, probing scales of 2--4\,AU --- significantly smaller than that which can be resolved with VLBI.

The {\it combination} of both a resolved radio jet and sensitive, high time-resolution photometry at multiple radio frequencies, together with simultaneous X-ray observations, can provide an enhanced and unique probe of the structure and geometry of BH XRB jets. As the X-ray variability from the accretion flow propagates downstream in the jet, we can observe it being manifested in the jet.  Since we see emission from the surface of optical depth unity at any given frequency, this is seen first at optical/infrared frequencies from closest to the jet base, and moves to lower frequency with time as the fluctuations propagate downstream.  This has already been demonstrated via correlations between optical/infrared and X-ray variability \citep{casella_10,kalamkar_16}, but not to date in the mm or radio bands.  With simultaneous, high-sensitivity observations at multiple mm wavelengths (where the variations are stronger and less smoothed out than in the cm band), we can determine the time delay between frequencies as a perturbation propagates downstream. If the characteristic height along the jet is also known from the VLBI observations, along with the exact launch time from simultaneous X-ray observations, these then give a direct measure of the jet speed.  Together with constraints on the jet radius from the characteristic variability timescales at the different frequencies, we can also probe the jet geometry, in particular the opening angle, which is typically unknown in BH XRBs \citep{miller-jones_06}.  Thus, simultaneous observations at multiple mm bands can provide one of the few feasible means for directly measuring the jet speed and geometry.

\begin{figure}[htbp]
\begin{center}
\includegraphics[width=1.0\textwidth]{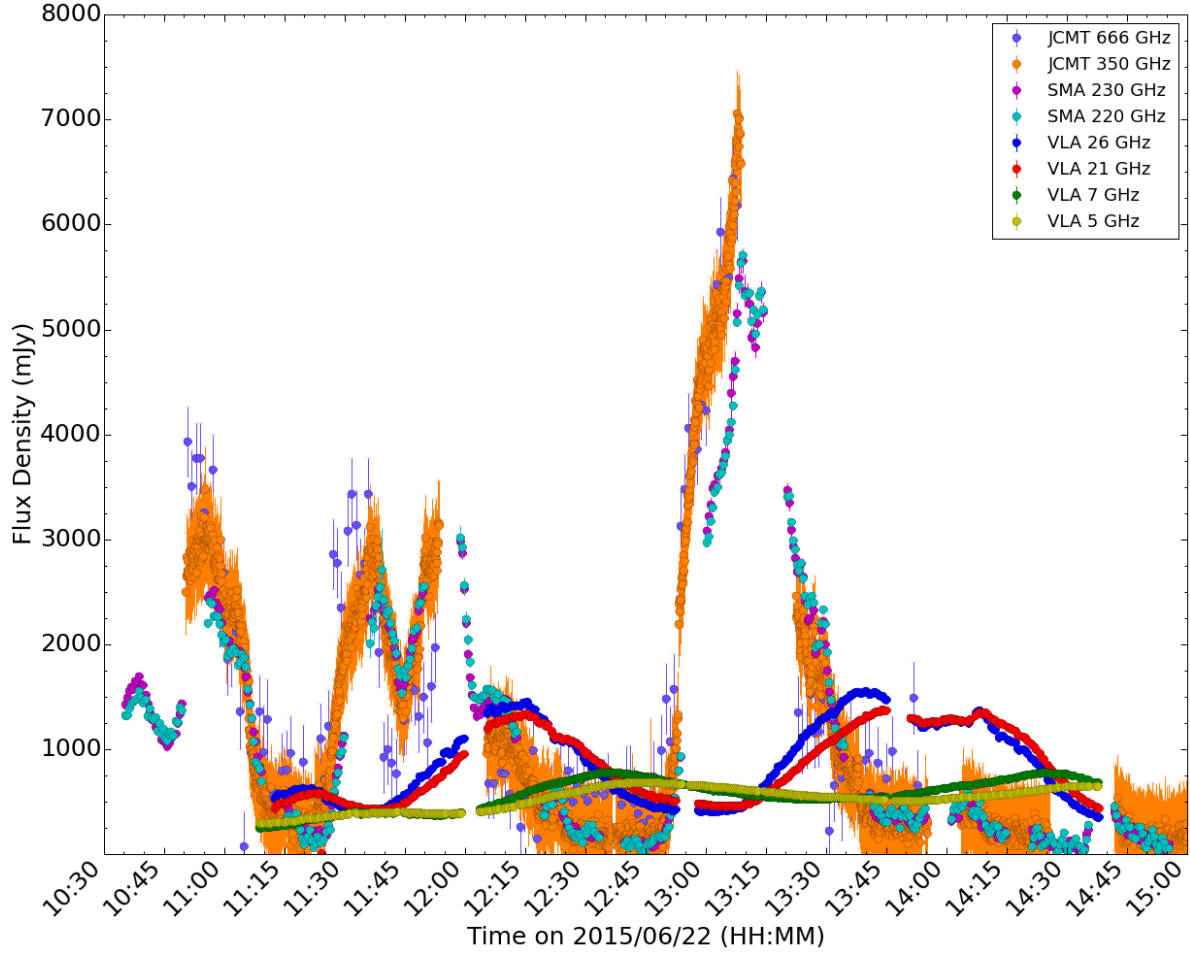}
\end{center}
\caption{Simultaneous multi-frequency light curves of the 2015 outburst of the BH XRB V404 Cygni, taken using the SMA, JCMT, and the JVLA operating in sub-array mode \citep{tetarenko_17}. Rapid variability corresponding to the ejection, expansion and fading of multiple jet knots (directly imaged at 2\,cm with the VLBA; Miller-Jones et al.\ in prep.) is observed.  The higher-frequency light curves peak earlier, at higher flux densities, and are less smoothed out than their counterparts at lower frequencies. This trend continues from the radio bands all the way up to the sub-mm bands, as evidenced by the JCMT observations at frequencies as high as 666\,GHz 
\citep[figure taken from][]{tetarenko_17}.}
\label{fig:v404}
\end{figure}

%
We have started trials of this VLBI$+$multi-band photometry technique using observations of bright transient ejection events from the 2015 outburst of the BH XRB V404 Cyg  \citep[Miller-Jones et al., in prep,][]{tetarenko_17}
The combination of VLBI imaging and Markov chain Monte Carlo  modeling of our multi-frequency light curves (Fig.~\ref{fig:v404}) is providing a powerful probe of the geometry, speed, structure, and energetics of the discrete ejecta that were launched during these flaring events.  This technique will subsequently be extended to the compact jets seen in sufficiently bright BH XRB outbursts.  However, the existing observations have required a combination of the JVLA in sub-array mode to provide the multi-frequency photometry, and the VLBA to provide the imaging.  The adoption of a simultaneous multi-wavelength mm-VLBI capability would enable such observations on a single instrument, thereby removing both the double-jeopardy and logistical complications involved in scheduling strictly-simultaneous observations on two different facilities.
Clearly this combination of high-resolution spatial and time domain analysis will benefit immeasurably from observations with continuous, contemporaneous mm-wave systems such as that provided by the KVN, once the capability is extended to longer and more sensitive global baselines.

\section{Concluding Remarks}


We have presented a few highlights of the science cases which would be advanced by the wide-spread adoption of recording systems capable of allowing simultaneous mm-wavelength VLBI. These are summarised in Table \ref{tab:sum}. 

The science goals cover a range of spectral line and continuum VLBI science, such as AGN polarimetric studies, and alignment of maser transitions to probe different physical conditions around massive star forming regions and AGB stars.  
Registration between the different frequencies is essential for the scientific interpretation in these fields. New methods and technologies have allowed for registration of the mm-wavelengths.
Similarly, time-sensitive observations of multiple frequencies in, for example, X-ray binaries are best delivered by systems capable of observing a wide range of frequencies simultaneously.
There are a multitude of possible technological solutions that could deliver these capabilities, and these were explored in the ERATec meeting, October 2015 \citep{bolli_eratec}. VLBA style fast switching has been demonstrated successfully for VLBI, but has significant observational overheads and is very unlikely to succeed at frequencies greater than 86GHz. KVN style receivers are large but have much reduced overheads and have worked at all KVN frequencies (i.e. up to 130GHz). However, a compact three-band single-Dewar receiver has recently been developed, which would be compatible with most single frequency receiver slots. Sub-arraying solutions are being investigated at the ATCA. These have demonstrated FPT and are expected to successfully demonstrate SFPR shortly, once the polarisation issues are resolved. Both ALMA and JVLA also can support sub-arraying modes. Other solutions are possible, for example there were exciting reports on innovative meta-material solutions in the ERATec meeting.

The cases presented in this paper represent the benefits that will accrue to existing scientific projects. The outcome of new technologies is always that they enable new measurements and open new scientific applications, so those listed here will only be a small subset of the expected fields of study. 

\begin{table*}[ht]
    \centering \begin{footnotesize}
    
\begin{tabular}{c|c|c}
\hline 
\hline 
Science Case (section) & Domain  & Requirements \\ 
\hline 
Weak Sources (\ref{sec:weak}) & Continuum VLBI  & mm-band phase stabilisation \\
\hline 
Spatial Astrometry  (\ref{sec:astrometry}) & VLBI  & mm-band phase referencing \\
\hline 
Faraday Rotation Studies (\ref{sec:faraday}) \& & Continuum VLBI  & mm-band frequency  \\
Spectral Index  (\ref{sec:core})  &   &  phase referencing \\
\hline 
Alignment of Maser transitions   & Line VLBI for Star-forming  & mm-band frequency \\
  (\ref{sec:evolved} \& \ref{sec:massive} )        &  \& Evolved Star regions  & phase referencing \\
\hline 
X-ray Binaries  (\ref{sec:xrayb}) & Temporal studies  & Simultaneous mm-band observing \\
\hline 
\end{tabular}
    \end{footnotesize}
\caption{Summary table for the science goals described in this review, along with the research domain which is relevant and the instrumental requirements for delivery of such capability.} \label{tab:sum}
\end{table*}
\paragraph{Acknowledgements:}
Figure 1 is reproduced by permission of POS. Figures 2 to 5, 8 and 10 are reproduced by permission of the AAS. Figure 6 and 7 are reproduced by permission of A\&A. Figure 9 is reproduced by permission of MNRAS.  TJ, MR, JG, RD, VB and LM wrote Section 2.1, 2.2, 2.3, 2.4, 3.1 and 3.2 respectively. JMJ, AT and GS contributed Section 4. We thank the referee for the helpful comments, which have improved the document.

\begin{footnotesize}

\end{footnotesize}


\begin{thebibliography}{164}
\expandafter\ifx\csname natexlab\endcsname\relax\def\natexlab#1{#1}\fi
\expandafter\ifx\csname url\endcsname\relax
  \def\url#1{\texttt{#1}}\fi
\expandafter\ifx\csname urlprefix\endcsname\relax\def\urlprefix{URL }\fi
\providecommand{\eprint}[2][]{\url{#2}}
\providecommand{\bibinfo}[2]{#2}
\ifx\xfnm\relax \def\xfnm[#1]{\unskip,\space#1}\fi
\bibitem[{Asada et~al.(2008a)Asada, Inoue, Kameno and Nagai}]{Asada:2008dq}
\bibinfo{author}{Asada, K.}, \bibinfo{author}{Inoue, M.},
  \bibinfo{author}{Kameno, S.}, \bibinfo{author}{Nagai, H.},
  \bibinfo{year}{2008}a.
\newblock \bibinfo{title}{{Time Variation of the Rotation Measure Gradient in
  the 3C 273 Jet}}.
\newblock \bibinfo{journal}{ApJ} \bibinfo{volume}{675},
  \bibinfo{pages}{79--82}.
\bibitem[{Asada et~al.(2008b)Asada, Inoue, Nakamura, Kameno and
  Nagai}]{Asada:2008gb}
\bibinfo{author}{Asada, K.}, \bibinfo{author}{Inoue, M.},
  \bibinfo{author}{Nakamura, M.}, \bibinfo{author}{Kameno, S.},
  \bibinfo{author}{Nagai, H.}, \bibinfo{year}{2008}b.
\newblock \bibinfo{title}{{Multifrequency Polarimetry of the NRAO 140 Jet:
  Possible Detection of a Helical Magnetic Field and Constraints on Its Pitch
  Angle}}.
\newblock \bibinfo{journal}{ApJ} \bibinfo{volume}{682},
  \bibinfo{pages}{798--802}.
\bibitem[{Asada et~al.(2002)Asada, Inoue, Uchida, Kameno, Fujisawa, Iguchi and
  Mutoh}]{Asada:2002to}
\bibinfo{author}{Asada, K.}, \bibinfo{author}{Inoue, M.},
  \bibinfo{author}{Uchida, Y.}, \bibinfo{author}{Kameno, S.},
  \bibinfo{author}{Fujisawa, K.}, \bibinfo{author}{Iguchi, S.},
  \bibinfo{author}{Mutoh, M.}, \bibinfo{year}{2002}.
\newblock \bibinfo{title}{{A Helical Magnetic Field in the Jet of 3C 273}}.
\newblock \bibinfo{journal}{PASJ} \bibinfo{volume}{54}, \bibinfo{pages}{L39}.
\bibitem[{Asada et~al.(2010)Asada, Nakamura, Inoue, Kameno and
  Nagai}]{Asada:2010dn}
\bibinfo{author}{Asada, K.}, \bibinfo{author}{Nakamura, M.},
  \bibinfo{author}{Inoue, M.}, \bibinfo{author}{Kameno, S.},
  \bibinfo{author}{Nagai, H.}, \bibinfo{year}{2010}.
\newblock \bibinfo{title}{{Multi-frequency Polarimetry toward S5 0836+710: A
  Possible Spine-Sheath Structure for the Jet}}.
\newblock \bibinfo{journal}{ApJ} \bibinfo{volume}{720},
  \bibinfo{pages}{41--45}.
\bibitem[{{Asaki} et~al.(2016){Asaki}, {Matsushita}, {Fomalont}, {Corder},
  {Nyman}, {Dent}, {Philips}, {Hirota}, {Takahashi}, {Vila-Vilaro}, {Nikolic},
  {Hunter}, {Remijan} and {Vlahakis}}]{2016SPIE.9906E..5UA}
\bibinfo{author}{{Asaki}, Y.}, \bibinfo{author}{{Matsushita}, S.},
  \bibinfo{author}{{Fomalont}, E.B.}, \bibinfo{author}{{Corder}, S.A.},
  \bibinfo{author}{{Nyman}, L.{\AA}.}, \bibinfo{author}{{Dent}, W.R.F.},
  \bibinfo{author}{{Philips}, N.M.}, \bibinfo{author}{{Hirota}, A.},
  \bibinfo{author}{{Takahashi}, S.}, \bibinfo{author}{{Vila-Vilaro}, B.},
  \bibinfo{author}{{Nikolic}, B.}, \bibinfo{author}{{Hunter}, T.R.},
  \bibinfo{author}{{Remijan}, A.}, \bibinfo{author}{{Vlahakis}, C.},
  \bibinfo{year}{2016}.
\newblock \bibinfo{title}{{ALMA long baseline phase calibration using phase
  referencing}}, in: \bibinfo{booktitle}{Ground-based and Airborne Telescopes
  VI}, p. \bibinfo{pages}{99065U}.
\bibitem[{{Asaki} et~al.(1996){Asaki}, {Saito}, {Kawabe}, {Morita} and
  {Sasao}}]{asaki_fpt_96}
\bibinfo{author}{{Asaki}, Y.}, \bibinfo{author}{{Saito}, M.},
  \bibinfo{author}{{Kawabe}, R.}, \bibinfo{author}{{Morita}, K.I.},
  \bibinfo{author}{{Sasao}, T.}, \bibinfo{year}{1996}.
\newblock \bibinfo{title}{{Phase compensation experiments with the paired
  antennas method}}.
\newblock \bibinfo{journal}{Radio Science} \bibinfo{volume}{31},
  \bibinfo{pages}{1615--1626}.
\bibitem[{{Balick} and {Frank}(2002)}]{balick_02}
\bibinfo{author}{{Balick}, B.}, \bibinfo{author}{{Frank}, A.},
  \bibinfo{year}{2002}.
\newblock \bibinfo{title}{{Shapes and Shaping of Planetary Nebulae}}.
\newblock \bibinfo{journal}{ARAA} \bibinfo{volume}{40},
  \bibinfo{pages}{439--486}.
\bibitem[{Bartolini et~al.(2016)Bartolini, Bolli, Keller, Dodson, Lindqvist,
  K-H., Hafok, Rioja, Orlati, Zanichelli, Braschi, Comoretto, Nesti, Panella
  and Stagni}]{bolli_eratec}
\bibinfo{author}{Bartolini, M.}, \bibinfo{author}{Bolli, P.},
  \bibinfo{author}{Keller, R.}, \bibinfo{author}{Dodson, R.},
  \bibinfo{author}{Lindqvist, M.}, \bibinfo{author}{K-H., M.},
  \bibinfo{author}{Hafok, H.}, \bibinfo{author}{Rioja, M.},
  \bibinfo{author}{Orlati, A.}, \bibinfo{author}{Zanichelli, A.},
  \bibinfo{author}{Braschi, P.}, \bibinfo{author}{Comoretto, G.},
  \bibinfo{author}{Nesti, R.}, \bibinfo{author}{Panella, D.},
  \bibinfo{author}{Stagni, M.}, \bibinfo{year}{2016}.
\newblock \bibinfo{title}{{Report on the 4th RadioNet3 European Radio Astronomy
  Technical Forum Workshop on MULTI-FREQUENCY MM-WAVE RADIO TELESCOPES OTHER
  SOFTWARE CONTROLLED OPERATIONS}}.
\newblock \bibinfo{journal}{Colle di Galileo} \bibinfo{volume}{5},
  \bibinfo{pages}{33--37}.
\bibitem[{Blandford and K{\"o}nigl(1979)}]{Blandford:1979eg}
\bibinfo{author}{Blandford, R.D.}, \bibinfo{author}{K{\"o}nigl, A.},
  \bibinfo{year}{1979}.
\newblock \bibinfo{title}{{Relativistic jets as compact radio sources}}.
\newblock \bibinfo{journal}{ApJ} \bibinfo{volume}{232}, \bibinfo{pages}{34}.
\bibitem[{{Bonnell} et~al.(2003){Bonnell}, {Bate} and {Vine}}]{Bon03}
\bibinfo{author}{{Bonnell}, I.A.}, \bibinfo{author}{{Bate}, M.R.},
  \bibinfo{author}{{Vine}, S.G.}, \bibinfo{year}{2003}.
\newblock \bibinfo{title}{{The hierarchical formation of a stellar cluster}}.
\newblock \bibinfo{journal}{\mnras} \bibinfo{volume}{343},
  \bibinfo{pages}{413--418}.
\newblock \eprint{astro-ph/0305082}.
\bibitem[{{Bonnell} et~al.(1998){Bonnell}, {Bate} and {Zinnecker}}]{Bon98}
\bibinfo{author}{{Bonnell}, I.A.}, \bibinfo{author}{{Bate}, M.R.},
  \bibinfo{author}{{Zinnecker}, H.}, \bibinfo{year}{1998}.
\newblock \bibinfo{title}{{On the formation of massive stars}}.
\newblock \bibinfo{journal}{\mnras} \bibinfo{volume}{298},
  \bibinfo{pages}{93--102}.
\newblock \eprint{astro-ph/9802332}.
\bibitem[{{Bujarrabal}(1994)}]{bujarrabal_94}
\bibinfo{author}{{Bujarrabal}, V.}, \bibinfo{year}{1994}.
\newblock \bibinfo{title}{{Numerical calculations of SiO maser emission II.
  Angular extent}}.
\newblock \bibinfo{journal}{\aap} \bibinfo{volume}{285}.
\bibitem[{{Bujarrabal} et~al.(2001){Bujarrabal}, {Castro-Carrizo}, {Alcolea}
  and {S{\'a}nchez Contreras}}]{bujarrabal_01}
\bibinfo{author}{{Bujarrabal}, V.}, \bibinfo{author}{{Castro-Carrizo}, A.},
  \bibinfo{author}{{Alcolea}, J.}, \bibinfo{author}{{S{\'a}nchez Contreras},
  C.}, \bibinfo{year}{2001}.
\newblock \bibinfo{title}{{Mass, linear momentum and kinetic energy of bipolar
  flows in protoplanetary nebulae}}.
\newblock \bibinfo{journal}{A\&A} \bibinfo{volume}{377},
  \bibinfo{pages}{868--897}.
\bibitem[{{Bujarrabal} and {Nguyen-Q-Rieu}(1981)}]{bujarrabal_81}
\bibinfo{author}{{Bujarrabal}, V.}, \bibinfo{author}{{Nguyen-Q-Rieu}},
  \bibinfo{year}{1981}.
\newblock \bibinfo{title}{{Collisional and radiative excitation of SiO
  masers}}.
\newblock \bibinfo{journal}{A\&A} \bibinfo{volume}{102},
  \bibinfo{pages}{65--72}.
\bibitem[{{Carilli} and {Holdaway}(1999)}]{carilli_99}
\bibinfo{author}{{Carilli}, C.L.}, \bibinfo{author}{{Holdaway}, M.A.},
  \bibinfo{year}{1999}.
\newblock \bibinfo{title}{{Tropospheric phase calibration in millimeter
  interferometry}}.
\newblock \bibinfo{journal}{Radio Science} \bibinfo{volume}{34},
  \bibinfo{pages}{817--840}.
\newblock \eprint{astro-ph/9904248}.
\bibitem[{Casadio et~al.(2015)Casadio, G{\'o}mez, Jorstad, Marscher, Larionov,
  Smith, Gurwell, L{\"a}hteenm{\"a}ki, Agudo, Molina, Bala, Joshi, Taylor,
  Williamson, Arkharov, Blinov, Borman, Di~Paola, Grishina, Hagen-Thorn, Itoh,
  Kopatskaya, Larionova, Larionova, Morozova, Rastorgueva-Foi, Sergeev,
  Tornikoski, Troitsky, Thum and Wiesemeyer}]{2015ApJ...813...51C}
\bibinfo{author}{Casadio, C.}, \bibinfo{author}{G{\'o}mez, J.L.},
  \bibinfo{author}{Jorstad, S.G.}, \bibinfo{author}{Marscher, A.P.},
  \bibinfo{author}{Larionov, V.M.}, \bibinfo{author}{Smith, P.S.},
  \bibinfo{author}{Gurwell, M.A.}, \bibinfo{author}{L{\"a}hteenm{\"a}ki, A.},
  \bibinfo{author}{Agudo, I.}, \bibinfo{author}{Molina, S.N.},
  \bibinfo{author}{Bala, V.}, \bibinfo{author}{Joshi, M.},
  \bibinfo{author}{Taylor, B.}, \bibinfo{author}{Williamson, K.E.},
  \bibinfo{author}{Arkharov, A.A.}, \bibinfo{author}{Blinov, D.A.},
  \bibinfo{author}{Borman, G.A.}, \bibinfo{author}{Di~Paola, A.},
  \bibinfo{author}{Grishina, T.S.}, \bibinfo{author}{Hagen-Thorn, V.A.},
  \bibinfo{author}{Itoh, R.}, \bibinfo{author}{Kopatskaya, E.N.},
  \bibinfo{author}{Larionova, E.G.}, \bibinfo{author}{Larionova, L.V.},
  \bibinfo{author}{Morozova, D.A.}, \bibinfo{author}{Rastorgueva-Foi, E.},
  \bibinfo{author}{Sergeev, S.G.}, \bibinfo{author}{Tornikoski, M.},
  \bibinfo{author}{Troitsky, I.S.}, \bibinfo{author}{Thum, C.},
  \bibinfo{author}{Wiesemeyer, H.}, \bibinfo{year}{2015}.
\newblock \bibinfo{title}{{A Multi-wavelength Polarimetric Study of the Blazar
  CTA 102 during a Gamma-Ray Flare in 2012}}.
\newblock \bibinfo{journal}{ApJ} \bibinfo{volume}{813}, \bibinfo{pages}{51}.
\bibitem[{{Casella} et~al.(2010){Casella}, {Maccarone}, {O'Brien}, {Fender},
  {Russell}, {van der Klis}, {Pe'Er}, {Maitra}, {Altamirano}, {Belloni},
  {Kanbach}, {Klein-Wolt}, {Mason}, {Soleri}, {Stefanescu}, {Wiersema} and
  {Wijnands}}]{casella_10}
\bibinfo{author}{{Casella}, P.}, \bibinfo{author}{{Maccarone}, T.J.},
  \bibinfo{author}{{O'Brien}, K.}, \bibinfo{author}{{Fender}, R.P.},
  \bibinfo{author}{{Russell}, D.M.}, \bibinfo{author}{{van der Klis}, M.},
  \bibinfo{author}{{Pe'Er}, A.}, \bibinfo{author}{{Maitra}, D.},
  \bibinfo{author}{{Altamirano}, D.}, \bibinfo{author}{{Belloni}, T.},
  \bibinfo{author}{{Kanbach}, G.}, \bibinfo{author}{{Klein-Wolt}, M.},
  \bibinfo{author}{{Mason}, E.}, \bibinfo{author}{{Soleri}, P.},
  \bibinfo{author}{{Stefanescu}, A.}, \bibinfo{author}{{Wiersema}, K.},
  \bibinfo{author}{{Wijnands}, R.}, \bibinfo{year}{2010}.
\newblock \bibinfo{title}{{Fast infrared variability from a relativistic jet in
  GX 339-4}}.
\newblock \bibinfo{journal}{\mnras} \bibinfo{volume}{404},
  \bibinfo{pages}{L21--L25}.
\newblock \eprint{1002.1233}.
\bibitem[{Chatterjee et~al.(2011)Chatterjee, Marscher, Jorstad, Markowitz,
  Rivers, Rothschild, McHardy, Aller, Aller, L{\"a}hteenm{\"a}ki, Tornikoski,
  Harrison, Agudo, G{\'o}mez, Taylor and Gurwell}]{2011ApJ...734...43C}
\bibinfo{author}{Chatterjee, R.}, \bibinfo{author}{Marscher, A.P.},
  \bibinfo{author}{Jorstad, S.G.}, \bibinfo{author}{Markowitz, A.},
  \bibinfo{author}{Rivers, E.}, \bibinfo{author}{Rothschild, R.E.},
  \bibinfo{author}{McHardy, I.M.}, \bibinfo{author}{Aller, M.F.},
  \bibinfo{author}{Aller, H.D.}, \bibinfo{author}{L{\"a}hteenm{\"a}ki, A.},
  \bibinfo{author}{Tornikoski, M.}, \bibinfo{author}{Harrison, B.},
  \bibinfo{author}{Agudo, I.}, \bibinfo{author}{G{\'o}mez, J.L.},
  \bibinfo{author}{Taylor, B.W.}, \bibinfo{author}{Gurwell, M.},
  \bibinfo{year}{2011}.
\newblock \bibinfo{title}{{Connection Between the Accretion Disk and Jet in the
  Radio Galaxy 3C 111}}.
\newblock \bibinfo{journal}{ApJ} \bibinfo{volume}{734}, \bibinfo{pages}{43}.
\bibitem[{{Cragg} et~al.(2005){Cragg}, {Sobolev} and {Godfrey}}]{cragg_05}
\bibinfo{author}{{Cragg}, D.M.}, \bibinfo{author}{{Sobolev}, A.M.},
  \bibinfo{author}{{Godfrey}, P.D.}, \bibinfo{year}{2005}.
\newblock \bibinfo{title}{{Models of class II methanol masers based on improved
  molecular data}}.
\newblock \bibinfo{journal}{\mnras} \bibinfo{volume}{360},
  \bibinfo{pages}{533--545}.
\newblock \eprint{astro-ph/0504194}.
\bibitem[{Croke and Gabuzda(2008)}]{Croke:2008bn}
\bibinfo{author}{Croke, S.M.}, \bibinfo{author}{Gabuzda, D.C.},
  \bibinfo{year}{2008}.
\newblock \bibinfo{title}{{Aligning VLBI images of active galactic nuclei at
  different frequencies}}.
\newblock \bibinfo{journal}{MNRAS} \bibinfo{volume}{386},
  \bibinfo{pages}{619--626}.
\bibitem[{{Daly} and {Marscher}(1988)}]{daly_88}
\bibinfo{author}{{Daly}, R.A.}, \bibinfo{author}{{Marscher}, A.P.},
  \bibinfo{year}{1988}.
\newblock \bibinfo{title}{{The gasdynamics of compact relativistic jets}}.
\newblock \bibinfo{journal}{\apj} \bibinfo{volume}{334},
  \bibinfo{pages}{539--551}.
\bibitem[{{Deller} and {Middelberg}(2014)}]{deller_14}
\bibinfo{author}{{Deller}, A.T.}, \bibinfo{author}{{Middelberg}, E.},
  \bibinfo{year}{2014}.
\newblock \bibinfo{title}{{mJIVE-20: A Survey for Compact mJy Radio Objects
  with the Very Long Baseline Array}}.
\newblock \bibinfo{journal}{\aj} \bibinfo{volume}{147}, \bibinfo{pages}{14}.
\newblock \eprint{1310.8191}.
\bibitem[{{Desmurs} et~al.(2007){Desmurs}, {Alcolea}, {Bujarrabal},
  {S{\'a}nchez Contreras} and {Colomer}}]{desmurs_07}
\bibinfo{author}{{Desmurs}, J.F.}, \bibinfo{author}{{Alcolea}, J.},
  \bibinfo{author}{{Bujarrabal}, V.}, \bibinfo{author}{{S{\'a}nchez Contreras},
  C.}, \bibinfo{author}{{Colomer}, F.}, \bibinfo{year}{2007}.
\newblock \bibinfo{title}{{Water vapor and silicon monoxide maser observations
  in the protoplanetary nebula OH\,231.8+4.2}}.
\newblock \bibinfo{journal}{A\&A} \bibinfo{volume}{468},
  \bibinfo{pages}{189--192}.
\newblock \eprint{arXiv:0704.2166}.
\bibitem[{{Desmurs} et~al.(2000){Desmurs}, {Bujarrabal}, {Colomer} and
  {Alcolea}}]{desmurs_00}
\bibinfo{author}{{Desmurs}, J.F.}, \bibinfo{author}{{Bujarrabal}, V.},
  \bibinfo{author}{{Colomer}, F.}, \bibinfo{author}{{Alcolea}, J.},
  \bibinfo{year}{2000}.
\newblock \bibinfo{title}{{VLBA observations of SiO masers: arguments in favor
  of radiative pumping mechanisms}}.
\newblock \bibinfo{journal}{A\&A} \bibinfo{volume}{360},
  \bibinfo{pages}{189--195}.
\bibitem[{{Desmurs} et~al.(2014){Desmurs}, {Bujarrabal}, {Lindqvist},
  {Alcolea}, {Soria-Ruiz} and {Bergman}}]{desmurs_14}
\bibinfo{author}{{Desmurs}, J.F.}, \bibinfo{author}{{Bujarrabal}, V.},
  \bibinfo{author}{{Lindqvist}, M.}, \bibinfo{author}{{Alcolea}, J.},
  \bibinfo{author}{{Soria-Ruiz}, R.}, \bibinfo{author}{{Bergman}, P.},
  \bibinfo{year}{2014}.
\newblock \bibinfo{title}{{SiO masers from AGB stars in the vibrationally
  excited v = 1, v = 2, and v = 3 states}}.
\newblock \bibinfo{journal}{A\&A} \bibinfo{volume}{565}, \bibinfo{pages}{A127}.
\bibitem[{{Dhawan} et~al.(2000){Dhawan}, {Mirabel} and
  {Rodr{\'{\i}}guez}}]{dhawan_00}
\bibinfo{author}{{Dhawan}, V.}, \bibinfo{author}{{Mirabel}, I.F.},
  \bibinfo{author}{{Rodr{\'{\i}}guez}, L.F.}, \bibinfo{year}{2000}.
\newblock \bibinfo{title}{{AU-Scale Synchrotron Jets and Superluminal Ejecta in
  GRS 1915+105}}.
\newblock \bibinfo{journal}{\apj} \bibinfo{volume}{543},
  \bibinfo{pages}{373--385}.
\newblock \eprint{astro-ph/0006086}.
\bibitem[{{Diamond} et~al.(1994){Diamond}, {Kemball}, {Junor}, {Zensus},
  {Benson} and {Dhawan}}]{diamond_94}
\bibinfo{author}{{Diamond}, P.J.}, \bibinfo{author}{{Kemball}, A.J.},
  \bibinfo{author}{{Junor}, W.}, \bibinfo{author}{{Zensus}, A.},
  \bibinfo{author}{{Benson}, J.}, \bibinfo{author}{{Dhawan}, V.},
  \bibinfo{year}{1994}.
\newblock \bibinfo{title}{{Observation of a ring structure in SiO maser
  emission from late-type stars}}.
\newblock \bibinfo{journal}{ApJL} \bibinfo{volume}{430},
  \bibinfo{pages}{L61--L64}.
\bibitem[{{Dodson} et~al.(2008){Dodson}, {Fomalont}, {Wiik}, {Horiuchi},
  {Hirabayashi}, {Edwards}, {Murata}, {Asaki}, {Moellenbrock}, {Scott},
  {Taylor}, {Gurvits}, {Paragi}, {Frey}, {Shen}, {Lovell}, {Tingay}, {Rioja},
  {Fodor}, {Lister}, {Mosoni}, {Coldwell}, {Piner} and {Yang}}]{dodson_08}
\bibinfo{author}{{Dodson}, R.}, \bibinfo{author}{{Fomalont}, E.B.},
  \bibinfo{author}{{Wiik}, K.}, \bibinfo{author}{{Horiuchi}, S.},
  \bibinfo{author}{{Hirabayashi}, H.}, \bibinfo{author}{{Edwards}, P.G.},
  \bibinfo{author}{{Murata}, Y.}, \bibinfo{author}{{Asaki}, Y.},
  \bibinfo{author}{{Moellenbrock}, G.A.}, \bibinfo{author}{{Scott}, W.K.},
  \bibinfo{author}{{Taylor}, A.R.}, \bibinfo{author}{{Gurvits}, L.I.},
  \bibinfo{author}{{Paragi}, Z.}, \bibinfo{author}{{Frey}, S.},
  \bibinfo{author}{{Shen}, Z.Q.}, \bibinfo{author}{{Lovell}, J.E.J.},
  \bibinfo{author}{{Tingay}, S.J.}, \bibinfo{author}{{Rioja}, M.J.},
  \bibinfo{author}{{Fodor}, S.}, \bibinfo{author}{{Lister}, M.L.},
  \bibinfo{author}{{Mosoni}, L.}, \bibinfo{author}{{Coldwell}, G.},
  \bibinfo{author}{{Piner}, B.G.}, \bibinfo{author}{{Yang}, J.},
  \bibinfo{year}{2008}.
\newblock \bibinfo{title}{{The VSOP 5 GHz Active Galactic Nucleus Survey. V.
  Imaging Results for the Remaining 140 Sources}}.
\newblock \bibinfo{journal}{\apjs} \bibinfo{volume}{175},
  \bibinfo{pages}{314--355}.
\newblock \eprint{0710.5707}.
\bibitem[{{Dodson} et~al.(2017a){Dodson}, {Rioja}, {Bujarrabal}, Cho, Choi,
  Youngjoo and Kim}]{dodson_17}
\bibinfo{author}{{Dodson}, R.}, \bibinfo{author}{{Rioja}, M.},
  \bibinfo{author}{{Bujarrabal}, V.}, \bibinfo{author}{Cho, S.},
  \bibinfo{author}{Choi, Y.}, \bibinfo{author}{Youngjoo, Y.},
  \bibinfo{author}{Kim, J.}, \bibinfo{year}{2017}a.
\newblock \bibinfo{title}{{Registration of H$_2$O and SiO masers in the
  Calabash Nebula, to confirm the Planetary Nebula paradigm}}.
\newblock \bibinfo{journal}{Submitted to MNRAS} .
\bibitem[{{Dodson} and {Rioja}(2009)}]{vlba_31}
\bibinfo{author}{{Dodson}, R.}, \bibinfo{author}{{Rioja}, M.J.},
  \bibinfo{year}{2009}.
\newblock \bibinfo{title}{{VLBA Scientific Memorandum n. 31: Astrometric
  calibration of mm-VLBI using ''Source/Frequency Phase Referenced''
  observations}}.
\newblock \bibinfo{type}{Technical Report}. NRAO.
\newblock \eprint{arXiv:0910.1159}.
\bibitem[{{Dodson} et~al.(2014){Dodson}, {Rioja}, {Jung}, {Sohn}, {Byun},
  {Cho}, {Lee}, {Kim}, {Kim}, {Oh}, {Han}, {Je}, {Chung}, {Wi}, {Kang}, {Lee},
  {Chung}, {Kim}, {Kim}, {Lee}, {Roh}, {Oh}, {Yeom}, {Song} and
  {Kang}}]{dodson_14}
\bibinfo{author}{{Dodson}, R.}, \bibinfo{author}{{Rioja}, M.J.},
  \bibinfo{author}{{Jung}, T.H.}, \bibinfo{author}{{Sohn}, B.W.},
  \bibinfo{author}{{Byun}, D.Y.}, \bibinfo{author}{{Cho}, S.H.},
  \bibinfo{author}{{Lee}, S.S.}, \bibinfo{author}{{Kim}, J.},
  \bibinfo{author}{{Kim}, K.T.}, \bibinfo{author}{{Oh}, C.S.},
  \bibinfo{author}{{Han}, S.T.}, \bibinfo{author}{{Je}, D.H.},
  \bibinfo{author}{{Chung}, M.H.}, \bibinfo{author}{{Wi}, S.O.},
  \bibinfo{author}{{Kang}, J.}, \bibinfo{author}{{Lee}, J.W.},
  \bibinfo{author}{{Chung}, H.}, \bibinfo{author}{{Kim}, H.R.},
  \bibinfo{author}{{Kim}, H.G.}, \bibinfo{author}{{Lee}, C.H.},
  \bibinfo{author}{{Roh}, D.G.}, \bibinfo{author}{{Oh}, S.J.},
  \bibinfo{author}{{Yeom}, J.H.}, \bibinfo{author}{{Song}, M.G.},
  \bibinfo{author}{{Kang}, Y.W.}, \bibinfo{year}{2014}.
\newblock \bibinfo{title}{{Astrometrically Registered Simultaneous Observations
  of the 22 GHz H$_{2}$O and 43 GHz SiO Masers toward R Leonis Minoris Using
  KVN and Source/Frequency Phase Referencing}}.
\newblock \bibinfo{journal}{AJ} \bibinfo{volume}{148}, \bibinfo{pages}{97}.
\newblock \eprint{arXiv:1408.3513}.
\bibitem[{{Dodson} et~al.(2017b){Dodson}, {Rioja}, {Molina} and
  {G{\'o}mez}}]{dodson_16}
\bibinfo{author}{{Dodson}, R.}, \bibinfo{author}{{Rioja}, M.J.},
  \bibinfo{author}{{Molina}, S.N.}, \bibinfo{author}{{G{\'o}mez}, J.L.},
  \bibinfo{year}{2017}b.
\newblock \bibinfo{title}{{High-precision Astrometric Millimeter Very Long
  Baseline Interferometry Using a New Method for Multi-frequency Calibration}}.
\newblock \bibinfo{journal}{\apj} \bibinfo{volume}{834}, \bibinfo{pages}{177}.
\newblock \eprint{1612.02958}.
\bibitem[{Doeleman et~al.(2008)Doeleman, Weintroub, Rogers, Plambeck, Freund,
  Tilanus, Friberg, Ziurys, Moran, Corey, Young, Smythe, Titus, Marrone,
  Cappallo, Bock, Bower, Chamberlin, Davis, Krichbaum, Lamb, Maness, Niell,
  Roy, Strittmatter, Werthimer, Whitney and Woody}]{Doeleman:2008ek}
\bibinfo{author}{Doeleman, S.S.}, \bibinfo{author}{Weintroub, J.},
  \bibinfo{author}{Rogers, A.E.E.}, \bibinfo{author}{Plambeck, R.},
  \bibinfo{author}{Freund, R.}, \bibinfo{author}{Tilanus, R.P.J.},
  \bibinfo{author}{Friberg, P.}, \bibinfo{author}{Ziurys, L.M.},
  \bibinfo{author}{Moran, J.M.}, \bibinfo{author}{Corey, B.},
  \bibinfo{author}{Young, K.H.}, \bibinfo{author}{Smythe, D.L.},
  \bibinfo{author}{Titus, M.}, \bibinfo{author}{Marrone, D.P.},
  \bibinfo{author}{Cappallo, R.J.}, \bibinfo{author}{Bock, D.C.J.},
  \bibinfo{author}{Bower, G.C.}, \bibinfo{author}{Chamberlin, R.},
  \bibinfo{author}{Davis, G.R.}, \bibinfo{author}{Krichbaum, T.P.},
  \bibinfo{author}{Lamb, J.}, \bibinfo{author}{Maness, H.},
  \bibinfo{author}{Niell, A.E.}, \bibinfo{author}{Roy, A.},
  \bibinfo{author}{Strittmatter, P.}, \bibinfo{author}{Werthimer, D.},
  \bibinfo{author}{Whitney, A.R.}, \bibinfo{author}{Woody, D.},
  \bibinfo{year}{2008}.
\newblock \bibinfo{title}{{Event-horizon-scale structure in the supermassive
  black hole candidate at the Galactic Centre}}.
\newblock \bibinfo{journal}{Nature} \bibinfo{volume}{455},
  \bibinfo{pages}{78--80}.
\bibitem[{{Ellingsen} et~al.(2012){Ellingsen}, {Sobolev}, {Cragg} and
  {Godfrey}}]{ellingsen_12}
\bibinfo{author}{{Ellingsen}, S.P.}, \bibinfo{author}{{Sobolev}, A.M.},
  \bibinfo{author}{{Cragg}, D.M.}, \bibinfo{author}{{Godfrey}, P.D.},
  \bibinfo{year}{2012}.
\newblock \bibinfo{title}{{Discovery of Two New Class II Methanol Maser
  Transitions in G 345.01+1.79}}.
\newblock \bibinfo{journal}{\apjl} \bibinfo{volume}{759}, \bibinfo{pages}{L5}.
\newblock \eprint{arXiv:1209.5744}.
\bibitem[{{Fender}(2003)}]{fender_03}
\bibinfo{author}{{Fender}, R.P.}, \bibinfo{year}{2003}.
\newblock \bibinfo{title}{{Uses and limitations of relativistic jet proper
  motions: lessons from Galactic microquasars}}.
\newblock \bibinfo{journal}{\mnras} \bibinfo{volume}{340},
  \bibinfo{pages}{1353--1358}.
\newblock \eprint{astro-ph/0301225}.
\bibitem[{Fender et~al.(2004)Fender, Belloni and Gallo}]{Fender:2004dt}
\bibinfo{author}{Fender, R.P.}, \bibinfo{author}{Belloni, T.},
  \bibinfo{author}{Gallo, E.}, \bibinfo{year}{2004}.
\newblock \bibinfo{title}{{Towards a unified model for black hole X-ray binary
  jets}}.
\newblock \bibinfo{journal}{MNRAS} \bibinfo{volume}{355},
  \bibinfo{pages}{1105--1118}.
\bibitem[{{Fish} et~al.(2011){Fish}, {Doeleman}, {Beaudoin}, {Blundell},
  {Bolin}, {Bower}, {Chamberlin}, {Freund}, {Friberg}, {Gurwell}, {Honma},
  {Inoue}, {Krichbaum}, {Lamb}, {Marrone}, {Moran}, {Oyama}, {Plambeck},
  {Primiani}, {Rogers}, {Smythe}, {SooHoo}, {Strittmatter}, {Tilanus}, {Titus},
  {Weintroub}, {Wright}, {Woody}, {Young} and {Ziurys}}]{fish_11}
\bibinfo{author}{{Fish}, V.L.}, \bibinfo{author}{{Doeleman}, S.S.},
  \bibinfo{author}{{Beaudoin}, C.}, \bibinfo{author}{{Blundell}, R.},
  \bibinfo{author}{{Bolin}, D.E.}, \bibinfo{author}{{Bower}, G.C.},
  \bibinfo{author}{{Chamberlin}, R.}, \bibinfo{author}{{Freund}, R.},
  \bibinfo{author}{{Friberg}, P.}, \bibinfo{author}{{Gurwell}, M.A.},
  \bibinfo{author}{{Honma}, M.}, \bibinfo{author}{{Inoue}, M.},
  \bibinfo{author}{{Krichbaum}, T.P.}, \bibinfo{author}{{Lamb}, J.},
  \bibinfo{author}{{Marrone}, D.P.}, \bibinfo{author}{{Moran}, J.M.},
  \bibinfo{author}{{Oyama}, T.}, \bibinfo{author}{{Plambeck}, R.},
  \bibinfo{author}{{Primiani}, R.}, \bibinfo{author}{{Rogers}, A.E.E.},
  \bibinfo{author}{{Smythe}, D.L.}, \bibinfo{author}{{SooHoo}, J.},
  \bibinfo{author}{{Strittmatter}, P.}, \bibinfo{author}{{Tilanus}, R.P.J.},
  \bibinfo{author}{{Titus}, M.}, \bibinfo{author}{{Weintroub}, J.},
  \bibinfo{author}{{Wright}, M.}, \bibinfo{author}{{Woody}, D.},
  \bibinfo{author}{{Young}, K.H.}, \bibinfo{author}{{Ziurys}, L.M.},
  \bibinfo{year}{2011}.
\newblock \bibinfo{title}{{1.3 mm Wavelength VLBI of Sagittarius A*: Detection
  of Time-variable Emission on Event Horizon Scales}}.
\newblock \bibinfo{journal}{\apjl} \bibinfo{volume}{727}, \bibinfo{pages}{L36}.
\newblock \eprint{1011.2472}.
\bibitem[{{Fomalont} et~al.(2014){Fomalont}, {van Kempen}, {Kneissl},
  {Marcelino}, {Barkats}, {Corder}, {Cortes}, {Hills}, {Lucas}, {Manning} and
  {Peck}}]{Fomalont_14}
\bibinfo{author}{{Fomalont}, E.}, \bibinfo{author}{{van Kempen}, T.},
  \bibinfo{author}{{Kneissl}, R.}, \bibinfo{author}{{Marcelino}, N.},
  \bibinfo{author}{{Barkats}, D.}, \bibinfo{author}{{Corder}, S.},
  \bibinfo{author}{{Cortes}, P.}, \bibinfo{author}{{Hills}, R.},
  \bibinfo{author}{{Lucas}, R.}, \bibinfo{author}{{Manning}, A.},
  \bibinfo{author}{{Peck}, A.}, \bibinfo{year}{2014}.
\newblock \bibinfo{title}{{The Calibration of ALMA using Radio Sources}}.
\newblock \bibinfo{journal}{The Messenger} \bibinfo{volume}{155},
  \bibinfo{pages}{19--22}.
\bibitem[{{Fomalont} et~al.(2000){Fomalont}, {Frey}, {Paragi}, {Gurvits},
  {Scott}, {Taylor}, {Edwards} and {Hirabayashi}}]{fomalont_00}
\bibinfo{author}{{Fomalont}, E.B.}, \bibinfo{author}{{Frey}, S.},
  \bibinfo{author}{{Paragi}, Z.}, \bibinfo{author}{{Gurvits}, L.I.},
  \bibinfo{author}{{Scott}, W.K.}, \bibinfo{author}{{Taylor}, A.R.},
  \bibinfo{author}{{Edwards}, P.G.}, \bibinfo{author}{{Hirabayashi}, H.},
  \bibinfo{year}{2000}.
\newblock \bibinfo{title}{{The VSOP 5 GHz Continuum Survey: The Prelaunch VLBA
  Observations}}.
\newblock \bibinfo{journal}{\apjs} \bibinfo{volume}{131},
  \bibinfo{pages}{95--183}.
\bibitem[{{Frank} and {Blackman}(2004)}]{frank_04}
\bibinfo{author}{{Frank}, A.}, \bibinfo{author}{{Blackman}, E.G.},
  \bibinfo{year}{2004}.
\newblock \bibinfo{title}{{Application of Magnetohydrodynamic Disk Wind
  Solutions to Planetary and Protoplanetary Nebulae}}.
\newblock \bibinfo{journal}{ApJ} \bibinfo{volume}{614},
  \bibinfo{pages}{737--744}.
\newblock \eprint{astro-ph/0207447}.
\bibitem[{{Fromm} et~al.(2015){Fromm}, {Perucho}, {Ros}, {Savolainen} and
  {Zensus}}]{fromm_15}
\bibinfo{author}{{Fromm}, C.M.}, \bibinfo{author}{{Perucho}, M.},
  \bibinfo{author}{{Ros}, E.}, \bibinfo{author}{{Savolainen}, T.},
  \bibinfo{author}{{Zensus}, J.A.}, \bibinfo{year}{2015}.
\newblock \bibinfo{title}{{On the location of the supermassive black hole in
  CTA 102}}.
\newblock \bibinfo{journal}{\aap} \bibinfo{volume}{576}, \bibinfo{pages}{A43}.
\newblock \eprint{arXiv:1412.1317}.
\bibitem[{Fromm et~al.(2015)Fromm, Perucho, Ros, Savolainen and
  Zensus}]{2015A&A...576A..43F}
\bibinfo{author}{Fromm, C.M.}, \bibinfo{author}{Perucho, M.},
  \bibinfo{author}{Ros, E.}, \bibinfo{author}{Savolainen, T.},
  \bibinfo{author}{Zensus, J.A.}, \bibinfo{year}{2015}.
\newblock \bibinfo{title}{{On the location of the supermassive black hole in
  CTA 102}}.
\newblock \bibinfo{journal}{A{\&}A} \bibinfo{volume}{576},
  \bibinfo{pages}{A43}.
\bibitem[{Fromm et~al.(2013)Fromm, Ros, Perucho, Savolainen, Mimica, Kadler,
  Lobanov and Zensus}]{Fromm:2013en}
\bibinfo{author}{Fromm, C.M.}, \bibinfo{author}{Ros, E.},
  \bibinfo{author}{Perucho, M.}, \bibinfo{author}{Savolainen, T.},
  \bibinfo{author}{Mimica, P.}, \bibinfo{author}{Kadler, M.},
  \bibinfo{author}{Lobanov, A.P.}, \bibinfo{author}{Zensus, J.A.},
  \bibinfo{year}{2013}.
\newblock \bibinfo{title}{{Catching the radio flare in CTA 102. III. Core-shift
  and spectral analysis}}.
\newblock \bibinfo{journal}{A{\&}A} \bibinfo{volume}{557},
  \bibinfo{pages}{105}.
\bibitem[{Gabuzda et~al.(2004)Gabuzda, Murray and Cronin}]{Gabuzda:2004dt}
\bibinfo{author}{Gabuzda, D.C.}, \bibinfo{author}{Murray, {\'E}.},
  \bibinfo{author}{Cronin, P.}, \bibinfo{year}{2004}.
\newblock \bibinfo{title}{{Helical magnetic fields associated with the
  relativistic jets of four BL Lac objects}}.
\newblock \bibinfo{journal}{MNRAS} \bibinfo{volume}{351},
  \bibinfo{pages}{L89--L93}.
\bibitem[{Gabuzda et~al.(2001)Gabuzda, Pushkarev and Garnich}]{Gabuzda:2001cu}
\bibinfo{author}{Gabuzda, D.C.}, \bibinfo{author}{Pushkarev, A.B.},
  \bibinfo{author}{Garnich, N.N.}, \bibinfo{year}{2001}.
\newblock \bibinfo{title}{{Unusual radio properties of the BL Lac object
  0820+225}}.
\newblock \bibinfo{journal}{MNRAS} \bibinfo{volume}{327},
  \bibinfo{pages}{1--9}.
\bibitem[{{Garrett}(2005)}]{garret_micro}
\bibinfo{author}{{Garrett}, M.A.}, \bibinfo{year}{2005}.
\newblock \bibinfo{title}{{Deep field surveys - A radio view}}, in:
  \bibinfo{editor}{{Gurvits}, L.I.}, \bibinfo{editor}{{Frey}, S.},
  \bibinfo{editor}{{Rawlings}, S.} (Eds.), \bibinfo{booktitle}{EAS Publications
  Series}, pp. \bibinfo{pages}{73--91}.
\newblock \eprint{astro-ph/0409197}.
\bibitem[{{Goddi} et~al.(2009){Goddi}, {Greenhill}, {Chandler}, {Humphreys},
  {Matthews} and {Gray}}]{God09}
\bibinfo{author}{{Goddi}, C.}, \bibinfo{author}{{Greenhill}, L.J.},
  \bibinfo{author}{{Chandler}, C.J.}, \bibinfo{author}{{Humphreys}, E.M.L.},
  \bibinfo{author}{{Matthews}, L.D.}, \bibinfo{author}{{Gray}, M.D.},
  \bibinfo{year}{2009}.
\newblock \bibinfo{title}{{Maser Emission from SiO Isotopologues Traces the
  Innermost 100 AU Around Radio Source I in Orion
  Becklin-Neugebauer/Kleinmann-Low}}.
\newblock \bibinfo{journal}{\apj} \bibinfo{volume}{698},
  \bibinfo{pages}{1165--1173}.
\newblock \eprint{arXiv:0904.1373}.
\bibitem[{G{\'o}mez et~al.(1993)G{\'o}mez, Alberdi and
  Marcaide}]{1993A&A...274...55G}
\bibinfo{author}{G{\'o}mez, J.L.}, \bibinfo{author}{Alberdi, A.},
  \bibinfo{author}{Marcaide, J.M.}, \bibinfo{year}{1993}.
\newblock \bibinfo{title}{{Synchrotron Emission from Bent Shocked Relativistic
  Jets - Part One - Bent Relativistic Jets}}.
\newblock \bibinfo{journal}{A{\&}A} \bibinfo{volume}{274}, \bibinfo{pages}{55}.
\bibitem[{G{\'o}mez et~al.(2016)G{\'o}mez, Lobanov, Bruni, Kovalev, Marscher,
  Jorstad, Mizuno, Bach, Sokolovsky, Anderson, Galindo, Kardashev and
  Lisakov}]{2016ApJ...817...96G}
\bibinfo{author}{G{\'o}mez, J.L.}, \bibinfo{author}{Lobanov, A.P.},
  \bibinfo{author}{Bruni, G.}, \bibinfo{author}{Kovalev, Y.Y.},
  \bibinfo{author}{Marscher, A.P.}, \bibinfo{author}{Jorstad, S.G.},
  \bibinfo{author}{Mizuno, Y.}, \bibinfo{author}{Bach, U.},
  \bibinfo{author}{Sokolovsky, K.V.}, \bibinfo{author}{Anderson, J.M.},
  \bibinfo{author}{Galindo, P.}, \bibinfo{author}{Kardashev, N.S.},
  \bibinfo{author}{Lisakov, M.M.}, \bibinfo{year}{2016}.
\newblock \bibinfo{title}{{Probing the Innermost Regions of AGN Jets and Their
  Magnetic Fields with RadioAstron. I. Imaging BL Lacertae at 21 Microarcsecond
  Resolution}}.
\newblock \bibinfo{journal}{ApJ} \bibinfo{volume}{817}, \bibinfo{pages}{96}.
\bibitem[{G{\'o}mez et~al.(2000)G{\'o}mez, Marscher, Alberdi, Jorstad and
  Garc{\'\i}a-Mir{\'o}}]{2000Sci...289.2317G}
\bibinfo{author}{G{\'o}mez, J.L.}, \bibinfo{author}{Marscher, A.P.},
  \bibinfo{author}{Alberdi, A.}, \bibinfo{author}{Jorstad, S.G.},
  \bibinfo{author}{Garc{\'\i}a-Mir{\'o}, C.}, \bibinfo{year}{2000}.
\newblock \bibinfo{title}{{Flashing Superluminal Components in the Jet of the
  Radio Galaxy 3C120}}.
\newblock \bibinfo{journal}{Science} \bibinfo{volume}{289},
  \bibinfo{pages}{2317--2320}.
\bibitem[{G{\'o}mez et~al.(2008)G{\'o}mez, Marscher, Jorstad, Agudo and
  Roca-Sogorb}]{2008ApJ...681L..69G}
\bibinfo{author}{G{\'o}mez, J.L.}, \bibinfo{author}{Marscher, A.P.},
  \bibinfo{author}{Jorstad, S.G.}, \bibinfo{author}{Agudo, I.},
  \bibinfo{author}{Roca-Sogorb, M.}, \bibinfo{year}{2008}.
\newblock \bibinfo{title}{{Faraday Rotation and Polarization Gradients in the
  Jet of 3C 120: Interaction with the External Medium and a Helical Magnetic
  Field?}}
\newblock \bibinfo{journal}{ApJ} \bibinfo{volume}{681},
  \bibinfo{pages}{L69--L72}.
\bibitem[{G{\'o}mez et~al.(1997a)G{\'o}mez, Marti, Marscher and
  Ibanez}]{Gomez:1997gq}
\bibinfo{author}{G{\'o}mez, J.L.}, \bibinfo{author}{Marti, J.M.},
  \bibinfo{author}{Marscher, A.P.}, \bibinfo{author}{Ibanez, J.M.},
  \bibinfo{year}{1997}a.
\newblock \bibinfo{title}{{Relativistic simulations of superluminal sources}}.
\newblock \bibinfo{journal}{Vistas in Astronomy} \bibinfo{volume}{41},
  \bibinfo{pages}{79--85}.
\bibitem[{G{\'o}mez et~al.(1997b)G{\'o}mez, Mart{\'\i}, Marscher,
  Ib{\'a}{\~n}ez and Alberdi}]{1997ApJ...482L..33G}
\bibinfo{author}{G{\'o}mez, J.L.}, \bibinfo{author}{Mart{\'\i}, J.M.},
  \bibinfo{author}{Marscher, A.P.}, \bibinfo{author}{Ib{\'a}{\~n}ez, J.M.},
  \bibinfo{author}{Alberdi, A.}, \bibinfo{year}{1997}b.
\newblock \bibinfo{title}{{Hydrodynamical Models of Superluminal Sources}}.
\newblock \bibinfo{journal}{ApJ} \bibinfo{volume}{482},
  \bibinfo{pages}{L33--L36}.
\bibitem[{{G{\'o}mez} et~al.(1995){G{\'o}mez}, {Marti}, {Marscher}, {Ibanez}
  and {Marcaide}}]{gomez_95}
\bibinfo{author}{{G{\'o}mez}, J.L.}, \bibinfo{author}{{Marti}, J.M.A.},
  \bibinfo{author}{{Marscher}, A.P.}, \bibinfo{author}{{Ibanez}, J.M.A.},
  \bibinfo{author}{{Marcaide}, J.M.}, \bibinfo{year}{1995}.
\newblock \bibinfo{title}{{Parsec-Scale Synchrotron Emission from Hydrodynamic
  Relativistic Jets in Active Galactic Nuclei}}.
\newblock \bibinfo{journal}{\apjl} \bibinfo{volume}{449}, \bibinfo{pages}{L19}.
\bibitem[{{Gray} et~al.(2016){Gray}, {Baudry}, {Richards}, {Humphreys},
  {Sobolev} and {Yates}}]{2016MNRAS.456..374G}
\bibinfo{author}{{Gray}, M.D.}, \bibinfo{author}{{Baudry}, A.},
  \bibinfo{author}{{Richards}, A.M.S.}, \bibinfo{author}{{Humphreys}, E.M.L.},
  \bibinfo{author}{{Sobolev}, A.M.}, \bibinfo{author}{{Yates}, J.A.},
  \bibinfo{year}{2016}.
\newblock \bibinfo{title}{{The physics of water masers observable with ALMA and
  SOFIA: model predictions for evolved stars}}.
\newblock \bibinfo{journal}{\mnras} \bibinfo{volume}{456},
  \bibinfo{pages}{374--404}.
\newblock \eprint{1510.06182}.
\bibitem[{{Gray} et~al.(2009){Gray}, {Wittkowski}, {Scholz}, {Humphreys},
  {Ohnaka} and {Boboltz}}]{gray_09}
\bibinfo{author}{{Gray}, M.D.}, \bibinfo{author}{{Wittkowski}, M.},
  \bibinfo{author}{{Scholz}, M.}, \bibinfo{author}{{Humphreys}, E.M.L.},
  \bibinfo{author}{{Ohnaka}, K.}, \bibinfo{author}{{Boboltz}, D.},
  \bibinfo{year}{2009}.
\newblock \bibinfo{title}{{SiO maser emission in Miras}}.
\newblock \bibinfo{journal}{MNRAS} \bibinfo{volume}{394},
  \bibinfo{pages}{51--66}.
\newblock \eprint{arXiv:0811.2770}.
\bibitem[{{Greenhill} et~al.(2013){Greenhill}, {Goddi}, {Chandler}, {Matthews}
  and {Humphreys}}]{Gre13}
\bibinfo{author}{{Greenhill}, L.J.}, \bibinfo{author}{{Goddi}, C.},
  \bibinfo{author}{{Chandler}, C.J.}, \bibinfo{author}{{Matthews}, L.D.},
  \bibinfo{author}{{Humphreys}, E.M.L.}, \bibinfo{year}{2013}.
\newblock \bibinfo{title}{{Dynamical Evidence for a Magnetocentrifugal Wind
  from a 20~M$_{\odot}$ Binary Young Stellar Object}}.
\newblock \bibinfo{journal}{\apjl} \bibinfo{volume}{770}, \bibinfo{pages}{L32}.
\newblock \eprint{arXiv:1305.4150}.
\bibitem[{{Guirado} et~al.(1995){Guirado}, {Marcaide}, {Alberdi}, {Elosegui},
  {Ratner}, {Shapiro}, {Kilger}, {Mantovani}, {Venturi}, {Rius}, {Ros},
  {Trigilio} and {Whitney}}]{guirado_95}
\bibinfo{author}{{Guirado}, J.C.}, \bibinfo{author}{{Marcaide}, J.M.},
  \bibinfo{author}{{Alberdi}, A.}, \bibinfo{author}{{Elosegui}, P.},
  \bibinfo{author}{{Ratner}, M.I.}, \bibinfo{author}{{Shapiro}, I.I.},
  \bibinfo{author}{{Kilger}, R.}, \bibinfo{author}{{Mantovani}, F.},
  \bibinfo{author}{{Venturi}, T.}, \bibinfo{author}{{Rius}, A.},
  \bibinfo{author}{{Ros}, E.}, \bibinfo{author}{{Trigilio}, C.},
  \bibinfo{author}{{Whitney}, A.R.}, \bibinfo{year}{1995}.
\newblock \bibinfo{title}{{Proper Motion of Components in 4C 39.25}}.
\newblock \bibinfo{journal}{\aj} \bibinfo{volume}{110}, \bibinfo{pages}{2586}.
\bibitem[{{Hada} et~al.(2011){Hada}, {Doi}, {Kino}, {Nagai}, {Hagiwara} and
  {Kawaguchi}}]{hada_11}
\bibinfo{author}{{Hada}, K.}, \bibinfo{author}{{Doi}, A.},
  \bibinfo{author}{{Kino}, M.}, \bibinfo{author}{{Nagai}, H.},
  \bibinfo{author}{{Hagiwara}, Y.}, \bibinfo{author}{{Kawaguchi}, N.},
  \bibinfo{year}{2011}.
\newblock \bibinfo{title}{{An origin of the radio jet in M87 at the location of
  the central black hole}}.
\newblock \bibinfo{journal}{Nat} \bibinfo{volume}{477},
  \bibinfo{pages}{185--187}.
\bibitem[{Hada et~al.(2011)Hada, Doi, Kino, Nagai, Hagiwara and
  Kawaguchi}]{Hada:2011im}
\bibinfo{author}{Hada, K.}, \bibinfo{author}{Doi, A.}, \bibinfo{author}{Kino,
  M.}, \bibinfo{author}{Nagai, H.}, \bibinfo{author}{Hagiwara, Y.},
  \bibinfo{author}{Kawaguchi, N.}, \bibinfo{year}{2011}.
\newblock \bibinfo{title}{{An origin of the radio jet in M87 at the location of
  the central black hole}}.
\newblock \bibinfo{journal}{Nature} \bibinfo{volume}{477},
  \bibinfo{pages}{185--187}.
\bibitem[{{Han} et~al.(2013){Han}, {Lee}, {Kang}, {Oh}, {Byun}, {Je}, {Chung},
  {Wi}, {Song}, {Kang}, {Lee}, {Kim}, {Sasao}, {Goldsmith} and
  {Wylde}}]{kvn_optics}
\bibinfo{author}{{Han}, S.T.}, \bibinfo{author}{{Lee}, J.W.},
  \bibinfo{author}{{Kang}, J.}, \bibinfo{author}{{Oh}, C.S.},
  \bibinfo{author}{{Byun}, D.Y.}, \bibinfo{author}{{Je}, D.H.},
  \bibinfo{author}{{Chung}, M.H.}, \bibinfo{author}{{Wi}, S.O.},
  \bibinfo{author}{{Song}, M.}, \bibinfo{author}{{Kang}, Y.W.},
  \bibinfo{author}{{Lee}, S.S.}, \bibinfo{author}{{Kim}, S.Y.},
  \bibinfo{author}{{Sasao}, T.}, \bibinfo{author}{{Goldsmith}, P.F.},
  \bibinfo{author}{{Wylde}, R.}, \bibinfo{year}{2013}.
\newblock \bibinfo{title}{{Korean VLBI Network Receiver Optics for Simultaneous
  Multifrequency Observation: Evaluation}}.
\newblock \bibinfo{journal}{PASP} \bibinfo{volume}{125},
  \bibinfo{pages}{539--547}.
\bibitem[{{Helmboldt} et~al.(2007){Helmboldt}, {Taylor}, {Tremblay},
  {Fassnacht}, {Walker}, {Myers}, {Sjouwerman}, {Pearson}, {Readhead},
  {Weintraub}, {Gehrels}, {Romani}, {Healey}, {Michelson}, {Blandford} and
  {Cotter}}]{helmboldt_07}
\bibinfo{author}{{Helmboldt}, J.F.}, \bibinfo{author}{{Taylor}, G.B.},
  \bibinfo{author}{{Tremblay}, S.}, \bibinfo{author}{{Fassnacht}, C.D.},
  \bibinfo{author}{{Walker}, R.C.}, \bibinfo{author}{{Myers}, S.T.},
  \bibinfo{author}{{Sjouwerman}, L.O.}, \bibinfo{author}{{Pearson}, T.J.},
  \bibinfo{author}{{Readhead}, A.C.S.}, \bibinfo{author}{{Weintraub}, L.},
  \bibinfo{author}{{Gehrels}, N.}, \bibinfo{author}{{Romani}, R.W.},
  \bibinfo{author}{{Healey}, S.}, \bibinfo{author}{{Michelson}, P.F.},
  \bibinfo{author}{{Blandford}, R.D.}, \bibinfo{author}{{Cotter}, G.},
  \bibinfo{year}{2007}.
\newblock \bibinfo{title}{{The VLBA Imaging and Polarimetry Survey at 5 GHz}}.
\newblock \bibinfo{journal}{\apj} \bibinfo{volume}{658},
  \bibinfo{pages}{203--216}.
\newblock \eprint{astro-ph/0611459}.
\bibitem[{{Hirota} et~al.(2017){Hirota}, {Machida}, {Matsushita}, {Motogi},
  {Matsumoto}, {Kim}, {Burns} and {Honma}}]{hirota_17}
\bibinfo{author}{{Hirota}, T.}, \bibinfo{author}{{Machida}, M.N.},
  \bibinfo{author}{{Matsushita}, Y.}, \bibinfo{author}{{Motogi}, K.},
  \bibinfo{author}{{Matsumoto}, N.}, \bibinfo{author}{{Kim}, M.K.},
  \bibinfo{author}{{Burns}, R.A.}, \bibinfo{author}{{Honma}, M.},
  \bibinfo{year}{2017}.
\newblock \bibinfo{title}{{Disk-driven rotating bipolar outflow in Orion Source
  I}}.
\newblock \bibinfo{journal}{Nature Astronomy} \bibinfo{volume}{1},
  \bibinfo{pages}{0146}.
\bibitem[{{Hirota} et~al.(2014){Hirota}, {Tsuboi}, {Kurono}, {Fujisawa},
  {Honma}, {Kim}, {Imai} and {Yonekura}}]{hirota_14b}
\bibinfo{author}{{Hirota}, T.}, \bibinfo{author}{{Tsuboi}, M.},
  \bibinfo{author}{{Kurono}, Y.}, \bibinfo{author}{{Fujisawa}, K.},
  \bibinfo{author}{{Honma}, M.}, \bibinfo{author}{{Kim}, M.K.},
  \bibinfo{author}{{Imai}, H.}, \bibinfo{author}{{Yonekura}, Y.},
  \bibinfo{year}{2014}.
\newblock \bibinfo{title}{{VERA and ALMA observations of the H$_{2}$O
  supermaser burst in Orion KL}}.
\newblock \bibinfo{journal}{\pasj} \bibinfo{volume}{66}, \bibinfo{pages}{106}.
\newblock \eprint{1407.2757}.
\bibitem[{{H{\"o}fner} et~al.(2003){H{\"o}fner}, {Gautschy-Loidl}, {Aringer}
  and {J{\o}rgensen}}]{hofner_03}
\bibinfo{author}{{H{\"o}fner}, S.}, \bibinfo{author}{{Gautschy-Loidl}, R.},
  \bibinfo{author}{{Aringer}, B.}, \bibinfo{author}{{J{\o}rgensen}, U.G.},
  \bibinfo{year}{2003}.
\newblock \bibinfo{title}{{Dynamic model atmospheres of AGB stars. III. Effects
  of frequency-dependent radiative transfer}}.
\newblock \bibinfo{journal}{A\&A} \bibinfo{volume}{399},
  \bibinfo{pages}{589--601}.
\bibitem[{{Hovatta} et~al.(2014){Hovatta}, {Aller}, {Aller}, {Clausen-Brown},
  {Homan}, {Kovalev}, {Lister}, {Pushkarev} and {Savolainen}}]{hovatta_14}
\bibinfo{author}{{Hovatta}, T.}, \bibinfo{author}{{Aller}, M.F.},
  \bibinfo{author}{{Aller}, H.D.}, \bibinfo{author}{{Clausen-Brown}, E.},
  \bibinfo{author}{{Homan}, D.C.}, \bibinfo{author}{{Kovalev}, Y.Y.},
  \bibinfo{author}{{Lister}, M.L.}, \bibinfo{author}{{Pushkarev}, A.B.},
  \bibinfo{author}{{Savolainen}, T.}, \bibinfo{year}{2014}.
\newblock \bibinfo{title}{{MOJAVE: Monitoring of Jets in Active Galactic Nuclei
  with VLBA Experiments. XI. Spectral Distributions}}.
\newblock \bibinfo{journal}{AJ} \bibinfo{volume}{147}, \bibinfo{pages}{143}.
\newblock \eprint{arXiv:1404.0014}.
\bibitem[{Hovatta et~al.(2012)Hovatta, Lister, Aller, Aller, Homan, Kovalev,
  Pushkarev and Savolainen}]{Hovatta:2012jv}
\bibinfo{author}{Hovatta, T.}, \bibinfo{author}{Lister, M.L.},
  \bibinfo{author}{Aller, M.F.}, \bibinfo{author}{Aller, H.D.},
  \bibinfo{author}{Homan, D.C.}, \bibinfo{author}{Kovalev, Y.Y.},
  \bibinfo{author}{Pushkarev, A.B.}, \bibinfo{author}{Savolainen, T.},
  \bibinfo{year}{2012}.
\newblock \bibinfo{title}{{MOJAVE: Monitoring of Jets in Active Galactic Nuclei
  with VLBA Experiments. VIII. Faraday Rotation in Parsec-scale AGN Jets}}.
\newblock \bibinfo{journal}{AJ} \bibinfo{volume}{144}, \bibinfo{pages}{105}.
\bibitem[{{Humphreys} et~al.(2002){Humphreys}, {Gray}, {Yates}, {Field},
  {Bowen} and {Diamond}}]{humphreys_02}
\bibinfo{author}{{Humphreys}, E.M.L.}, \bibinfo{author}{{Gray}, M.D.},
  \bibinfo{author}{{Yates}, J.A.}, \bibinfo{author}{{Field}, D.},
  \bibinfo{author}{{Bowen}, G.H.}, \bibinfo{author}{{Diamond}, P.J.},
  \bibinfo{year}{2002}.
\newblock \bibinfo{title}{{Numerical simulations of stellar SiO maser
  variability. Investigation of the effect of shocks}}.
\newblock \bibinfo{journal}{\aap} \bibinfo{volume}{386},
  \bibinfo{pages}{256--270}.
\newblock \eprint{astro-ph/0202426}.
\bibitem[{{Imai} et~al.(2010){Imai}, {Nakashima}, {Deguchi}, {Yamauchi},
  {Nakagawa} and {Nagayama}}]{v3_reference}
\bibinfo{author}{{Imai}, H.}, \bibinfo{author}{{Nakashima}, J.I.},
  \bibinfo{author}{{Deguchi}, S.}, \bibinfo{author}{{Yamauchi}, A.},
  \bibinfo{author}{{Nakagawa}, A.}, \bibinfo{author}{{Nagayama}, T.},
  \bibinfo{year}{2010}.
\newblock \bibinfo{title}{{Japanese VLBI Network Mapping of SiO v = 3 J = 1-0
  Maser Emission in W Hydrae}}.
\newblock \bibinfo{journal}{PASJ} \bibinfo{volume}{62},
  \bibinfo{pages}{431--439}.
\bibitem[{Jorstad et~al.(2013)Jorstad, Marscher, Smith, Larionov, Agudo,
  Gurwell, Wehrle, L{\"a}hteenm{\"a}ki, Nikolashvili, Schmidt, Arkharov,
  Blinov, Blumenthal, Casadio, Chigladze, Efimova, Eggen, G{\'o}mez, Grupe,
  Hagen-Thorn, Joshi, Kimeridze, Konstantinova, Kopatskaya, Kurtanidze,
  Kurtanidze, Larionova, Larionova, Sigua, MacDonald, Maune, McHardy, Miller,
  Molina, Morozova, Scott, Taylor, Tornikoski, Troitsky, Thum, Walker,
  Williamson, Sallum, Consiglio and Strelnitski}]{2013ApJ...773..147J}
\bibinfo{author}{Jorstad, S.G.}, \bibinfo{author}{Marscher, A.P.},
  \bibinfo{author}{Smith, P.S.}, \bibinfo{author}{Larionov, V.M.},
  \bibinfo{author}{Agudo, I.}, \bibinfo{author}{Gurwell, M.},
  \bibinfo{author}{Wehrle, A.E.}, \bibinfo{author}{L{\"a}hteenm{\"a}ki, A.},
  \bibinfo{author}{Nikolashvili, M.G.}, \bibinfo{author}{Schmidt, G.D.},
  \bibinfo{author}{Arkharov, A.A.}, \bibinfo{author}{Blinov, D.A.},
  \bibinfo{author}{Blumenthal, K.}, \bibinfo{author}{Casadio, C.},
  \bibinfo{author}{Chigladze, R.A.}, \bibinfo{author}{Efimova, N.V.},
  \bibinfo{author}{Eggen, J.R.}, \bibinfo{author}{G{\'o}mez, J.L.},
  \bibinfo{author}{Grupe, D.}, \bibinfo{author}{Hagen-Thorn, V.A.},
  \bibinfo{author}{Joshi, M.}, \bibinfo{author}{Kimeridze, G.N.},
  \bibinfo{author}{Konstantinova, T.S.}, \bibinfo{author}{Kopatskaya, E.N.},
  \bibinfo{author}{Kurtanidze, O.M.}, \bibinfo{author}{Kurtanidze, S.O.},
  \bibinfo{author}{Larionova, E.G.}, \bibinfo{author}{Larionova, L.V.},
  \bibinfo{author}{Sigua, L.A.}, \bibinfo{author}{MacDonald, N.R.},
  \bibinfo{author}{Maune, J.D.}, \bibinfo{author}{McHardy, I.M.},
  \bibinfo{author}{Miller, H.R.}, \bibinfo{author}{Molina, S.N.},
  \bibinfo{author}{Morozova, D.A.}, \bibinfo{author}{Scott, T.},
  \bibinfo{author}{Taylor, B.W.}, \bibinfo{author}{Tornikoski, M.},
  \bibinfo{author}{Troitsky, I.S.}, \bibinfo{author}{Thum, C.},
  \bibinfo{author}{Walker, G.}, \bibinfo{author}{Williamson, K.E.},
  \bibinfo{author}{Sallum, S.}, \bibinfo{author}{Consiglio, S.},
  \bibinfo{author}{Strelnitski, V.}, \bibinfo{year}{2013}.
\newblock \bibinfo{title}{{A Tight Connection between Gamma-Ray Outbursts and
  Parsec-scale Jet Activity in the Quasar 3C 454.3}}.
\newblock \bibinfo{journal}{ApJ} \bibinfo{volume}{773}, \bibinfo{pages}{147}.
\bibitem[{{Jung} et~al.(2015){Jung}, {Dodson}, {Han}, {Rioja}, {Byun}, {Honma},
  {Stevens}, {de Vincente} and {Sohn}}]{jung_15}
\bibinfo{author}{{Jung}, T.}, \bibinfo{author}{{Dodson}, R.},
  \bibinfo{author}{{Han}, S.T.}, \bibinfo{author}{{Rioja}, M.J.},
  \bibinfo{author}{{Byun}, D.Y.}, \bibinfo{author}{{Honma}, M.},
  \bibinfo{author}{{Stevens}, J.}, \bibinfo{author}{{de Vincente}, P.},
  \bibinfo{author}{{Sohn}, B.W.}, \bibinfo{year}{2015}.
\newblock \bibinfo{title}{{Measuring the Core Shift Effect in AGN Jets with the
  Extended Korean VLBI Network}}.
\newblock \bibinfo{journal}{Journal of Korean Astronomical Society}
  \bibinfo{volume}{48}, \bibinfo{pages}{277--284}.
\bibitem[{Jung et~al.(2012)Jung, Sohn and Byun}]{jung_12}
\bibinfo{author}{Jung, T.}, \bibinfo{author}{Sohn, B.W.},
  \bibinfo{author}{Byun, D.Y.}, \bibinfo{year}{2012}.
\newblock \bibinfo{title}{First simultaneous 4-frequency phase referencing test
  for mm-vlbi observation}, in: \bibinfo{booktitle}{Proceedings of the 11th
  European VLBI Network Symposium \& Users Meeting. 9-12 October, 2012.
  Bordeaux (France). Online at http://pos. sissa. it/cgi-bin/reader/conf. cgi?
  confid= 178, id. 60}, p.~\bibinfo{pages}{60}.
\bibitem[{{Jung} et~al.(2011){Jung}, {Sohn}, {Kobayashi}, {Sasao}, {Hirota},
  {Kameya}, {Choi} and {Chung}}]{jung_11}
\bibinfo{author}{{Jung}, T.}, \bibinfo{author}{{Sohn}, B.W.},
  \bibinfo{author}{{Kobayashi}, H.}, \bibinfo{author}{{Sasao}, T.},
  \bibinfo{author}{{Hirota}, T.}, \bibinfo{author}{{Kameya}, O.},
  \bibinfo{author}{{Choi}, Y.K.}, \bibinfo{author}{{Chung}, H.S.},
  \bibinfo{year}{2011}.
\newblock \bibinfo{title}{{First Simultaneous Dual-Frequency Phase Referencing
  VLBI Observation with VERA}}.
\newblock \bibinfo{journal}{\pasj} \bibinfo{volume}{63},
  \bibinfo{pages}{375--385}.
\bibitem[{{Kalamkar} et~al.(2016){Kalamkar}, {Casella}, {Uttley}, {O'Brien},
  {Russell}, {Maccarone}, {van der Klis} and {Vincentelli}}]{kalamkar_16}
\bibinfo{author}{{Kalamkar}, M.}, \bibinfo{author}{{Casella}, P.},
  \bibinfo{author}{{Uttley}, P.}, \bibinfo{author}{{O'Brien}, K.},
  \bibinfo{author}{{Russell}, D.}, \bibinfo{author}{{Maccarone}, T.},
  \bibinfo{author}{{van der Klis}, M.}, \bibinfo{author}{{Vincentelli}, F.},
  \bibinfo{year}{2016}.
\newblock \bibinfo{title}{{Detection of the first infra-red quasi-periodic
  oscillation in a black hole X-ray binary}}.
\newblock \bibinfo{journal}{\mnras} \bibinfo{volume}{460},
  \bibinfo{pages}{3284--3291}.
\newblock \eprint{1510.08907}.
\bibitem[{{Kamohara} et~al.(2010){Kamohara}, {Bujarrabal}, {Honma}, {Nakagawa},
  {Matsumoto}, {Oyama}, {Hirota}, {Imai}, {Shibata}, {Kobayashi}, {Sato} and
  {Ueno}}]{kamohara_10}
\bibinfo{author}{{Kamohara}, R.}, \bibinfo{author}{{Bujarrabal}, V.},
  \bibinfo{author}{{Honma}, M.}, \bibinfo{author}{{Nakagawa}, A.},
  \bibinfo{author}{{Matsumoto}, N.}, \bibinfo{author}{{Oyama}, T.},
  \bibinfo{author}{{Hirota}, T.}, \bibinfo{author}{{Imai}, H.},
  \bibinfo{author}{{Shibata}, K.M.}, \bibinfo{author}{{Kobayashi}, H.},
  \bibinfo{author}{{Sato}, K.}, \bibinfo{author}{{Ueno}, Y.},
  \bibinfo{year}{2010}.
\newblock \bibinfo{title}{{VERA observations of SiO maser emission from R
  Aquarii}}.
\newblock \bibinfo{journal}{A\&A} \bibinfo{volume}{510}, \bibinfo{pages}{A69}.
\bibitem[{{Kim} et~al.(2004){Kim}, {Han}, {Sohn}, {Oh}, {Je}, {Wi} and
  {Song}}]{early_kvn}
\bibinfo{author}{{Kim}, H.G.}, \bibinfo{author}{{Han}, S.T.},
  \bibinfo{author}{{Sohn}, B.W.}, \bibinfo{author}{{Oh}, S.J.},
  \bibinfo{author}{{Je}, D.H.}, \bibinfo{author}{{Wi}, S.O.},
  \bibinfo{author}{{Song}, M.G.}, \bibinfo{year}{2004}.
\newblock \bibinfo{title}{{Construction of the Korean VLBI Network (KVN)}}, in:
  \bibinfo{editor}{{Bachiller}, R.}, \bibinfo{editor}{{Colomer}, F.},
  \bibinfo{editor}{{Desmurs}, J.F.}, \bibinfo{editor}{{de Vicente}, P.} (Eds.),
  \bibinfo{booktitle}{European VLBI Network on New Developments in VLBI Science
  and Technology}, pp. \bibinfo{pages}{281--284}.
\newblock \eprint{astro-ph/0412689}.
\bibitem[{K{\"o}nigl(1981)}]{Konigl:1981kx}
\bibinfo{author}{K{\"o}nigl, A.}, \bibinfo{year}{1981}.
\newblock \bibinfo{title}{{Relativistic jets as X-ray and gamma-ray sources}}.
\newblock \bibinfo{journal}{ApJ} \bibinfo{volume}{243}, \bibinfo{pages}{700}.
\bibitem[{{Kovalev} et~al.(2005){Kovalev}, {Kellermann}, {Lister}, {Homan},
  {Vermeulen}, {Cohen}, {Ros}, {Kadler}, {Lobanov}, {Zensus}, {Kardashev},
  {Gurvits}, {Aller} and {Aller}}]{kovalev_05}
\bibinfo{author}{{Kovalev}, Y.Y.}, \bibinfo{author}{{Kellermann}, K.I.},
  \bibinfo{author}{{Lister}, M.L.}, \bibinfo{author}{{Homan}, D.C.},
  \bibinfo{author}{{Vermeulen}, R.C.}, \bibinfo{author}{{Cohen}, M.H.},
  \bibinfo{author}{{Ros}, E.}, \bibinfo{author}{{Kadler}, M.},
  \bibinfo{author}{{Lobanov}, A.P.}, \bibinfo{author}{{Zensus}, J.A.},
  \bibinfo{author}{{Kardashev}, N.S.}, \bibinfo{author}{{Gurvits}, L.I.},
  \bibinfo{author}{{Aller}, M.F.}, \bibinfo{author}{{Aller}, H.D.},
  \bibinfo{year}{2005}.
\newblock \bibinfo{title}{{Sub-Milliarcsecond Imaging of Quasars and Active
  Galactic Nuclei. IV. Fine-Scale Structure}}.
\newblock \bibinfo{journal}{\aj} \bibinfo{volume}{130},
  \bibinfo{pages}{2473--2505}.
\newblock \eprint{astro-ph/0505536}.
\bibitem[{{Kovalev} et~al.(2008){Kovalev}, {Lobanov}, {Pushkarev} and
  {Zensus}}]{kovalev_08}
\bibinfo{author}{{Kovalev}, Y.Y.}, \bibinfo{author}{{Lobanov}, A.P.},
  \bibinfo{author}{{Pushkarev}, A.B.}, \bibinfo{author}{{Zensus}, J.A.},
  \bibinfo{year}{2008}.
\newblock \bibinfo{title}{{Opacity in compact extragalactic radio sources and
  its effect on astrophysical and astrometric studies}}.
\newblock \bibinfo{journal}{\aap} \bibinfo{volume}{483},
  \bibinfo{pages}{759--768}.
\newblock \eprint{0802.2970}.
\bibitem[{{Kovalev} et~al.(2007){Kovalev}, {Petrov}, {Fomalont} and
  {Gordon}}]{kovalev_07}
\bibinfo{author}{{Kovalev}, Y.Y.}, \bibinfo{author}{{Petrov}, L.},
  \bibinfo{author}{{Fomalont}, E.B.}, \bibinfo{author}{{Gordon}, D.},
  \bibinfo{year}{2007}.
\newblock \bibinfo{title}{{The Fifth VLBA Calibrator Survey: VCS5}}.
\newblock \bibinfo{journal}{\aj} \bibinfo{volume}{133},
  \bibinfo{pages}{1236--1242}.
\newblock \eprint{astro-ph/0607524}.
\bibitem[{{Krichbaum} et~al.(2014){Krichbaum}, {Roy}, {Lu}, {Zensus}, {Fish},
  {Doeleman} and {Event Horizon Telescope (EHT)
  Collaboration}}]{krichbaum_mmvlbi}
\bibinfo{author}{{Krichbaum}, T.P.}, \bibinfo{author}{{Roy}, A.},
  \bibinfo{author}{{Lu}, R.S.}, \bibinfo{author}{{Zensus}, J.A.},
  \bibinfo{author}{{Fish}, V.}, \bibinfo{author}{{Doeleman}, S.},
  \bibinfo{author}{{Event Horizon Telescope (EHT) Collaboration}},
  \bibinfo{year}{2014}.
\newblock \bibinfo{title}{{Millimiter VLBI observations: Black Hole Physics and
  the Origin of Jets}}, in: \bibinfo{booktitle}{Proceedings of the 12th
  European VLBI Network Symposium and Users Meeting (EVN 2014). 7-10 October
  2014. Cagliari, Italy. Online at
  http://pos.sissa.it/cgi-bin/reader/conf.cgi?confid=230,id.13},
  p.~\bibinfo{pages}{13}.
\bibitem[{{Krumholz} et~al.(2009){Krumholz}, {Klein}, {McKee}, {Offner} and
  {Cunningham}}]{Kru09}
\bibinfo{author}{{Krumholz}, M.R.}, \bibinfo{author}{{Klein}, R.I.},
  \bibinfo{author}{{McKee}, C.F.}, \bibinfo{author}{{Offner}, S.S.R.},
  \bibinfo{author}{{Cunningham}, A.J.}, \bibinfo{year}{2009}.
\newblock \bibinfo{title}{{The Formation of Massive Star Systems by
  Accretion}}.
\newblock \bibinfo{journal}{Science} \bibinfo{volume}{323},
  \bibinfo{pages}{754--}.
\newblock \eprint{arXiv:0901.3157}.
\bibitem[{{Kuiper} et~al.(2010){Kuiper}, {Klahr}, {Beuther} and
  {Henning}}]{Kui10}
\bibinfo{author}{{Kuiper}, R.}, \bibinfo{author}{{Klahr}, H.},
  \bibinfo{author}{{Beuther}, H.}, \bibinfo{author}{{Henning}, T.},
  \bibinfo{year}{2010}.
\newblock \bibinfo{title}{{Circumventing the Radiation Pressure Barrier in the
  Formation of Massive Stars via Disk Accretion}}.
\newblock \bibinfo{journal}{\apj} \bibinfo{volume}{722},
  \bibinfo{pages}{1556--1576}.
\newblock \eprint{arXiv:1008.4516}.
\bibitem[{{Kutkin} et~al.(2014){Kutkin}, {Sokolovsky}, {Lisakov}, {Kovalev},
  {Savolainen}, {Voytsik}, {Lobanov}, {Aller}, {Aller}, {Lahteenmaki},
  {Tornikoski}, {Volvach} and {Volvach}}]{kutkin_14}
\bibinfo{author}{{Kutkin}, A.M.}, \bibinfo{author}{{Sokolovsky}, K.V.},
  \bibinfo{author}{{Lisakov}, M.M.}, \bibinfo{author}{{Kovalev}, Y.Y.},
  \bibinfo{author}{{Savolainen}, T.}, \bibinfo{author}{{Voytsik}, P.A.},
  \bibinfo{author}{{Lobanov}, A.P.}, \bibinfo{author}{{Aller}, H.D.},
  \bibinfo{author}{{Aller}, M.F.}, \bibinfo{author}{{Lahteenmaki}, A.},
  \bibinfo{author}{{Tornikoski}, M.}, \bibinfo{author}{{Volvach}, A.E.},
  \bibinfo{author}{{Volvach}, L.N.}, \bibinfo{year}{2014}.
\newblock \bibinfo{title}{{The core shift effect in the blazar 3C 454.3}}.
\newblock \bibinfo{journal}{\mnras} \bibinfo{volume}{437},
  \bibinfo{pages}{3396--3404}.
\newblock \eprint{1307.4100}.
\bibitem[{Laing(1981)}]{Laing:1981bx}
\bibinfo{author}{Laing, R.A.}, \bibinfo{year}{1981}.
\newblock \bibinfo{title}{{Magnetic fields in extragalactic radio sources}}.
\newblock \bibinfo{journal}{ApJ} \bibinfo{volume}{248}, \bibinfo{pages}{87}.
\bibitem[{{Lanyi} et~al.(2010){Lanyi}, {Boboltz}, {Charlot}, {Fey}, {Fomalont},
  {Geldzahler}, {Gordon}, {Jacobs}, {Ma}, {Naudet}, {Romney}, {Sovers} and
  {Zhang}}]{lanyi_10}
\bibinfo{author}{{Lanyi}, G.E.}, \bibinfo{author}{{Boboltz}, D.A.},
  \bibinfo{author}{{Charlot}, P.}, \bibinfo{author}{{Fey}, A.L.},
  \bibinfo{author}{{Fomalont}, E.B.}, \bibinfo{author}{{Geldzahler}, B.J.},
  \bibinfo{author}{{Gordon}, D.}, \bibinfo{author}{{Jacobs}, C.S.},
  \bibinfo{author}{{Ma}, C.}, \bibinfo{author}{{Naudet}, C.J.},
  \bibinfo{author}{{Romney}, J.D.}, \bibinfo{author}{{Sovers}, O.J.},
  \bibinfo{author}{{Zhang}, L.D.}, \bibinfo{year}{2010}.
\newblock \bibinfo{title}{{The Celestial Reference Frame at 24 and 43 GHz. I.
  Astrometry}}.
\newblock \bibinfo{journal}{\aj} \bibinfo{volume}{139},
  \bibinfo{pages}{1695--1712}.
\bibitem[{{Lara} et~al.(1994){Lara}, {Alberdi}, {Marcaide} and
  {Muxlow}}]{lara_94}
\bibinfo{author}{{Lara}, L.}, \bibinfo{author}{{Alberdi}, A.},
  \bibinfo{author}{{Marcaide}, J.M.}, \bibinfo{author}{{Muxlow}, T.W.B.},
  \bibinfo{year}{1994}.
\newblock \bibinfo{title}{{The quasar 3C395 revisited: new VLBI observations
  and numerical simulations}}.
\newblock \bibinfo{journal}{\aap} \bibinfo{volume}{285},
  \bibinfo{pages}{393--403}.
\bibitem[{{Lee} et~al.(2008){Lee}, {Lobanov}, {Krichbaum}, {Witzel}, {Zensus},
  {Bremer}, {Greve} and {Grewing}}]{sslee_86}
\bibinfo{author}{{Lee}, S.S.}, \bibinfo{author}{{Lobanov}, A.P.},
  \bibinfo{author}{{Krichbaum}, T.P.}, \bibinfo{author}{{Witzel}, A.},
  \bibinfo{author}{{Zensus}, A.}, \bibinfo{author}{{Bremer}, M.},
  \bibinfo{author}{{Greve}, A.}, \bibinfo{author}{{Grewing}, M.},
  \bibinfo{year}{2008}.
\newblock \bibinfo{title}{{A Global 86 GHz VLBI Survey of Compact Radio
  Sources}}.
\newblock \bibinfo{journal}{AJ} \bibinfo{volume}{136},
  \bibinfo{pages}{159--180}.
\newblock \eprint{arXiv:0803.4035}.
\bibitem[{{Lee} et~al.(2014){Lee}, {Petrov}, {Byun}, {Kim}, {Jung}, {Song},
  {Oh}, {Roh}, {Je}, {Wi}, {Sohn}, {Oh}, {Kim}, {Yeom}, {Chung}, {Kang}, {Han},
  {Lee}, {Kim}, {Chung}, {Kim}, {Ryoung Kim}, {Kang} and {Cho}}]{sslee_14}
\bibinfo{author}{{Lee}, S.S.}, \bibinfo{author}{{Petrov}, L.},
  \bibinfo{author}{{Byun}, D.Y.}, \bibinfo{author}{{Kim}, J.},
  \bibinfo{author}{{Jung}, T.}, \bibinfo{author}{{Song}, M.G.},
  \bibinfo{author}{{Oh}, C.S.}, \bibinfo{author}{{Roh}, D.G.},
  \bibinfo{author}{{Je}, D.H.}, \bibinfo{author}{{Wi}, S.O.},
  \bibinfo{author}{{Sohn}, B.W.}, \bibinfo{author}{{Oh}, S.J.},
  \bibinfo{author}{{Kim}, K.T.}, \bibinfo{author}{{Yeom}, J.H.},
  \bibinfo{author}{{Chung}, M.H.}, \bibinfo{author}{{Kang}, J.},
  \bibinfo{author}{{Han}, S.T.}, \bibinfo{author}{{Lee}, J.W.},
  \bibinfo{author}{{Kim}, B.G.}, \bibinfo{author}{{Chung}, H.},
  \bibinfo{author}{{Kim}, H.G.}, \bibinfo{author}{{Ryoung Kim}, H.},
  \bibinfo{author}{{Kang}, Y.W.}, \bibinfo{author}{{Cho}, S.H.},
  \bibinfo{year}{2014}.
\newblock \bibinfo{title}{{Early Science with the Korean VLBI Network:
  Evaluation of System Performance}}.
\newblock \bibinfo{journal}{AJ} \bibinfo{volume}{147}, \bibinfo{pages}{77}.
\bibitem[{{Lister} and {Homan}(2005)}]{lister_05}
\bibinfo{author}{{Lister}, M.L.}, \bibinfo{author}{{Homan}, D.C.},
  \bibinfo{year}{2005}.
\newblock \bibinfo{title}{{MOJAVE: Monitoring of Jets in Active Galactic Nuclei
  with VLBA Experiments. I. First-Epoch 15 GHz Linear Polarization Images}}.
\newblock \bibinfo{journal}{\aj} \bibinfo{volume}{130},
  \bibinfo{pages}{1389--1417}.
\newblock \eprint{astro-ph/0503152}.
\bibitem[{Lobanov(1998)}]{Lobanov:1998vr}
\bibinfo{author}{Lobanov, A.P.}, \bibinfo{year}{1998}.
\newblock \bibinfo{title}{{Ultracompact jets in active galactic nuclei}}.
\newblock \bibinfo{journal}{A{\&}A} \bibinfo{volume}{330}, \bibinfo{pages}{79}.
\bibitem[{{Lockett} and {Elitzur}(1992)}]{lockett_92}
\bibinfo{author}{{Lockett}, P.}, \bibinfo{author}{{Elitzur}, M.},
  \bibinfo{year}{1992}.
\newblock \bibinfo{title}{{Modeling SiO maser emission from late-type stars}}.
\newblock \bibinfo{journal}{ApJ} \bibinfo{volume}{399},
  \bibinfo{pages}{704--713}.
\bibitem[{Marscher et~al.(2008)Marscher, Jorstad, D'Arcangelo, Smith, Williams,
  Larionov, Oh, Olmstead, Aller, Aller, McHardy, L{\"a}hteenm{\"a}ki,
  Tornikoski, Valtaoja, Hagen-Thorn, Kopatskaya, Gear, Tosti, Kurtanidze,
  Nikolashvili, Sigua, Miller and Ryle}]{Marscher:2008ii}
\bibinfo{author}{Marscher, A.P.}, \bibinfo{author}{Jorstad, S.G.},
  \bibinfo{author}{D'Arcangelo, F.D.}, \bibinfo{author}{Smith, P.S.},
  \bibinfo{author}{Williams, G.G.}, \bibinfo{author}{Larionov, V.M.},
  \bibinfo{author}{Oh, H.}, \bibinfo{author}{Olmstead, A.R.},
  \bibinfo{author}{Aller, M.F.}, \bibinfo{author}{Aller, H.D.},
  \bibinfo{author}{McHardy, I.M.}, \bibinfo{author}{L{\"a}hteenm{\"a}ki, A.},
  \bibinfo{author}{Tornikoski, M.}, \bibinfo{author}{Valtaoja, E.},
  \bibinfo{author}{Hagen-Thorn, V.A.}, \bibinfo{author}{Kopatskaya, E.N.},
  \bibinfo{author}{Gear, W.K.}, \bibinfo{author}{Tosti, G.},
  \bibinfo{author}{Kurtanidze, O.M.}, \bibinfo{author}{Nikolashvili, M.},
  \bibinfo{author}{Sigua, L.}, \bibinfo{author}{Miller, H.R.},
  \bibinfo{author}{Ryle, W.T.}, \bibinfo{year}{2008}.
\newblock \bibinfo{title}{{The inner jet of an active galactic nucleus as
  revealed by a radio-to-$\gamma$-ray outburst}}.
\newblock \bibinfo{journal}{Nature} \bibinfo{volume}{452},
  \bibinfo{pages}{966--969}.
\bibitem[{Marscher et~al.(2002)Marscher, Jorstad, G{\'o}mez, Aller,
  Ter{\"a}sranta, Lister and Stirling}]{2002Natur.417..625M}
\bibinfo{author}{Marscher, A.P.}, \bibinfo{author}{Jorstad, S.G.},
  \bibinfo{author}{G{\'o}mez, J.L.}, \bibinfo{author}{Aller, M.F.},
  \bibinfo{author}{Ter{\"a}sranta, H.}, \bibinfo{author}{Lister, M.L.},
  \bibinfo{author}{Stirling, A.M.}, \bibinfo{year}{2002}.
\newblock \bibinfo{title}{{Observational evidence for the accretion-disk origin
  for a radio jet in an active galaxy}}.
\newblock \bibinfo{journal}{Nature} \bibinfo{volume}{417},
  \bibinfo{pages}{625--627}.
\bibitem[{Marscher et~al.(2010)Marscher, Jorstad, Larionov, Aller, Aller,
  L{\"a}hteenm{\"a}ki, Agudo, Smith, Gurwell, Hagen-Thorn, Konstantinova,
  Larionova, Larionova, Melnichuk, Blinov, Kopatskaya, Troitsky, Tornikoski,
  Hovatta, Schmidt, D'Arcangelo, Bhattarai, Taylor, Olmstead, Manne-Nicholas,
  Roca-Sogorb, G{\'o}mez, McHardy, Kurtanidze, Nikolashvili, Kimeridze and
  Sigua}]{2010ApJ...710L.126M}
\bibinfo{author}{Marscher, A.P.}, \bibinfo{author}{Jorstad, S.G.},
  \bibinfo{author}{Larionov, V.M.}, \bibinfo{author}{Aller, M.F.},
  \bibinfo{author}{Aller, H.D.}, \bibinfo{author}{L{\"a}hteenm{\"a}ki, A.},
  \bibinfo{author}{Agudo, I.}, \bibinfo{author}{Smith, P.S.},
  \bibinfo{author}{Gurwell, M.}, \bibinfo{author}{Hagen-Thorn, V.A.},
  \bibinfo{author}{Konstantinova, T.S.}, \bibinfo{author}{Larionova, E.G.},
  \bibinfo{author}{Larionova, L.V.}, \bibinfo{author}{Melnichuk, D.A.},
  \bibinfo{author}{Blinov, D.A.}, \bibinfo{author}{Kopatskaya, E.N.},
  \bibinfo{author}{Troitsky, I.S.}, \bibinfo{author}{Tornikoski, M.},
  \bibinfo{author}{Hovatta, T.}, \bibinfo{author}{Schmidt, G.D.},
  \bibinfo{author}{D'Arcangelo, F.D.}, \bibinfo{author}{Bhattarai, D.},
  \bibinfo{author}{Taylor, B.}, \bibinfo{author}{Olmstead, A.R.},
  \bibinfo{author}{Manne-Nicholas, E.}, \bibinfo{author}{Roca-Sogorb, M.},
  \bibinfo{author}{G{\'o}mez, J.L.}, \bibinfo{author}{McHardy, I.M.},
  \bibinfo{author}{Kurtanidze, O.}, \bibinfo{author}{Nikolashvili, M.G.},
  \bibinfo{author}{Kimeridze, G.N.}, \bibinfo{author}{Sigua, L.A.},
  \bibinfo{year}{2010}.
\newblock \bibinfo{title}{{Probing the Inner Jet of the Quasar PKS 1510-089
  with Multi-Waveband Monitoring During Strong Gamma-Ray Activity}}.
\newblock \bibinfo{journal}{ApJ} \bibinfo{volume}{710},
  \bibinfo{pages}{L126--L131}.
\bibitem[{{Marti-Vidal} et~al.(2016){Marti-Vidal}, {Abellan}, {Marcaide},
  {Guirado}, {Perez-Torres} and {Ros}}]{marti-vidal_16}
\bibinfo{author}{{Marti-Vidal}, I.}, \bibinfo{author}{{Abellan}, F.J.},
  \bibinfo{author}{{Marcaide}, J.M.}, \bibinfo{author}{{Guirado}, J.C.},
  \bibinfo{author}{{Perez-Torres}, M.A.}, \bibinfo{author}{{Ros}, E.},
  \bibinfo{year}{2016}.
\newblock \bibinfo{title}{{Absolute kinematics of radio-source components in
  the complete S5 polar cap sample. IV. Proper motions of the radio cores over
  a decade and spectral properties}}.
\newblock \bibinfo{journal}{ArXiv e-prints} \eprint{arXiv:1607.05089}.
\bibitem[{{Mart{\'{\i}}-Vidal} et~al.(2011){Mart{\'{\i}}-Vidal}, {Marcaide},
  {Alberdi}, {P{\'e}rez-Torres}, {Ros} and {Guirado}}]{martividal_11}
\bibinfo{author}{{Mart{\'{\i}}-Vidal}, I.}, \bibinfo{author}{{Marcaide}, J.M.},
  \bibinfo{author}{{Alberdi}, A.}, \bibinfo{author}{{P{\'e}rez-Torres}, M.A.},
  \bibinfo{author}{{Ros}, E.}, \bibinfo{author}{{Guirado}, J.C.},
  \bibinfo{year}{2011}.
\newblock \bibinfo{title}{{Detection of jet precession in the active nucleus of
  M 81}}.
\newblock \bibinfo{journal}{\aap} \bibinfo{volume}{533}, \bibinfo{pages}{A111}.
\newblock \eprint{1107.0704}.
\bibitem[{{Matthews} and {Crew}(2015)}]{alma_pp}
\bibinfo{author}{{Matthews}, L.}, \bibinfo{author}{{Crew}, G.},
  \bibinfo{year}{2015}.
\newblock \bibinfo{title}{{Summary of the First ALMA Phasing Project (APP)
  Commissioning and Science Verification Mission: 2015 January 6-13}}.
\newblock \bibinfo{type}{Technical Report}. NRAO.
\bibitem[{{Matthews} et~al.(2010){Matthews}, {Greenhill}, {Goddi}, {Chandler},
  {Humphreys} and {Kunz}}]{Mat10}
\bibinfo{author}{{Matthews}, L.D.}, \bibinfo{author}{{Greenhill}, L.J.},
  \bibinfo{author}{{Goddi}, C.}, \bibinfo{author}{{Chandler}, C.J.},
  \bibinfo{author}{{Humphreys}, E.M.L.}, \bibinfo{author}{{Kunz}, M.W.},
  \bibinfo{year}{2010}.
\newblock \bibinfo{title}{{A Feature Movie of SiO Emission 20-100 AU from the
  Massive Young Stellar Object Orion Source I}}.
\newblock \bibinfo{journal}{\apj} \bibinfo{volume}{708},
  \bibinfo{pages}{80--92}.
\newblock \eprint{arXiv:0911.2473}.
\bibitem[{McKinney et~al.(2013)McKinney, Tchekhovskoy and
  Blandford}]{McKinney:2013fd}
\bibinfo{author}{McKinney, J.C.}, \bibinfo{author}{Tchekhovskoy, A.},
  \bibinfo{author}{Blandford, R.D.}, \bibinfo{year}{2013}.
\newblock \bibinfo{title}{{Alignment of Magnetized Accretion Disks and
  Relativistic Jets with Spinning Black Holes}}.
\newblock \bibinfo{journal}{Science} \bibinfo{volume}{339},
  \bibinfo{pages}{49--52}.
\bibitem[{{Middelberg} et~al.(2005){Middelberg}, {Roy}, {Walker} and
  {Falcke}}]{middelberg_05}
\bibinfo{author}{{Middelberg}, E.}, \bibinfo{author}{{Roy}, A.L.},
  \bibinfo{author}{{Walker}, R.C.}, \bibinfo{author}{{Falcke}, H.},
  \bibinfo{year}{2005}.
\newblock \bibinfo{title}{{VLBI observations of weak sources using fast
  frequency switching}}.
\newblock \bibinfo{journal}{A\&A} \bibinfo{volume}{433},
  \bibinfo{pages}{897--909}.
\newblock \eprint{astro-ph/0412564}.
\bibitem[{{Miller-Jones} et~al.(2006){Miller-Jones}, {Fender} and
  {Nakar}}]{miller-jones_06}
\bibinfo{author}{{Miller-Jones}, J.C.A.}, \bibinfo{author}{{Fender}, R.P.},
  \bibinfo{author}{{Nakar}, E.}, \bibinfo{year}{2006}.
\newblock \bibinfo{title}{{Opening angles, Lorentz factors and confinement of
  X-ray binary jets}}.
\newblock \bibinfo{journal}{\mnras} \bibinfo{volume}{367},
  \bibinfo{pages}{1432--1440}.
\newblock \eprint{astro-ph/0601482}.
\bibitem[{{Miller-Jones} et~al.(2009){Miller-Jones}, {Rupen}, {T{\"u}rler},
  {Lindfors}, {Blundell} and {Pooley}}]{miller-jones_09}
\bibinfo{author}{{Miller-Jones}, J.C.A.}, \bibinfo{author}{{Rupen}, M.P.},
  \bibinfo{author}{{T{\"u}rler}, M.}, \bibinfo{author}{{Lindfors}, E.J.},
  \bibinfo{author}{{Blundell}, K.M.}, \bibinfo{author}{{Pooley}, G.G.},
  \bibinfo{year}{2009}.
\newblock \bibinfo{title}{{Opacity effects and shock-in-jet modelling of
  low-level activity in Cygnus X-3}}.
\newblock \bibinfo{journal}{\mnras} \bibinfo{volume}{394},
  \bibinfo{pages}{309--322}.
\newblock \eprint{0811.3377}.
\bibitem[{{Mimica} et~al.(2009){Mimica}, {Aloy}, {Agudo}, {Mart{\'{\i}}},
  {G{\'o}mez} and {Miralles}}]{mimica_09}
\bibinfo{author}{{Mimica}, P.}, \bibinfo{author}{{Aloy}, M.A.},
  \bibinfo{author}{{Agudo}, I.}, \bibinfo{author}{{Mart{\'{\i}}}, J.M.},
  \bibinfo{author}{{G{\'o}mez}, J.L.}, \bibinfo{author}{{Miralles}, J.A.},
  \bibinfo{year}{2009}.
\newblock \bibinfo{title}{{Spectral Evolution of Superluminal Components in
  Parsec-Scale Jets}}.
\newblock \bibinfo{journal}{\apj} \bibinfo{volume}{696},
  \bibinfo{pages}{1142--1163}.
\newblock \eprint{arXiv:0811.1143}.
\bibitem[{Mirabel and Rodr{\'\i}guez(1994)}]{Mirabel:1994gl}
\bibinfo{author}{Mirabel, I.F.}, \bibinfo{author}{Rodr{\'\i}guez, L.F.},
  \bibinfo{year}{1994}.
\newblock \bibinfo{title}{{A superluminal source in the Galaxy}}.
\newblock \bibinfo{journal}{Nature} \bibinfo{volume}{371},
  \bibinfo{pages}{46--48}.
\bibitem[{Mirabel and Rodriguez(1999)}]{1999ARA&A..37..409M}
\bibinfo{author}{Mirabel, I.F.}, \bibinfo{author}{Rodriguez, L.F.},
  \bibinfo{year}{1999}.
\newblock \bibinfo{title}{{Sources of Relativistic Jets in the Galaxy}}.
\newblock \bibinfo{journal}{Annu. Rev. Astro. Astrophys.} \bibinfo{volume}{37},
  \bibinfo{pages}{409--443}.
\bibitem[{{Miyoshi} et~al.(1994){Miyoshi}, {Matsumoto}, {Kameno}, {Takaba} and
  {Lwata}}]{miyoshi_94}
\bibinfo{author}{{Miyoshi}, M.}, \bibinfo{author}{{Matsumoto}, K.},
  \bibinfo{author}{{Kameno}, S.}, \bibinfo{author}{{Takaba}, H.},
  \bibinfo{author}{{Lwata}, T.}, \bibinfo{year}{1994}.
\newblock \bibinfo{title}{{Collisional pumping of SiO masers in evolved
  stars}}.
\newblock \bibinfo{journal}{Nature} \bibinfo{volume}{371},
  \bibinfo{pages}{395--397}.
\bibitem[{Molina et~al.(2014)Molina, Agudo, G{\'o}mez, Krichbaum,
  Mart{\'\i}-Vidal and Roy}]{2014A&A...566A..26M}
\bibinfo{author}{Molina, S.N.}, \bibinfo{author}{Agudo, I.},
  \bibinfo{author}{G{\'o}mez, J.L.}, \bibinfo{author}{Krichbaum, T.P.},
  \bibinfo{author}{Mart{\'\i}-Vidal, I.}, \bibinfo{author}{Roy, A.L.},
  \bibinfo{year}{2014}.
\newblock \bibinfo{title}{{Evidence of internal rotation and a helical magnetic
  field in the jet of the quasar NRAO 150}}.
\newblock \bibinfo{journal}{A{\&}A} \bibinfo{volume}{566},
  \bibinfo{pages}{A26}.
\bibitem[{Momjian et~al.(2003)Momjian, Romney, Carilli and
  Troland}]{momjian2003sensitive}
\bibinfo{author}{Momjian, E.}, \bibinfo{author}{Romney, J.D.},
  \bibinfo{author}{Carilli, C.L.}, \bibinfo{author}{Troland, T.H.},
  \bibinfo{year}{2003}.
\newblock \bibinfo{title}{Sensitive vlbi continuum and hi absorption
  observations of ngc 7674: First scientific observations with the combined
  array vlba, vla, and arecibo}.
\newblock \bibinfo{journal}{The Astrophysical Journal} \bibinfo{volume}{597},
  \bibinfo{pages}{809}.
\bibitem[{{Moscadelli} et~al.(2011){Moscadelli}, {Cesaroni}, {Rioja}, {Dodson}
  and {Reid}}]{moscadelli_11}
\bibinfo{author}{{Moscadelli}, L.}, \bibinfo{author}{{Cesaroni}, R.},
  \bibinfo{author}{{Rioja}, M.J.}, \bibinfo{author}{{Dodson}, R.},
  \bibinfo{author}{{Reid}, M.J.}, \bibinfo{year}{2011}.
\newblock \bibinfo{title}{{Methanol and water masers in IRAS 20126+4104: the
  distance, the disk, and the jet}}.
\newblock \bibinfo{journal}{\aap} \bibinfo{volume}{526}, \bibinfo{pages}{A66}.
\newblock \eprint{1011.4816}.
\bibitem[{{Neufeld} and {Melnick}(1991)}]{neufeld_91}
\bibinfo{author}{{Neufeld}, D.A.}, \bibinfo{author}{{Melnick}, G.J.},
  \bibinfo{year}{1991}.
\newblock \bibinfo{title}{{Excitation of millimeter and submillimeter water
  masers}}.
\newblock \bibinfo{journal}{\apj} \bibinfo{volume}{368},
  \bibinfo{pages}{215--230}.
\bibitem[{{Neufeld} et~al.(2013){Neufeld}, {Wu}, {Kraus}, {Menten}, {Tolls},
  {Melnick} and {Nagy}}]{neufeld_13}
\bibinfo{author}{{Neufeld}, D.A.}, \bibinfo{author}{{Wu}, Y.},
  \bibinfo{author}{{Kraus}, A.}, \bibinfo{author}{{Menten}, K.M.},
  \bibinfo{author}{{Tolls}, V.}, \bibinfo{author}{{Melnick}, G.J.},
  \bibinfo{author}{{Nagy}, Z.}, \bibinfo{year}{2013}.
\newblock \bibinfo{title}{{Herschel/HIFI Observations of a New Interstellar
  Water Maser: The 5$_{32}$-4$_{41}$ Transition at 620.701 GHz}}.
\newblock \bibinfo{journal}{\apj} \bibinfo{volume}{769}, \bibinfo{pages}{48}.
\newblock \eprint{arXiv:1304.0910}.
\bibitem[{{Oh} et~al.(2011){Oh}, {Roh}, {Wajima}, {Kawaguchi}, {Byun}, {Yeom},
  {Je}, {Han}, {Iguchi}, {Kawakami}, {Ozeki}, {Kobayashi}, {Sasao}, {Sohn},
  {Kim}, {Miyazaki}, {Oyama} and {Kurayama}}]{kvn_backend}
\bibinfo{author}{{Oh}, S.J.}, \bibinfo{author}{{Roh}, D.G.},
  \bibinfo{author}{{Wajima}, K.}, \bibinfo{author}{{Kawaguchi}, N.},
  \bibinfo{author}{{Byun}, D.Y.}, \bibinfo{author}{{Yeom}, J.H.},
  \bibinfo{author}{{Je}, D.H.}, \bibinfo{author}{{Han}, S.T.},
  \bibinfo{author}{{Iguchi}, S.}, \bibinfo{author}{{Kawakami}, K.},
  \bibinfo{author}{{Ozeki}, K.}, \bibinfo{author}{{Kobayashi}, H.},
  \bibinfo{author}{{Sasao}, T.}, \bibinfo{author}{{Sohn}, B.},
  \bibinfo{author}{{Kim}, J.}, \bibinfo{author}{{Miyazaki}, A.},
  \bibinfo{author}{{Oyama}, T.}, \bibinfo{author}{{Kurayama}, T.},
  \bibinfo{year}{2011}.
\newblock \bibinfo{title}{{Design and Development of a High-Speed
  Data-Acquisition System for the Korean VLBI Network}}.
\newblock \bibinfo{journal}{PASJ} \bibinfo{volume}{63},
  \bibinfo{pages}{1229--1242}.
\bibitem[{{Ojha} et~al.(2005){Ojha}, {Fey}, {Charlot}, {Jauncey}, {Johnston},
  {Reynolds}, {Tzioumis}, {Quick}, {Nicolson}, {Ellingsen}, {McCulloch} and
  {Koyama}}]{ojha_05}
\bibinfo{author}{{Ojha}, R.}, \bibinfo{author}{{Fey}, A.L.},
  \bibinfo{author}{{Charlot}, P.}, \bibinfo{author}{{Jauncey}, D.L.},
  \bibinfo{author}{{Johnston}, K.J.}, \bibinfo{author}{{Reynolds}, J.E.},
  \bibinfo{author}{{Tzioumis}, A.K.}, \bibinfo{author}{{Quick}, J.F.H.},
  \bibinfo{author}{{Nicolson}, G.D.}, \bibinfo{author}{{Ellingsen}, S.P.},
  \bibinfo{author}{{McCulloch}, P.M.}, \bibinfo{author}{{Koyama}, Y.},
  \bibinfo{year}{2005}.
\newblock \bibinfo{title}{{VLBI Observations of Southern Hemisphere ICRF
  Sources. II. Astrometric Suitability Based on Intrinsic Structure}}.
\newblock \bibinfo{journal}{\aj} \bibinfo{volume}{130},
  \bibinfo{pages}{2529--2540}.
\bibitem[{{Ojha} et~al.(2004){Ojha}, {Fey}, {Johnston}, {Jauncey}, {Reynolds},
  {Tzioumis}, {Quick}, {Nicolson}, {Ellingsen}, {Dodson} and
  {McCulloch}}]{ojha_04}
\bibinfo{author}{{Ojha}, R.}, \bibinfo{author}{{Fey}, A.L.},
  \bibinfo{author}{{Johnston}, K.J.}, \bibinfo{author}{{Jauncey}, D.L.},
  \bibinfo{author}{{Reynolds}, J.E.}, \bibinfo{author}{{Tzioumis}, A.K.},
  \bibinfo{author}{{Quick}, J.F.H.}, \bibinfo{author}{{Nicolson}, G.D.},
  \bibinfo{author}{{Ellingsen}, S.P.}, \bibinfo{author}{{Dodson}, R.G.},
  \bibinfo{author}{{McCulloch}, P.M.}, \bibinfo{year}{2004}.
\newblock \bibinfo{title}{{VLBI Observations of Southern Hemisphere ICRF
  Sources. I.}}
\newblock \bibinfo{journal}{\aj} \bibinfo{volume}{127},
  \bibinfo{pages}{3609--3621}.
\bibitem[{{Ojha} et~al.(2010){Ojha}, {Kadler}, {B{\"o}ck}, {Booth}, {Dutka},
  {Edwards}, {Fey}, {Fuhrmann}, {Gaume}, {Hase}, {Horiuchi}, {Jauncey},
  {Johnston}, {Katz}, {Lister}, {Lovell}, {M{\"u}ller}, {Pl{\"o}tz}, {Quick},
  {Ros}, {Taylor}, {Thompson}, {Tingay}, {Tosti}, {Tzioumis}, {Wilms} and
  {Zensus}}]{ojha_10}
\bibinfo{author}{{Ojha}, R.}, \bibinfo{author}{{Kadler}, M.},
  \bibinfo{author}{{B{\"o}ck}, M.}, \bibinfo{author}{{Booth}, R.},
  \bibinfo{author}{{Dutka}, M.S.}, \bibinfo{author}{{Edwards}, P.G.},
  \bibinfo{author}{{Fey}, A.L.}, \bibinfo{author}{{Fuhrmann}, L.},
  \bibinfo{author}{{Gaume}, R.A.}, \bibinfo{author}{{Hase}, H.},
  \bibinfo{author}{{Horiuchi}, S.}, \bibinfo{author}{{Jauncey}, D.L.},
  \bibinfo{author}{{Johnston}, K.J.}, \bibinfo{author}{{Katz}, U.},
  \bibinfo{author}{{Lister}, M.}, \bibinfo{author}{{Lovell}, J.E.J.},
  \bibinfo{author}{{M{\"u}ller}, C.}, \bibinfo{author}{{Pl{\"o}tz}, C.},
  \bibinfo{author}{{Quick}, J.F.H.}, \bibinfo{author}{{Ros}, E.},
  \bibinfo{author}{{Taylor}, G.B.}, \bibinfo{author}{{Thompson}, D.J.},
  \bibinfo{author}{{Tingay}, S.J.}, \bibinfo{author}{{Tosti}, G.},
  \bibinfo{author}{{Tzioumis}, A.K.}, \bibinfo{author}{{Wilms}, J.},
  \bibinfo{author}{{Zensus}, J.A.}, \bibinfo{year}{2010}.
\newblock \bibinfo{title}{{TANAMI: tracking active galactic nuclei with austral
  milliarcsecond interferometry . I. First-epoch 8.4 GHz images}}.
\newblock \bibinfo{journal}{\aap} \bibinfo{volume}{519}, \bibinfo{pages}{A45}.
\newblock \eprint{1005.4432}.
\bibitem[{{Olofsson} et~al.(1981){Olofsson}, {Rydbeck}, {Lane} and
  {Predmore}}]{olofsson_81}
\bibinfo{author}{{Olofsson}, H.}, \bibinfo{author}{{Rydbeck}, O.E.H.},
  \bibinfo{author}{{Lane}, A.P.}, \bibinfo{author}{{Predmore}, C.R.},
  \bibinfo{year}{1981}.
\newblock \bibinfo{title}{{Detection of the SiO /v = 2, J = 2-1/ maser}}.
\newblock \bibinfo{journal}{ApJL} \bibinfo{volume}{247},
  \bibinfo{pages}{L81--L84}.
\bibitem[{{O'Sullivan} and {Gabuzda}(2009)}]{osullivan_09}
\bibinfo{author}{{O'Sullivan}, S.P.}, \bibinfo{author}{{Gabuzda}, D.C.},
  \bibinfo{year}{2009}.
\newblock \bibinfo{title}{{Magnetic field strength and spectral distribution of
  six parsec-scale active galactic nuclei jets}}.
\newblock \bibinfo{journal}{\mnras} \bibinfo{volume}{400},
  \bibinfo{pages}{26--42}.
\newblock \eprint{0907.5211}.
\bibitem[{O'Sullivan and Gabuzda(2009)}]{OSullivan:2009bx}
\bibinfo{author}{O'Sullivan, S.P.}, \bibinfo{author}{Gabuzda, D.C.},
  \bibinfo{year}{2009}.
\newblock \bibinfo{title}{{Three-dimensional magnetic field structure of six
  parsec-scale active galactic nuclei jets}}.
\newblock \bibinfo{journal}{MNRAS} \bibinfo{volume}{393},
  \bibinfo{pages}{429--456}.
\bibitem[{{Perucho} et~al.(2010){Perucho}, {Bosch-Ramon} and
  {Khangulyan}}]{perucho_10}
\bibinfo{author}{{Perucho}, M.}, \bibinfo{author}{{Bosch-Ramon}, V.},
  \bibinfo{author}{{Khangulyan}, D.}, \bibinfo{year}{2010}.
\newblock \bibinfo{title}{{3D simulations of wind-jet interaction in massive
  X-ray binaries}}.
\newblock \bibinfo{journal}{\aap} \bibinfo{volume}{512}, \bibinfo{pages}{L4}.
\newblock \eprint{arXiv:1002.4562}.
\bibitem[{{Petrov}(2016)}]{petrov_16}
\bibinfo{author}{{Petrov}, L.}, \bibinfo{year}{2016}.
\newblock \bibinfo{title}{{VLBA Calibrator Survey 9 (VCS-9)}}.
\newblock \bibinfo{journal}{ArXiv e-prints} \eprint{1610.04951}.
\bibitem[{{Petrov} et~al.(2007){Petrov}, {Hirota}, {Honma}, {Shibata}, {Jike}
  and {Kobayashi}}]{petrov_07}
\bibinfo{author}{{Petrov}, L.}, \bibinfo{author}{{Hirota}, T.},
  \bibinfo{author}{{Honma}, M.}, \bibinfo{author}{{Shibata}, K.M.},
  \bibinfo{author}{{Jike}, T.}, \bibinfo{author}{{Kobayashi}, H.},
  \bibinfo{year}{2007}.
\newblock \bibinfo{title}{{VERA 22 GHz Fringe Search Survey}}.
\newblock \bibinfo{journal}{\aj} \bibinfo{volume}{133},
  \bibinfo{pages}{2487--2494}.
\newblock \eprint{astro-ph/0609557}.
\bibitem[{{Petrov} et~al.(2008){Petrov}, {Kovalev}, {Fomalont} and
  {Gordon}}]{petrov_08}
\bibinfo{author}{{Petrov}, L.}, \bibinfo{author}{{Kovalev}, Y.Y.},
  \bibinfo{author}{{Fomalont}, E.B.}, \bibinfo{author}{{Gordon}, D.},
  \bibinfo{year}{2008}.
\newblock \bibinfo{title}{{The Sixth VLBA Calibrator Survey: VCS6}}.
\newblock \bibinfo{journal}{\aj} \bibinfo{volume}{136},
  \bibinfo{pages}{580--585}.
\newblock \eprint{0801.3895}.
\bibitem[{{Petrov} et~al.(2012){Petrov}, {Lee}, {Kim}, {Jung}, {Oh}, {Sohn},
  {Byun}, {Chung}, {Je}, {Wi}, {Song}, {Kang}, {Han}, {Lee}, {Kim}, {Chung} and
  {Kim}}]{petrov_12}
\bibinfo{author}{{Petrov}, L.}, \bibinfo{author}{{Lee}, S.S.},
  \bibinfo{author}{{Kim}, J.}, \bibinfo{author}{{Jung}, T.},
  \bibinfo{author}{{Oh}, J.}, \bibinfo{author}{{Sohn}, B.W.},
  \bibinfo{author}{{Byun}, D.Y.}, \bibinfo{author}{{Chung}, M.H.},
  \bibinfo{author}{{Je}, D.H.}, \bibinfo{author}{{Wi}, S.O.},
  \bibinfo{author}{{Song}, M.G.}, \bibinfo{author}{{Kang}, J.},
  \bibinfo{author}{{Han}, S.T.}, \bibinfo{author}{{Lee}, J.W.},
  \bibinfo{author}{{Kim}, B.G.}, \bibinfo{author}{{Chung}, H.},
  \bibinfo{author}{{Kim}, H.G.}, \bibinfo{year}{2012}.
\newblock \bibinfo{title}{{Early Science with the Korean VLBI Network: The
  QCAL-1 43 GHz Calibrator Survey}}.
\newblock \bibinfo{journal}{\aj} \bibinfo{volume}{144}, \bibinfo{pages}{150}.
\newblock \eprint{1207.5872}.
\bibitem[{{Pollack} et~al.(2003){Pollack}, {Taylor} and {Zavala}}]{pollack_03}
\bibinfo{author}{{Pollack}, L.K.}, \bibinfo{author}{{Taylor}, G.B.},
  \bibinfo{author}{{Zavala}, R.T.}, \bibinfo{year}{2003}.
\newblock \bibinfo{title}{{VLBI Polarimetry of 177 Sources from the
  Caltech-Jodrell Bank Flat-Spectrum Survey}}.
\newblock \bibinfo{journal}{\apj} \bibinfo{volume}{589},
  \bibinfo{pages}{733--751}.
\newblock \eprint{astro-ph/0302211}.
\bibitem[{{Porcas} and {Rioja}(2002)}]{porcas_02}
\bibinfo{author}{{Porcas}, R.W.}, \bibinfo{author}{{Rioja}, M.J.},
  \bibinfo{year}{2002}.
\newblock \bibinfo{title}{{VLBI phase-reference investigations at 86 GHz}}, in:
  \bibinfo{editor}{{Ros}, E.}, \bibinfo{editor}{{Porcas}, R.W.},
  \bibinfo{editor}{{Lobanov}, A.P.}, \bibinfo{editor}{{Zensus}, J.A.} (Eds.),
  \bibinfo{booktitle}{Proceedings of the 6th EVN Symposium},
  p.~\bibinfo{pages}{65}.
\bibitem[{Porth et~al.(2011)Porth, Fendt, Meliani and Vaidya}]{Porth:2011ev}
\bibinfo{author}{Porth, O.}, \bibinfo{author}{Fendt, C.},
  \bibinfo{author}{Meliani, Z.}, \bibinfo{author}{Vaidya, B.},
  \bibinfo{year}{2011}.
\newblock \bibinfo{title}{{Synchrotron Radiation of Self-collimating
  Relativistic Magnetohydrodynamic Jets}}.
\newblock \bibinfo{journal}{ApJ} \bibinfo{volume}{737}, \bibinfo{pages}{42}.
\bibitem[{{Pudritz} and {Banerjee}(2005)}]{Pud05}
\bibinfo{author}{{Pudritz}, R.E.}, \bibinfo{author}{{Banerjee}, R.},
  \bibinfo{year}{2005}.
\newblock \bibinfo{title}{{The disc-jet connection}}, in:
  \bibinfo{editor}{{Cesaroni}, R.}, \bibinfo{editor}{{Felli}, M.},
  \bibinfo{editor}{{Churchwell}, E.}, \bibinfo{editor}{{Walmsley}, M.} (Eds.),
  \bibinfo{booktitle}{Massive Star Birth: A Crossroads of Astrophysics}, pp.
  \bibinfo{pages}{163--173}.
\newblock \eprint{arXiv:astro-ph/0507268}.
\bibitem[{{Reid} and {Honma}(2014)}]{reid_micro}
\bibinfo{author}{{Reid}, M.J.}, \bibinfo{author}{{Honma}, M.},
  \bibinfo{year}{2014}.
\newblock \bibinfo{title}{{Microarcsecond Radio Astrometry}}.
\newblock \bibinfo{journal}{ARAA} \bibinfo{volume}{52},
  \bibinfo{pages}{339--372}.
\newblock \eprint{arXiv:1312.2871}.
\bibitem[{{Reid} et~al.(2007){Reid}, {Menten}, {Greenhill} and
  {Chandler}}]{Rei07}
\bibinfo{author}{{Reid}, M.J.}, \bibinfo{author}{{Menten}, K.M.},
  \bibinfo{author}{{Greenhill}, L.J.}, \bibinfo{author}{{Chandler}, C.J.},
  \bibinfo{year}{2007}.
\newblock \bibinfo{title}{{Imaging the Ionized Disk of the High-Mass Protostar
  Orion I}}.
\newblock \bibinfo{journal}{\apj} \bibinfo{volume}{664},
  \bibinfo{pages}{950--955}.
\newblock \eprint{arXiv:0704.2309}.
\bibitem[{{Rib{\'o}} et~al.(2004){Rib{\'o}}, {Dhawan} and {Mirabel}}]{ribo_04}
\bibinfo{author}{{Rib{\'o}}, M.}, \bibinfo{author}{{Dhawan}, V.},
  \bibinfo{author}{{Mirabel}, I.F.}, \bibinfo{year}{2004}.
\newblock \bibinfo{title}{{The asymmetric compact jet of GRS 1915+105}}, in:
  \bibinfo{editor}{{Bachiller}, R.}, \bibinfo{editor}{{Colomer}, F.},
  \bibinfo{editor}{{Desmurs}, J.F.}, \bibinfo{editor}{{de Vicente}, P.} (Eds.),
  \bibinfo{booktitle}{European VLBI Network on New Developments in VLBI Science
  and Technology}, pp. \bibinfo{pages}{111--112}.
\newblock \eprint{astro-ph/0412657}.
\bibitem[{{Richards} et~al.(2012){Richards}, {Etoka}, {Gray}, {Lekht},
  {Mendoza-Torres}, {Murakawa}, {Rudnitskij} and {Yates}}]{richards_12}
\bibinfo{author}{{Richards}, A.M.S.}, \bibinfo{author}{{Etoka}, S.},
  \bibinfo{author}{{Gray}, M.D.}, \bibinfo{author}{{Lekht}, E.E.},
  \bibinfo{author}{{Mendoza-Torres}, J.E.}, \bibinfo{author}{{Murakawa}, K.},
  \bibinfo{author}{{Rudnitskij}, G.}, \bibinfo{author}{{Yates}, J.A.},
  \bibinfo{year}{2012}.
\newblock \bibinfo{title}{{Evolved star water maser cloud size determined by
  star size}}.
\newblock \bibinfo{journal}{\aap} \bibinfo{volume}{546}, \bibinfo{pages}{A16}.
\newblock \eprint{arXiv:1207.2583}.
\bibitem[{{Richards} et~al.(2014){Richards}, {Impellizzeri}, {Humphreys},
  {Vlahakis}, {Vlemmings}, {Baudry}, {De Beck}, {Decin}, {Etoka}, {Gray},
  {Harper}, {Hunter}, {Kervella}, {Kerschbaum}, {McDonald}, {Melnick},
  {Muller}, {Neufeld}, {O'Gorman}, {Parfenov}, {Peck}, {Shinnaga}, {Sobolev},
  {Testi}, {Uscanga}, {Wootten}, {Yates} and {Zijlstra}}]{richards_14}
\bibinfo{author}{{Richards}, A.M.S.}, \bibinfo{author}{{Impellizzeri}, C.M.V.},
  \bibinfo{author}{{Humphreys}, E.M.}, \bibinfo{author}{{Vlahakis}, C.},
  \bibinfo{author}{{Vlemmings}, W.}, \bibinfo{author}{{Baudry}, A.},
  \bibinfo{author}{{De Beck}, E.}, \bibinfo{author}{{Decin}, L.},
  \bibinfo{author}{{Etoka}, S.}, \bibinfo{author}{{Gray}, M.D.},
  \bibinfo{author}{{Harper}, G.M.}, \bibinfo{author}{{Hunter}, T.R.},
  \bibinfo{author}{{Kervella}, P.}, \bibinfo{author}{{Kerschbaum}, F.},
  \bibinfo{author}{{McDonald}, I.}, \bibinfo{author}{{Melnick}, G.},
  \bibinfo{author}{{Muller}, S.}, \bibinfo{author}{{Neufeld}, D.},
  \bibinfo{author}{{O'Gorman}, E.}, \bibinfo{author}{{Parfenov}, S.Y.},
  \bibinfo{author}{{Peck}, A.B.}, \bibinfo{author}{{Shinnaga}, H.},
  \bibinfo{author}{{Sobolev}, A.M.}, \bibinfo{author}{{Testi}, L.},
  \bibinfo{author}{{Uscanga}, L.}, \bibinfo{author}{{Wootten}, A.},
  \bibinfo{author}{{Yates}, J.A.}, \bibinfo{author}{{Zijlstra}, A.},
  \bibinfo{year}{2014}.
\newblock \bibinfo{title}{{ALMA sub-mm maser and dust distribution of VY Canis
  Majoris}}.
\newblock \bibinfo{journal}{\aap} \bibinfo{volume}{572}, \bibinfo{pages}{L9}.
\newblock \eprint{1409.5497}.
\bibitem[{{Rioja} and {Dodson}(2011)}]{rioja_11a}
\bibinfo{author}{{Rioja}, M.}, \bibinfo{author}{{Dodson}, R.},
  \bibinfo{year}{2011}.
\newblock \bibinfo{title}{{High-precision Astrometric Millimeter Very Long
  Baseline Interferometry Using a New Method for Atmospheric Calibration}}.
\newblock \bibinfo{journal}{AJ} \bibinfo{volume}{141}, \bibinfo{pages}{114}.
\newblock \eprint{arXiv:1101.2051}.
\bibitem[{{Rioja} et~al.(2011){Rioja}, {Dodson}, {Malarecki} and
  {Asaki}}]{rioja_11b}
\bibinfo{author}{{Rioja}, M.}, \bibinfo{author}{{Dodson}, R.},
  \bibinfo{author}{{Malarecki}, J.}, \bibinfo{author}{{Asaki}, Y.},
  \bibinfo{year}{2011}.
\newblock \bibinfo{title}{{Exploration of Source Frequency Phase Referencing
  Techniques for Astrometry and Observations of Weak Sources with High
  Frequency Space Very Long Baseline Interferometry}}.
\newblock \bibinfo{journal}{AJ} \bibinfo{volume}{142}, \bibinfo{pages}{157}.
\newblock \eprint{arXiv:1110.0267}.
\bibitem[{{Rioja} et~al.(2015){Rioja}, {Dodson}, {Jung} and {Sohn}}]{rioja_15}
\bibinfo{author}{{Rioja}, M.J.}, \bibinfo{author}{{Dodson}, R.},
  \bibinfo{author}{{Jung}, T.}, \bibinfo{author}{{Sohn}, B.W.},
  \bibinfo{year}{2015}.
\newblock \bibinfo{title}{{The Power of Simultaneous Multi-Frequency
  Observations for mm-VLBI: Astrometry up to 130 GHz with the KVN}}.
\newblock \bibinfo{journal}{AJ} \bibinfo{volume}{150}, \bibinfo{pages}{202}.
\bibitem[{{Rioja} et~al.(2014){Rioja}, {Dodson}, {Jung}, {Sohn}, {Byun},
  {Agudo}, {Cho}, {Lee}, {Kim}, {Kim}, {Oh}, {Han}, {Je}, {Chung}, {Wi},
  {Kang}, {Lee}, {Chung}, {Ryoung Kim}, {Kim}, {Lee}, {Roh}, {Oh}, {Yeom},
  {Song} and {Kang}}]{rioja_14}
\bibinfo{author}{{Rioja}, M.J.}, \bibinfo{author}{{Dodson}, R.},
  \bibinfo{author}{{Jung}, T.}, \bibinfo{author}{{Sohn}, B.W.},
  \bibinfo{author}{{Byun}, D.Y.}, \bibinfo{author}{{Agudo}, I.},
  \bibinfo{author}{{Cho}, S.H.}, \bibinfo{author}{{Lee}, S.S.},
  \bibinfo{author}{{Kim}, J.}, \bibinfo{author}{{Kim}, K.T.},
  \bibinfo{author}{{Oh}, C.S.}, \bibinfo{author}{{Han}, S.T.},
  \bibinfo{author}{{Je}, D.H.}, \bibinfo{author}{{Chung}, M.H.},
  \bibinfo{author}{{Wi}, S.O.}, \bibinfo{author}{{Kang}, J.},
  \bibinfo{author}{{Lee}, J.W.}, \bibinfo{author}{{Chung}, H.},
  \bibinfo{author}{{Ryoung Kim}, H.}, \bibinfo{author}{{Kim}, H.G.},
  \bibinfo{author}{{Lee}, C.H.}, \bibinfo{author}{{Roh}, D.G.},
  \bibinfo{author}{{Oh}, S.J.}, \bibinfo{author}{{Yeom}, J.H.},
  \bibinfo{author}{{Song}, M.G.}, \bibinfo{author}{{Kang}, Y.W.},
  \bibinfo{year}{2014}.
\newblock \bibinfo{title}{{Verification of the Astrometric Performance of the
  Korean VLBI Network, Using Comparative SFPR Studies with the VLBA at 14/7
  mm}}.
\newblock \bibinfo{journal}{AJ} \bibinfo{volume}{148}, \bibinfo{pages}{84}.
\newblock \eprint{arXiv:1407.4604}.
\bibitem[{{Rioja} et~al.(2008){Rioja}, {Dodson}, {Kamohara}, {Colomer},
  {Bujarrabal} and {Kobayashi}}]{rioja_08}
\bibinfo{author}{{Rioja}, M.J.}, \bibinfo{author}{{Dodson}, R.},
  \bibinfo{author}{{Kamohara}, R.}, \bibinfo{author}{{Colomer}, F.},
  \bibinfo{author}{{Bujarrabal}, V.}, \bibinfo{author}{{Kobayashi}, H.},
  \bibinfo{year}{2008}.
\newblock \bibinfo{title}{{Relative Astrometry of the J = 1$\rightarrow$0, v =
  1 and v = 2 SiO Masers toward R Leonis Minoris Using VERA}}.
\newblock \bibinfo{journal}{PASJ} \bibinfo{volume}{60},
  \bibinfo{pages}{1031--1038}.
\newblock \eprint{arXiv:0811.3820}.
\bibitem[{{Rioja} and {Porcas}(1998)}]{rioja_98}
\bibinfo{author}{{Rioja}, M.J.}, \bibinfo{author}{{Porcas}, R.W.},
  \bibinfo{year}{1998}.
\newblock \bibinfo{title}{{Multi-Frequency VLBA+Effelsberg Observations of
  1038+528 A/B}}, in: \bibinfo{editor}{{Zensus}, J.A.},
  \bibinfo{editor}{{Taylor}, G.B.}, \bibinfo{editor}{{Wrobel}, J.M.} (Eds.),
  \bibinfo{booktitle}{IAU Colloq. 164: Radio Emission from Galactic and
  Extragalactic Compact Sources}, p.~\bibinfo{pages}{95}.
\bibitem[{{Ros} et~al.(2001){Ros}, {Marcaide}, {Guirado} and
  {P{\'e}rez-Torres}}]{ros_01}
\bibinfo{author}{{Ros}, E.}, \bibinfo{author}{{Marcaide}, J.M.},
  \bibinfo{author}{{Guirado}, J.C.}, \bibinfo{author}{{P{\'e}rez-Torres},
  M.A.}, \bibinfo{year}{2001}.
\newblock \bibinfo{title}{{Absolute kinematics of radio source components in
  the complete S5 polar cap sample. I. First and second epoch maps at 8.4
  GHz}}.
\newblock \bibinfo{journal}{A\&A} \bibinfo{volume}{376},
  \bibinfo{pages}{1090--1105}.
\newblock \eprint{astro-ph/0107155}.
\bibitem[{{S{\'a}nchez Contreras} et~al.(2002){S{\'a}nchez Contreras},
  {Desmurs}, {Bujarrabal}, {Alcolea} and {Colomer}}]{sanchez_02}
\bibinfo{author}{{S{\'a}nchez Contreras}, C.}, \bibinfo{author}{{Desmurs},
  J.F.}, \bibinfo{author}{{Bujarrabal}, V.}, \bibinfo{author}{{Alcolea}, J.},
  \bibinfo{author}{{Colomer}, F.}, \bibinfo{year}{2002}.
\newblock \bibinfo{title}{{Submilliarcsecond-resolution mapping of the 43 GHz
  SiO maser emission in the bipolar post-AGB nebula OH231.8+4.2}}.
\newblock \bibinfo{journal}{A\&A} \bibinfo{volume}{385},
  \bibinfo{pages}{L1--L4}.
\bibitem[{{Shen} et~al.(2005){Shen}, {Lo}, {Liang}, {Ho} and
  {Zhao}}]{sgra_size}
\bibinfo{author}{{Shen}, Z.Q.}, \bibinfo{author}{{Lo}, K.Y.},
  \bibinfo{author}{{Liang}, M.C.}, \bibinfo{author}{{Ho}, P.T.P.},
  \bibinfo{author}{{Zhao}, J.H.}, \bibinfo{year}{2005}.
\newblock \bibinfo{title}{{A size of \~{}1AU for the radio source Sgr A* at the
  centre of the Milky Way}}.
\newblock \bibinfo{journal}{\nat} \bibinfo{volume}{438},
  \bibinfo{pages}{62--64}.
\newblock \eprint{astro-ph/0512515}.
\bibitem[{{Shu}(1985)}]{Shu85}
\bibinfo{author}{{Shu}, F.H.}, \bibinfo{year}{1985}.
\newblock \bibinfo{title}{{Star formation in molecular clouds}}, in:
  \bibinfo{editor}{{van Woerden}, H.}, \bibinfo{editor}{{Allen}, R.J.},
  \bibinfo{editor}{{Burton}, W.B.} (Eds.), \bibinfo{booktitle}{The Milky Way
  Galaxy}, pp. \bibinfo{pages}{561--566}.
\bibitem[{{Soker}(2001)}]{soker_01}
\bibinfo{author}{{Soker}, N.}, \bibinfo{year}{2001}.
\newblock \bibinfo{title}{{Collimated Fast Winds in Wide Binary Progenitors of
  Planetary Nebulae}}.
\newblock \bibinfo{journal}{ApJ} \bibinfo{volume}{558},
  \bibinfo{pages}{157--164}.
\newblock \eprint{astro-ph/0102110}.
\bibitem[{{Sokolovsky} et~al.(2011){Sokolovsky}, {Kovalev}, {Pushkarev},
  {Mimica} and {Perucho}}]{sokolovsky_11}
\bibinfo{author}{{Sokolovsky}, K.V.}, \bibinfo{author}{{Kovalev}, Y.Y.},
  \bibinfo{author}{{Pushkarev}, A.B.}, \bibinfo{author}{{Mimica}, P.},
  \bibinfo{author}{{Perucho}, M.}, \bibinfo{year}{2011}.
\newblock \bibinfo{title}{{VLBI-selected sample of compact symmetric object
  candidates and frequency-dependent position of hotspots}}.
\newblock \bibinfo{journal}{\aap} \bibinfo{volume}{535}, \bibinfo{pages}{A24}.
\newblock \eprint{1107.0719}.
\bibitem[{{Soria-Ruiz} et~al.(2004){Soria-Ruiz}, {Alcolea}, {Colomer},
  {Bujarrabal}, {Desmurs}, {Marvel} and {Diamond}}]{soria-ruiz_04}
\bibinfo{author}{{Soria-Ruiz}, R.}, \bibinfo{author}{{Alcolea}, J.},
  \bibinfo{author}{{Colomer}, F.}, \bibinfo{author}{{Bujarrabal}, V.},
  \bibinfo{author}{{Desmurs}, J.F.}, \bibinfo{author}{{Marvel}, K.B.},
  \bibinfo{author}{{Diamond}, P.J.}, \bibinfo{year}{2004}.
\newblock \bibinfo{title}{{High resolution observations of SiO masers:
  Comparing the spatial distribution at 43 and 86 GHz}}.
\newblock \bibinfo{journal}{A\&A} \bibinfo{volume}{426},
  \bibinfo{pages}{131--144}.
\newblock \eprint{astro-ph/0409467}.
\bibitem[{{Stirling} et~al.(2001){Stirling}, {Spencer}, {de la Force},
  {Garrett}, {Fender} and {Ogley}}]{stirling_01}
\bibinfo{author}{{Stirling}, A.M.}, \bibinfo{author}{{Spencer}, R.E.},
  \bibinfo{author}{{de la Force}, C.J.}, \bibinfo{author}{{Garrett}, M.A.},
  \bibinfo{author}{{Fender}, R.P.}, \bibinfo{author}{{Ogley}, R.N.},
  \bibinfo{year}{2001}.
\newblock \bibinfo{title}{{A relativistic jet from Cygnus X-1 in the low/hard
  X-ray state}}.
\newblock \bibinfo{journal}{\mnras} \bibinfo{volume}{327},
  \bibinfo{pages}{1273--1278}.
\newblock \eprint{astro-ph/0107192}.
\bibitem[{{Taylor} et~al.(1996){Taylor}, {Vermeulen}, {Readhead}, {Pearson},
  {Henstock} and {Wilkinson}}]{taylor_96}
\bibinfo{author}{{Taylor}, G.B.}, \bibinfo{author}{{Vermeulen}, R.C.},
  \bibinfo{author}{{Readhead}, A.C.S.}, \bibinfo{author}{{Pearson}, T.J.},
  \bibinfo{author}{{Henstock}, D.R.}, \bibinfo{author}{{Wilkinson}, P.N.},
  \bibinfo{year}{1996}.
\newblock \bibinfo{title}{{A Complete Flux-Density--limited VLBI Survey of 293
  Flat-Spectrum Radio Sources}}.
\newblock \bibinfo{journal}{\apjs} \bibinfo{volume}{107}, \bibinfo{pages}{37}.
\bibitem[{Taylor and Zavala(2010)}]{Taylor:2010ih}
\bibinfo{author}{Taylor, G.B.}, \bibinfo{author}{Zavala, R.},
  \bibinfo{year}{2010}.
\newblock \bibinfo{title}{{Are There Rotation Measure Gradients Across Active
  Galactic Nuclei Jets?}}
\newblock \bibinfo{journal}{ApJ} \bibinfo{volume}{722},
  \bibinfo{pages}{L183--L187}.
\bibitem[{{Tetarenko} et~al.(2017){Tetarenko}, {Sivakoff}, {Miller-Jones},
  {Rosolowsky}, {Petitpas}, {Gurwell}, {Wouterloot}, {Fender}, {Heinz},
  {Maitra}, {Markoff}, {Migliari}, {Rupen}, {Rushton}, {Russell}, {Russell} and
  {Sarazin}}]{tetarenko_17}
\bibinfo{author}{{Tetarenko}, A.J.}, \bibinfo{author}{{Sivakoff}, G.R.},
  \bibinfo{author}{{Miller-Jones}, J.C.A.}, \bibinfo{author}{{Rosolowsky},
  E.W.}, \bibinfo{author}{{Petitpas}, G.}, \bibinfo{author}{{Gurwell}, M.},
  \bibinfo{author}{{Wouterloot}, J.}, \bibinfo{author}{{Fender}, R.},
  \bibinfo{author}{{Heinz}, S.}, \bibinfo{author}{{Maitra}, D.},
  \bibinfo{author}{{Markoff}, S.B.}, \bibinfo{author}{{Migliari}, S.},
  \bibinfo{author}{{Rupen}, M.P.}, \bibinfo{author}{{Rushton}, A.P.},
  \bibinfo{author}{{Russell}, D.M.}, \bibinfo{author}{{Russell}, T.D.},
  \bibinfo{author}{{Sarazin}, C.L.}, \bibinfo{year}{2017}.
\newblock \bibinfo{title}{{Extreme jet ejections from the black hole X-ray
  binary V404 Cygni}}.
\newblock \bibinfo{journal}{\mnras} \bibinfo{volume}{469},
  \bibinfo{pages}{3141--3162}.
\newblock \eprint{1704.08726}.
\bibitem[{{Thompson} et~al.(2007){Thompson}, {Moran} and {Swenson}}]{TMS}
\bibinfo{author}{{Thompson}, A.R.}, \bibinfo{author}{{Moran}, J.M.},
  \bibinfo{author}{{Swenson}, G.W.}, \bibinfo{year}{2007}.
\newblock \bibinfo{title}{{Interferometry and Synthesis in Radio Astronomy,
  John Wiley and Sons, 2007.}}
\bibitem[{{Tilanus} et~al.(2014){Tilanus}, {Krichbaum}, {Zensus}, {Baudry},
  {Bremer}, {Falcke}, {Giovannini}, {Laing}, {van Langevelde}, {Vlemmings},
  {Abraham}, {Afonso}, {Agudo}, {Alberdi}, {Alcolea}, {Altamirano}, {Asadi},
  {Assaf}, {Augusto}, {Baczko}, {Boeck}, {Boller}, {Bondi}, {Boone}, {Bourda},
  {Brajsa}, {Brand}, {Britzen}, {Bujarrabal}, {Cales}, {Casadio}, {Casasola},
  {Castangia}, {Cernicharo}, {Charlot}, {Chemin}, {Clenet}, {Colomer},
  {Combes}, {Cordes}, {Coriat}, {Cross}, {D'Ammando}, {Dallacasa}, {Desmurs},
  {Eatough}, {Eckart}, {Eisenacher}, {Etoka}, {Felix}, {Fender}, {Ferreira},
  {Freeland}, {Frey}, {Fromm}, {Fuhrmann}, {Gabanyi}, {Galvan-Madrid},
  {Giroletti}, {Goddi}, {Gomez}, {Gourgoulhon}, {Gray}, {di Gregorio},
  {Greimel}, {Grosso}, {Guirado}, {Hada}, {Hanslmeier}, {Henkel}, {Herpin},
  {Hess}, {Hodgson}, {Horns}, {Humphreys}, {Hutawarakorn Kramer}, {Ilyushin},
  {Impellizzeri}, {Ivanov}, {Juli{\~a}o}, {Kadler}, {Kerins}, {Klaassen}, {van
  't Klooster}, {Kording}, {Kozlov}, {Kramer}, {Kreikenbohm}, {Kurtanidze},
  {Lazio}, {Leite}, {Leitzinger}, {Lepine}, {Levshakov}, {Lico}, {Lindqvist},
  {Liuzzo}, {Lobanov}, {Lucas}, {Mannheim}, {Marcaide}, {Markoff},
  {Mart{\'{\i}}-Vidal}, {Martins}, {Masetti}, {Massardi}, {Menten}, {Messias},
  {Migliari}, {Mignano}, {Miller-Jones}, {Minniti}, {Molaro}, {Molina},
  {Monteiro}, {Moscadelli}, {Mueller}, {M{\"u}ller}, {Muller}, {Niederhofer},
  {Odert}, {Olofsson}, {Orienti}, {Paladino}, {Panessa}, {Paragi}, {Paumard},
  {Pedrosa}, {P{\'e}rez-Torres}, {Perrin}, {Perucho}, {Porquet}, {Prandoni},
  {Ransom}, {Reimers}, {Rejkuba}, {Rezzolla}, {Richards}, {Ros}, {Roy},
  {Rushton}, {Savolainen}, {Schulz}, {Silva}, {Sivakoff}, {Soria-Ruiz},
  {Soria}, {Spaans}, {Spencer}, {Stappers}, {Surcis}, {Tarchi}, {Temmer},
  {Thompson}, {Torrelles}, {Truestedt}, {Tudose}, {Venturi}, {Verbiest},
  {Vieira}, {Vielzeuf}, {Vincent}, {Wex}, {Wiik}, {Wiklind}, {Wilms},
  {Zackrisson} and {Zechlin}}]{tilanus_14}
\bibinfo{author}{{Tilanus}, R.P.J.}, \bibinfo{author}{{Krichbaum}, T.P.},
  \bibinfo{author}{{Zensus}, J.A.}, \bibinfo{author}{{Baudry}, A.},
  \bibinfo{author}{{Bremer}, M.}, \bibinfo{author}{{Falcke}, H.},
  \bibinfo{author}{{Giovannini}, G.}, \bibinfo{author}{{Laing}, R.},
  \bibinfo{author}{{van Langevelde}, H.J.}, \bibinfo{author}{{Vlemmings}, W.},
  \bibinfo{author}{{Abraham}, Z.}, \bibinfo{author}{{Afonso}, J.},
  \bibinfo{author}{{Agudo}, I.}, \bibinfo{author}{{Alberdi}, A.},
  \bibinfo{author}{{Alcolea}, J.}, \bibinfo{author}{{Altamirano}, D.},
  \bibinfo{author}{{Asadi}, S.}, \bibinfo{author}{{Assaf}, K.},
  \bibinfo{author}{{Augusto}, P.}, \bibinfo{author}{{Baczko}, A.},
  \bibinfo{author}{{Boeck}, M.}, \bibinfo{author}{{Boller}, T.},
  \bibinfo{author}{{Bondi}, M.}, \bibinfo{author}{{Boone}, F.},
  \bibinfo{author}{{Bourda}, G.}, \bibinfo{author}{{Brajsa}, R.},
  \bibinfo{author}{{Brand}, J.}, \bibinfo{author}{{Britzen}, S.},
  \bibinfo{author}{{Bujarrabal}, V.}, \bibinfo{author}{{Cales}, S.},
  \bibinfo{author}{{Casadio}, C.}, \bibinfo{author}{{Casasola}, V.},
  \bibinfo{author}{{Castangia}, P.}, \bibinfo{author}{{Cernicharo}, J.},
  \bibinfo{author}{{Charlot}, P.}, \bibinfo{author}{{Chemin}, L.},
  \bibinfo{author}{{Clenet}, Y.}, \bibinfo{author}{{Colomer}, F.},
  \bibinfo{author}{{Combes}, F.}, \bibinfo{author}{{Cordes}, J.},
  \bibinfo{author}{{Coriat}, M.}, \bibinfo{author}{{Cross}, N.},
  \bibinfo{author}{{D'Ammando}, F.}, \bibinfo{author}{{Dallacasa}, D.},
  \bibinfo{author}{{Desmurs}, J.}, \bibinfo{author}{{Eatough}, R.},
  \bibinfo{author}{{Eckart}, A.}, \bibinfo{author}{{Eisenacher}, D.},
  \bibinfo{author}{{Etoka}, S.}, \bibinfo{author}{{Felix}, M.},
  \bibinfo{author}{{Fender}, R.}, \bibinfo{author}{{Ferreira}, M.},
  \bibinfo{author}{{Freeland}, E.}, \bibinfo{author}{{Frey}, S.},
  \bibinfo{author}{{Fromm}, C.}, \bibinfo{author}{{Fuhrmann}, L.},
  \bibinfo{author}{{Gabanyi}, K.}, \bibinfo{author}{{Galvan-Madrid}, R.},
  \bibinfo{author}{{Giroletti}, M.}, \bibinfo{author}{{Goddi}, C.},
  \bibinfo{author}{{Gomez}, J.}, \bibinfo{author}{{Gourgoulhon}, E.},
  \bibinfo{author}{{Gray}, M.}, \bibinfo{author}{{di Gregorio}, I.},
  \bibinfo{author}{{Greimel}, R.}, \bibinfo{author}{{Grosso}, N.},
  \bibinfo{author}{{Guirado}, J.}, \bibinfo{author}{{Hada}, K.},
  \bibinfo{author}{{Hanslmeier}, A.}, \bibinfo{author}{{Henkel}, C.},
  \bibinfo{author}{{Herpin}, F.}, \bibinfo{author}{{Hess}, P.},
  \bibinfo{author}{{Hodgson}, J.}, \bibinfo{author}{{Horns}, D.},
  \bibinfo{author}{{Humphreys}, E.}, \bibinfo{author}{{Hutawarakorn Kramer},
  B.}, \bibinfo{author}{{Ilyushin}, V.}, \bibinfo{author}{{Impellizzeri}, V.},
  \bibinfo{author}{{Ivanov}, V.}, \bibinfo{author}{{Juli{\~a}o}, M.},
  \bibinfo{author}{{Kadler}, M.}, \bibinfo{author}{{Kerins}, E.},
  \bibinfo{author}{{Klaassen}, P.}, \bibinfo{author}{{van 't Klooster}, K.},
  \bibinfo{author}{{Kording}, E.}, \bibinfo{author}{{Kozlov}, M.},
  \bibinfo{author}{{Kramer}, M.}, \bibinfo{author}{{Kreikenbohm}, A.},
  \bibinfo{author}{{Kurtanidze}, O.}, \bibinfo{author}{{Lazio}, J.},
  \bibinfo{author}{{Leite}, A.}, \bibinfo{author}{{Leitzinger}, M.},
  \bibinfo{author}{{Lepine}, J.}, \bibinfo{author}{{Levshakov}, S.},
  \bibinfo{author}{{Lico}, R.}, \bibinfo{author}{{Lindqvist}, M.},
  \bibinfo{author}{{Liuzzo}, E.}, \bibinfo{author}{{Lobanov}, A.},
  \bibinfo{author}{{Lucas}, P.}, \bibinfo{author}{{Mannheim}, K.},
  \bibinfo{author}{{Marcaide}, J.}, \bibinfo{author}{{Markoff}, S.},
  \bibinfo{author}{{Mart{\'{\i}}-Vidal}, I.}, \bibinfo{author}{{Martins}, C.},
  \bibinfo{author}{{Masetti}, N.}, \bibinfo{author}{{Massardi}, M.},
  \bibinfo{author}{{Menten}, K.}, \bibinfo{author}{{Messias}, H.},
  \bibinfo{author}{{Migliari}, S.}, \bibinfo{author}{{Mignano}, A.},
  \bibinfo{author}{{Miller-Jones}, J.}, \bibinfo{author}{{Minniti}, D.},
  \bibinfo{author}{{Molaro}, P.}, \bibinfo{author}{{Molina}, S.},
  \bibinfo{author}{{Monteiro}, A.}, \bibinfo{author}{{Moscadelli}, L.},
  \bibinfo{author}{{Mueller}, C.}, \bibinfo{author}{{M{\"u}ller}, A.},
  \bibinfo{author}{{Muller}, S.}, \bibinfo{author}{{Niederhofer}, F.},
  \bibinfo{author}{{Odert}, P.}, \bibinfo{author}{{Olofsson}, H.},
  \bibinfo{author}{{Orienti}, M.}, \bibinfo{author}{{Paladino}, R.},
  \bibinfo{author}{{Panessa}, F.}, \bibinfo{author}{{Paragi}, Z.},
  \bibinfo{author}{{Paumard}, T.}, \bibinfo{author}{{Pedrosa}, P.},
  \bibinfo{author}{{P{\'e}rez-Torres}, M.}, \bibinfo{author}{{Perrin}, G.},
  \bibinfo{author}{{Perucho}, M.}, \bibinfo{author}{{Porquet}, D.},
  \bibinfo{author}{{Prandoni}, I.}, \bibinfo{author}{{Ransom}, S.},
  \bibinfo{author}{{Reimers}, D.}, \bibinfo{author}{{Rejkuba}, M.},
  \bibinfo{author}{{Rezzolla}, L.}, \bibinfo{author}{{Richards}, A.},
  \bibinfo{author}{{Ros}, E.}, \bibinfo{author}{{Roy}, A.},
  \bibinfo{author}{{Rushton}, A.}, \bibinfo{author}{{Savolainen}, T.},
  \bibinfo{author}{{Schulz}, R.}, \bibinfo{author}{{Silva}, M.},
  \bibinfo{author}{{Sivakoff}, G.}, \bibinfo{author}{{Soria-Ruiz}, R.},
  \bibinfo{author}{{Soria}, R.}, \bibinfo{author}{{Spaans}, M.},
  \bibinfo{author}{{Spencer}, R.}, \bibinfo{author}{{Stappers}, B.},
  \bibinfo{author}{{Surcis}, G.}, \bibinfo{author}{{Tarchi}, A.},
  \bibinfo{author}{{Temmer}, M.}, \bibinfo{author}{{Thompson}, M.},
  \bibinfo{author}{{Torrelles}, J.}, \bibinfo{author}{{Truestedt}, J.},
  \bibinfo{author}{{Tudose}, V.}, \bibinfo{author}{{Venturi}, T.},
  \bibinfo{author}{{Verbiest}, J.}, \bibinfo{author}{{Vieira}, J.},
  \bibinfo{author}{{Vielzeuf}, P.}, \bibinfo{author}{{Vincent}, F.},
  \bibinfo{author}{{Wex}, N.}, \bibinfo{author}{{Wiik}, K.},
  \bibinfo{author}{{Wiklind}, T.}, \bibinfo{author}{{Wilms}, J.},
  \bibinfo{author}{{Zackrisson}, E.}, \bibinfo{author}{{Zechlin}, H.},
  \bibinfo{year}{2014}.
\newblock \bibinfo{title}{{Future mmVLBI Research with ALMA: A European
  vision}}.
\newblock \bibinfo{journal}{ArXiv e-prints} \eprint{1406.4650}.
\bibitem[{{Tingay} et~al.(1995){Tingay}, {Jauncey}, {Preston}, {Reynolds},
  {Meier}, {Murphy}, {Tzioumis}, {McKay}, {Kesteven}, {Lovell},
  {Campbell-Wilson}, {Ellingsen}, {Gough}, {Hunstead}, {Jonos}, {McCulloch},
  {Migenes}, {Quick}, {Sinclair} and {Smits}}]{tingay_95}
\bibinfo{author}{{Tingay}, S.J.}, \bibinfo{author}{{Jauncey}, D.L.},
  \bibinfo{author}{{Preston}, R.A.}, \bibinfo{author}{{Reynolds}, J.E.},
  \bibinfo{author}{{Meier}, D.L.}, \bibinfo{author}{{Murphy}, D.W.},
  \bibinfo{author}{{Tzioumis}, A.K.}, \bibinfo{author}{{McKay}, D.J.},
  \bibinfo{author}{{Kesteven}, M.J.}, \bibinfo{author}{{Lovell}, J.E.J.},
  \bibinfo{author}{{Campbell-Wilson}, D.}, \bibinfo{author}{{Ellingsen}, S.P.},
  \bibinfo{author}{{Gough}, R.}, \bibinfo{author}{{Hunstead}, R.W.},
  \bibinfo{author}{{Jonos}, D.L.}, \bibinfo{author}{{McCulloch}, P.M.},
  \bibinfo{author}{{Migenes}, V.}, \bibinfo{author}{{Quick}, J.},
  \bibinfo{author}{{Sinclair}, M.W.}, \bibinfo{author}{{Smits}, D.},
  \bibinfo{year}{1995}.
\newblock \bibinfo{title}{{Relativistic motion in a nearby bright X-ray
  source}}.
\newblock \bibinfo{journal}{Nature} \bibinfo{volume}{374},
  \bibinfo{pages}{141--143}.
\bibitem[{{Vaidya} and {Goddi}(2013)}]{Vai13}
\bibinfo{author}{{Vaidya}, B.}, \bibinfo{author}{{Goddi}, C.},
  \bibinfo{year}{2013}.
\newblock \bibinfo{title}{{MHD modelling of a disc wind from a high-mass
  protobinary: the case of Orion Source I}}.
\newblock \bibinfo{journal}{\mnras} \bibinfo{volume}{429},
  \bibinfo{pages}{L50--L54}.
\newblock \eprint{arXiv:1210.7775}.
\bibitem[{{van Langevelde} et~al.(2005){van Langevelde}, {Pihlstr{\"o}m} and
  {Beasley}}]{2005Ap&SS.295..249V}
\bibinfo{author}{{van Langevelde}, H.J.}, \bibinfo{author}{{Pihlstr{\"o}m},
  Y.}, \bibinfo{author}{{Beasley}, A.}, \bibinfo{year}{2005}.
\newblock \bibinfo{title}{{Molecular Absorption in Cen a on Vlbi Scales}}.
\newblock \bibinfo{journal}{\apss} \bibinfo{volume}{295},
  \bibinfo{pages}{249--255}.
\newblock \eprint{astro-ph/0409147}.
\bibitem[{{van Leeuwen}(2007)}]{hip07}
\bibinfo{editor}{{van Leeuwen}, F.} (Ed.), \bibinfo{year}{2007}.
\newblock \bibinfo{title}{{Hipparcos, the New Reduction of the Raw Data}}.
  volume \bibinfo{volume}{350} of \textit{\bibinfo{series}{Astrophysics and
  Space Science Library}}.
\bibitem[{Walker et~al.(2000)Walker, Dhawan, Romney, Kellermann and
  Vermeulen}]{Walker:2000gb}
\bibinfo{author}{Walker, R.C.}, \bibinfo{author}{Dhawan, V.},
  \bibinfo{author}{Romney, J.D.}, \bibinfo{author}{Kellermann, K.I.},
  \bibinfo{author}{Vermeulen, R.C.}, \bibinfo{year}{2000}.
\newblock \bibinfo{title}{{VLBA Absorption Imaging of Ionized Gas Associated
  with the Accretion Disk in NGC 1275}}.
\newblock \bibinfo{journal}{ApJ} \bibinfo{volume}{530},
  \bibinfo{pages}{233--244}.
\bibitem[{{Wolfire} and {Cassinelli}(1987)}]{Wol87}
\bibinfo{author}{{Wolfire}, M.G.}, \bibinfo{author}{{Cassinelli}, J.P.},
  \bibinfo{year}{1987}.
\newblock \bibinfo{title}{{Conditions for the formation of massive stars}}.
\newblock \bibinfo{journal}{\apj} \bibinfo{volume}{319},
  \bibinfo{pages}{850--867}.
\bibitem[{{Wrobel}(1993)}]{1993AJ....106..444W}
\bibinfo{author}{{Wrobel}, J.M.}, \bibinfo{year}{1993}.
\newblock \bibinfo{title}{{Faraday rotation measures and intrinsic polarization
  position angles of very long baseline interferometry core-jet sources}}.
\newblock \bibinfo{journal}{\aj} \bibinfo{volume}{106},
  \bibinfo{pages}{444--454}.
\bibitem[{{Yoon} et~al.(2017){Yoon}, {Cho}, {Yun}, Choi, Dodson, Rioja, Kim,
  Kim, Yang, H. and Byun}]{cho_17}
\bibinfo{author}{{Yoon}, D.}, \bibinfo{author}{{Cho}, S.},
  \bibinfo{author}{{Yun}, Y.}, \bibinfo{author}{Choi, Y.},
  \bibinfo{author}{Dodson, R.}, \bibinfo{author}{Rioja, M.},
  \bibinfo{author}{Kim, J.}, \bibinfo{author}{Kim, D.}, \bibinfo{author}{Yang,
  H.}, \bibinfo{author}{H., I.}, \bibinfo{author}{Byun, D.},
  \bibinfo{year}{2017}.
\newblock \bibinfo{title}{{Simultaneous VLBI observations of H2O and SiO masers
  toward VX Sagittarii using Korean VLBI Network}}.
\newblock \bibinfo{journal}{Submitted to Nature} .
\bibitem[{{Yun} et~al.(2016){Yun}, {Cho}, {Imai}, {Kim}, {Asaki}, {Chibueze},
  {Choi}, {Dodson}, {Kim}, {Kusuno}, {Matsumoto}, {Min}, {Oyadomari}, {Rioja},
  {Yoon}, {Byun}, {Chung}, {Chung}, {Hagiwara}, {Han}, {Han}, {Hirota},
  {Honma}, {Hwang}, {Je}, {Jike}, {Jung}, {Jung}, {Kang}, {Kang}, {Kang},
  {Kan-ya}, {Kanaguchi}, {Kawaguchi}, {Kim}, {Ryoung Kim}, {Kim}, {Kim}, {Kim},
  {Kim}, {Kobayashi}, {Kono}, {Kurayama}, {Lee}, {Lee}, {Lee}, {Lee}, {Lee},
  {Lee}, {Lyo}, {Minh}, {Oh}, {Oh}, {Oyama}, {Roh}, {Sawada-Satoh}, {Shibata},
  {Sohn}, {Song}, {Tamura}, {Wi} and {Yeom}}]{yun_16}
\bibinfo{author}{{Yun}, Y.}, \bibinfo{author}{{Cho}, S.H.},
  \bibinfo{author}{{Imai}, H.}, \bibinfo{author}{{Kim}, J.},
  \bibinfo{author}{{Asaki}, Y.}, \bibinfo{author}{{Chibueze}, J.O.},
  \bibinfo{author}{{Choi}, Y.K.}, \bibinfo{author}{{Dodson}, R.},
  \bibinfo{author}{{Kim}, D.J.}, \bibinfo{author}{{Kusuno}, K.},
  \bibinfo{author}{{Matsumoto}, N.}, \bibinfo{author}{{Min}, C.},
  \bibinfo{author}{{Oyadomari}, M.}, \bibinfo{author}{{Rioja}, M.J.},
  \bibinfo{author}{{Yoon}, D.H.}, \bibinfo{author}{{Byun}, D.Y.},
  \bibinfo{author}{{Chung}, H.}, \bibinfo{author}{{Chung}, M.H.},
  \bibinfo{author}{{Hagiwara}, Y.}, \bibinfo{author}{{Han}, M.H.},
  \bibinfo{author}{{Han}, S.T.}, \bibinfo{author}{{Hirota}, T.},
  \bibinfo{author}{{Honma}, M.}, \bibinfo{author}{{Hwang}, J.W.},
  \bibinfo{author}{{Je}, D.H.}, \bibinfo{author}{{Jike}, T.},
  \bibinfo{author}{{Jung}, D.K.}, \bibinfo{author}{{Jung}, T.},
  \bibinfo{author}{{Kang}, J.H.}, \bibinfo{author}{{Kang}, J.},
  \bibinfo{author}{{Kang}, Y.W.}, \bibinfo{author}{{Kan-ya}, Y.},
  \bibinfo{author}{{Kanaguchi}, M.}, \bibinfo{author}{{Kawaguchi}, N.},
  \bibinfo{author}{{Kim}, B.G.}, \bibinfo{author}{{Ryoung Kim}, H.},
  \bibinfo{author}{{Kim}, H.G.}, \bibinfo{author}{{Kim}, J.},
  \bibinfo{author}{{Kim}, K.T.}, \bibinfo{author}{{Kim}, M.},
  \bibinfo{author}{{Kobayashi}, H.}, \bibinfo{author}{{Kono}, Y.},
  \bibinfo{author}{{Kurayama}, T.}, \bibinfo{author}{{Lee}, C.},
  \bibinfo{author}{{Lee}, J.}, \bibinfo{author}{{Lee}, J.A.},
  \bibinfo{author}{{Lee}, J.W.}, \bibinfo{author}{{Lee}, S.H.},
  \bibinfo{author}{{Lee}, S.S.}, \bibinfo{author}{{Lyo}, A.R.},
  \bibinfo{author}{{Minh}, Y.C.}, \bibinfo{author}{{Oh}, C.},
  \bibinfo{author}{{Oh}, S.J.}, \bibinfo{author}{{Oyama}, T.},
  \bibinfo{author}{{Roh}, D.G.}, \bibinfo{author}{{Sawada-Satoh}, S.},
  \bibinfo{author}{{Shibata}, K.M.}, \bibinfo{author}{{Sohn}, B.W.},
  \bibinfo{author}{{Song}, M.G.}, \bibinfo{author}{{Tamura}, Y.},
  \bibinfo{author}{{Wi}, S.O.}, \bibinfo{author}{{Yeom}, J.H.},
  \bibinfo{year}{2016}.
\newblock \bibinfo{title}{{SiO Masers around WX Psc Mapped with the KVN and
  VERA Array (KaVA)}}.
\newblock \bibinfo{journal}{\apj} \bibinfo{volume}{822}, \bibinfo{pages}{3}.
\bibitem[{Zamaninasab et~al.(2014)Zamaninasab, Clausen-Brown, Savolainen and
  Tchekhovskoy}]{Zamaninasab:2014hs}
\bibinfo{author}{Zamaninasab, M.}, \bibinfo{author}{Clausen-Brown, E.},
  \bibinfo{author}{Savolainen, T.}, \bibinfo{author}{Tchekhovskoy, A.},
  \bibinfo{year}{2014}.
\newblock \bibinfo{title}{{Dynamically important magnetic fields near accreting
  supermassive black holes}}.
\newblock \bibinfo{journal}{Nature} \bibinfo{volume}{510},
  \bibinfo{pages}{126--128}.
\bibitem[{Zavala and Taylor(2004)}]{Zavala:2004ga}
\bibinfo{author}{Zavala, R.T.}, \bibinfo{author}{Taylor, G.B.},
  \bibinfo{year}{2004}.
\newblock \bibinfo{title}{{A View through Faraday's Fog. II. Parsec-Scale
  Rotation Measures in 40 Active Galactic Nuclei}}.
\newblock \bibinfo{journal}{ApJ} \bibinfo{volume}{612},
  \bibinfo{pages}{749--779}.
\bibitem[{Zavala and Taylor(2005)}]{Zavala:2005im}
\bibinfo{author}{Zavala, R.T.}, \bibinfo{author}{Taylor, G.B.},
  \bibinfo{year}{2005}.
\newblock \bibinfo{title}{{Faraday Rotation Measure Gradients from a Helical
  Magnetic Field in 3C 273}}.
\newblock \bibinfo{journal}{ApJ} \bibinfo{volume}{626},
  \bibinfo{pages}{L73--L76}.

\end{thebibliography}
\end{document}